\documentclass[reqno,11pt]{amsart} 
\usepackage{amsmath}
\usepackage{amssymb}
\usepackage{amsthm}
\usepackage{verbatim}
\usepackage{xcolor}
\usepackage{graphics}

\newcommand\NoBlackBoxes{\global\overfullrule0pt}

\NoBlackBoxes
\theoremstyle{plain}

\def\E{{\mathbb E}}
\def\R{{\mathbb R}}
\def\C{{\mathbb C}}
\def\P{{\mathbb P}}
\def\Z{{\mathbb Z}}

\def\var{{\rm Var}}
\def\Var{{\rm Var}}

\def\Var{{\rm Var}}

\def\ep{\varepsilon}
\def\phi{\varphi}

\def\bee{\begin{eqnarray*}}
\def\ene{\end{eqnarray*}}
\def\be{\begin{equation}}
\def\en{\end{equation}}
	
\begin{document}

	
\title{R\'enyi Divergences in Central Limit \\ 
	Theorems: Old and New}

\author{S. G. Bobkov$^{1}$}
\thanks{1) 
School of Mathematics, University of Minnesota, Minneapolis, MN, USA,
bobkov@math.umn.edu. 
}

\author{G. P. Chistyakov$^{2}$}
\thanks{2) Faculty of Mathematics,
Bielefeld University, Germany,
chistyak@math-uni.bielefeld.de.
}

\author{F. G\"otze$^{2}$}
\thanks{2) Faculty of Mathematics,
Bielefeld University, Germany,
goetze@math-uni.bielefeld.de.
}

\subjclass[2010]
{Primary 60E, 60F} 
\keywords{Central limit theorem, R\'enyi divergence, subgaussian distributions}

\begin{abstract}
We give an overview of various results and methods related to 
information-theoretic distances of R\'enyi type in the light of their 
applications to the central limit theorem (CLT).  The first part 
(Sections 1-9) is devoted to the total variation and the Kullback-Leibler
distance (relative entropy).  In the second part (Sections 10-15) we discuss
general properties of R\'enyi and Tsallis divergences of order $\alpha>1$,
and then in the third part (Sections 16-21) we turn to the CLT and 
non-uniform local limit theorems with respect to these strong distances. 
In the fourth part (Sections 22-31), we discuss recent results on 
strictly subgaussian distributions and describe necessary and sufficient 
conditions which ensure the validity 
of the CLT with respect to the R\'enyi divergence of infinite order. 
\end{abstract}

\maketitle
\markboth{S. G. Bobkov  G.P. Chistyakov  F. G\"otze}{R\'enyi Divergences and CLT}


\thispagestyle{empty}

{\sl Contents}:

\vskip2mm
{\small
PART I

1.\, R\'enyi and Tsallis divergences. Basic definitions

2.\, Central limit theorem in total variation

3.\, Relative entropy with respect to normal laws

4.\, Bounds for relative entropy via other distances

5.\, Convexity and monotonicity along convolutions

6.\, Entropic central limit theorem and Orlicz spaces

7.\, Rates of convergence in the entropic CLT

8.\, Berry-Esseen bounds for total variation

9.\, Berry-Esseen bounds for relative entropy

\vskip2mm
PART II

10. R\'enyi and Tsallis divergences with respect to the normal law

11. Pearson's $\chi^2$-distance to the normal law

12. Exponential series and normal moments

13. Behavior of R\'enyi divergence under convolutions

14. Examples of Convolutions

15. Superadditivity of $\chi^2$ with respect to marginals

\vskip2mm
PART III

16. Edgeworth expansion for densities and truncated distances

17. Edgeworth expansion and CLT for R\'enyi divergences

18. Non-uniform local limit theorems

19. Comments on the proofs

20. Some examples and counter-examples

21. Sufficient Conditions for Convergence in $D_\alpha$

\vskip2mm
PART IV

22. Strictly subgaussian distributions

23. Zeros of characteristic functions

24. Examples of strictly subgaussian distributions

25. Laplace transforms with periodic components

26. Richter's theorem and its refinement

27. CLT in $D_\infty$ with rate of convergence

28. Action of Esscher transform on convolutions

29. Necessary and sufficient conditions

30. Characterization in the periodic case

31. The multidimensional case
}

\vskip5mm
\section{{\bf R\'enyi and Tsallis Divergences. Basic Definitions}}
\setcounter{equation}{0}

\noindent
Representing strong information-theoretical directional distances without 
the symmetry property, R\'enyi's divergences allow one to effectively explore
various approximation problems in Probability and Statistics
(not to mention Information Theory). They are defined in the most 
abstract setting and do not require any topological structure.
Let us start with basic notations and general relations.

Let $(\Omega,\mathfrak F)$ be a measurable space.
Given random elements $X$ and $Y$ in $\Omega$ with distributions
$P$ and $Q$ respectively, pick up a $\sigma$-finite measure $\mu$ on 
$(\Omega,\mathfrak F)$ such that $P$ and $Q$ are absolutely continuous
with respect to $\mu$ and have densities 
$$
p = \frac{dP}{d\mu},\quad q = \frac{dQ}{d\mu}.
$$
Given a parameter $\alpha>0$, $\alpha \neq 1$, the R\'enyi's divergence
of $P$ from $Q$ of order/index $\alpha$, called also the relative 
$\alpha$-entropy, is then defined by
\be
D_\alpha(X||Y) = D_\alpha(P||Q) = \frac{1}{\alpha - 1} \log 
\int \Big(\frac{p}{q}\Big)^\alpha\,q\, d\mu.
\en

This quantity is determined by the couple $(P,Q)$ only and
does not depend on the choice of the measure $\mu$
(one may take $\mu = P+Q$, for example).
A closely related functional with a similar property is the Tsallis distance
\be
T_\alpha(X||Y) = T_\alpha(P||Q) = \frac{1}{\alpha - 1} 
\bigg[\int \Big(\frac{p}{q}\Big)^\alpha\,q\, d\mu - 1\bigg].
\en

Clearly, $0 \leq D_\alpha \leq \infty$, and
$D_\alpha = 0$ if and only if $P=Q$, and similarly for $T_\alpha$. Since 
$$
T_\alpha = \frac{1}{\alpha - 1}\,\big[\exp\{(\alpha-1)\,D_\alpha\} - 1\big],
$$
we have $D_\alpha \leq T_\alpha$, and moreover --
both distances are of a similar order, when they are small.
Hence, approximation problems in $D_\alpha$ and $T_\alpha$
are equivalent.

For the range $0 < \alpha < 1$, the right-hand sides in (1.1)-(1.2)
are finite.  In this case, $D_\alpha$ and $T_\alpha$ are topologically 
equivalent to the total variation distance 
$$
\|P-Q\|_{\rm TV} = \int |p-q|\,d\mu
$$ 
between $P$ and $Q$, which may be seen from 
\be
\frac{\alpha}{2}\,\|P-Q\|_{\rm TV}^2 \leq D_\alpha(P||Q) \leq
\frac{1}{1 - \alpha}\,\|P-Q\|_{\rm TV}.
\en
Here, the upper bound is elementary, while the lower bound represents 
an extended Pinsker-type inequality due to Gilardoni \cite{G}.

The specific value $\alpha = \frac{1}{2}$ leads in (1.2)
to a function of the Hellinger metric $H(P,Q)$, namely
$$
H^2(P,Q) = \frac{1}{2} \int (\sqrt{p} - \sqrt{q})^2\,d\mu = 
\frac{1}{2}\,T_{1/2}(P||Q).
$$
It is symmetric in $(P,Q)$ and satisfies $0 \leq H(P,Q) \leq 1$.

The functions $\alpha \rightarrow D_\alpha$ and $\alpha \rightarrow T_\alpha$
are non-decreasing, so that one may naturally define these distances
for the value $\alpha=1$, by taking the limits
$D_1 = \lim_{\alpha \uparrow 1} D_\alpha$ and 
$T_1 = \lim_{\alpha \uparrow 1} T_\alpha$. In fact, $T_1 = D_1 = D$, where
\be
D(X||Y) = D(P||Q) = \int p\log \frac{p}{q}\, d\mu
\en
is the classical information divergence, also called the relative entropy, 
or Kullback-Leibler distance. For the finiteness of $D(P||Q)$, it is necessary,
although not sufficient in general, that $P$ is absolutely continuous with 
respect to $Q$. The latter is equivalent to the implication
$p(x) = 0 \Rightarrow q(x) = 0$ for $\mu$-almost all $x \in \Omega$.

Anyhow, (1.3) is extended to the value $\alpha=1$ by monotonicity, 
which yields the Pinsker-type inequality due to Csisc\'ar \cite{Cs1} and Kullback \cite{K}
$$
D(P||Q) \geq \frac{1}{2}\,\|P-Q\|_{\rm TV}^2.
$$
It may be strengthened in terms of weighted total variation 
distances. As was shown by Bolley and Villani \cite{B-V}, 
for any measurable function $w \geq 0$ on $\Omega$,
\be
\Big(\int |p-q|\,w\,d\mu\Big)^2 \leq c\, D(P||Q)
\en
with constant
$$
c = c_Q(w) = 2\, \Big(1 + \log \int e^{w^2} q\,d\mu\Big).
$$
When $w=1$, (1.5) yields the Pinsker-type inequality with 
an additional factor~2.

The orders/indexes $\alpha>1$ lead to much stronger R\'enyi/Tsallis distances,
that are defined by (1.1)-(1.2) with the assumption that the probability
measure $P$ is absolutely continuous with respect to $Q$; otherwise 
$D_\alpha(X||Y) = T_\alpha(X||Y) = \infty$. For example, in the
particular case $\alpha = 2$, we obtain the Pearson $\chi^2$-distance
$$
T_2(X||Y) = \chi^2(X,Y) = \int \frac{(p-q)^2}{q}\, d\mu.
$$

Since the monotonicity of the functions $D_\alpha$ and $T_\alpha$
continues to hold in the region $\alpha>1$, one may define 
the R\'enyi divergence of infinite order 
$D_\infty = \lim_{\alpha \uparrow \infty} D_\alpha$. It is easy to see that
\be
D_\infty(X||Y) = \log \,{\rm ess\,sup}\, \frac{p}{q}, 
\en
where the essential supremum is understood with respect to the 
measure $\mu$.

However, $\lim_{\alpha \uparrow \infty} T_\alpha(X||Y) = \infty$ when 
$P \neq Q$. Nevertheless, analogously to (1.1)-(1.2), it makes sense 
to consider the quantity
\be
T_\infty(X||Y) = {\rm ess\,sup}\, \frac{p-q}{q}
\en
and call it the Tsallis distance of $P$ from $Q$ of infinite order. Note 
that $T_\infty = \exp(D_\infty) - 1$, so, both distances are of a similar 
order, when they are small, and we still have $D_\infty \leq T_\infty$.

Similarly to the total variation, all R\'enyi and Tsallis distances 
satisfy the following contractivity property: If a map 
$S:\Omega \rightarrow \Omega$ is measurable, then for the images 
(distributions) $P_S = PS^{-1}$ and $Q_S = QS^{-1}$, we have
\be
D_\alpha(P_S||Q_S) \leq D_\alpha(P||Q), \quad
T_\alpha(P_S||Q_S) \leq T_\alpha(P||Q).
\en
Hence, these distances are invariant under isomorphisms
of measurable spaces: If $S$ is bijective and measurable 
together with its inverse $S^{-1}$, then
$$
D_\alpha(P_S||Q_S) = D_\alpha(P||Q), \quad
T_\alpha(P_S||Q_S) = T_\alpha(P||Q).
$$

The $\chi^2$-distance may also be regarded as a particular member 
of the family of Pearson-Vajda distances
$$
\chi_\alpha(X,Z) = \chi_\alpha(P,Q) = 
\int \Big|\frac{p-q}{q}\Big|^\alpha q\, d\mu, \quad \alpha \geq 1.
$$
Again, this quantity
does not depend on the choice of the dominating measure $\mu$.
The function $\chi_\alpha^{1/\alpha}$ is non-decreasing in $\alpha$, 
and when $\alpha = 1$, we arrive at the total variation distance between 
$P$ and $Q$. The distances $T_\alpha = T_\alpha(P||Q)$ and 
$\chi_\alpha = \chi_\alpha(P||Q)$ are metrically equivalent.
Namely, if $\alpha > 1$, we have
\be
T_\alpha \, \leq \, \frac{1}{\alpha - 1}\, 
\Big[\big(1 + \chi_\alpha^{1/\alpha}\big)^\alpha - 1\Big],
\en
and conversely,
\be
T_\alpha \, \geq \, \frac{3}{16}\,\min\{\chi_\alpha,\chi_\alpha^{2/\alpha}\} \ \ 
(1 < \alpha \leq 2), \ \quad \ 
T_\alpha\, \geq \, \alpha\, 3^{-\alpha}\, \chi_\alpha \ \ (\alpha \geq 2).
\en

For various properties and applications of these distances, we refer 
an interested reader to \cite{LC}, \cite{Cs2}, \cite{S}, \cite{D-C-T},
\cite{VE-H} and \cite{B-C-G6}.

\vskip7mm
\section{{\bf Central Limit Theorem in Total Variation}}
\setcounter{equation}{0}

\noindent
In information-theoretic variants of the central limit theorems,
one chooses for $Q$ the standard Gaussian measure on the
Euclidean space $\Omega = \R^d$ equipped with its Borel $\sigma$-algebra 
$\mathfrak F$ and the Euclidean norm $|\cdot|$, thus with density
$$
\varphi(x) = \frac{dQ(x)}{dx} = \frac{1}{(2\pi)^{d/2}}\,e^{-|x|^2/2},
\quad x \in \R^d,
$$
with respect to the Lebesgue measure $\mu_d$ on $\R^d$.
In the sequel, we denote by $Z$ a standard normal random vector
in $\R^d$, hence distributed according to $Q = P_Z$.

Let us start with a model of i.i.d. (independent identically distributed) 
random vectors $(X_k)_{k \geq 1}$ in $\R^d$ with a common 
distribution $P$ having mean zero and identity covariance
matrix. According to the classical central limit theorem, the normalized sums
$$
Z_n = \frac{X_1 + \dots + X_n}{\sqrt{n}}
$$
are weakly convergent in distribution as $n \rightarrow \infty$
to the standard normal law $P_Z$ on $\R^d$, which is often written as
$Z_n \Rightarrow Z$. The weak convergence means that
$$
\E\,u(Z_n) \rightarrow \E\,u(Z)
$$
as $n \rightarrow \infty$ for any bounded continuous function $u$
on $\R^d$.

Whether or not there is a convergence
in a stronger sense, including information-theoretic distances,
depends on the common distribution $P$ like in the following:

\vskip5mm
{\bf Theorem 2.1.} {\sl For any fixed $0 < \alpha < 1$, we have
\be
D_\alpha(Z_n||Z) \rightarrow 0 \quad {\sl as} \ n \rightarrow \infty,
\en
if and only if, for some $n$, the distribution $P_{Z_n}$ of $Z_n$
has a non-zero absolutely continuous component with respect to the
Lebesgue measure on $\R^d$.
}

\vskip5mm
In particular, if $P$ has a density, then all $P_{Z_n}$ have densities,
and (2.1) holds.

Theorem 2.1 is a reformulation of a result by Prokhorov \cite{Pr}
which provides a similar characterization for the convergence
\be
\|P_{Z_n} - P_Z\|_{\rm TV} \rightarrow 0
\en
(recall that $D_\alpha$ and the total variation
distance are topologically equivalent, as emphasized in (1.3)). 
Thus, (2.2) holds true, if and only if
$$
\|P_{Z_n} - P_Z\|_{\rm TV} < 2 \quad {\rm for \ some} \ n.
$$
Note that the total variation distance may take values in the interval
$[0,2]$, and the maximal possible value $\|P_{Z_n} - P_Z\|_{\rm TV} = 2$
is attained if and only if the measures $P_{Z_n}$ and $P_Z$ are
orthogonal, that is, when $P_{Z_n}$ is supported on a set in $\R^d$
of Lebesgue measure zero.

If $P_{Z_n}$ have densities $p_n$ for all large $n$, the properties 
(2.1)-(2.2) may be stated as the convergence of densities 
in the space $L^1(\R^d)$, i.e.
\be
\int |p_n(x) - \varphi(x)|\,dx \rightarrow 0 \quad 
{\rm as} \ n \rightarrow \infty.
\en

For the proof, Prokhorov introduced the method of decomposition
of densities which proved to be useful in many further investigations
of the CLT for strong distances. In particular, with this method
Ranga Rao and Varadarajan \cite{RR-V} showed that, if $Z_n$ 
have densities $p_n$ for all large $n$, then necessarily
\be
p_n(x) \rightarrow \varphi(x)
\en 
almost everywhere. Hence, (2.1)-(2.2) appear as a consequence 
of this pointwise convergence by applying Scheffe's lemma.

The question of when $Z_n$ have some densities for all large $n$ 
in terms of $P$ or in terms of its characteristic function (Fourier-Stieltjes)
transform
$$
f(t) = \E\,e^{i\left<t,X_1\right>} = \int e^{i\left<t,x\right>}\,dP(x),
\quad t \in \R^d,
$$
is rather delicate and is open in general. On the other hand, there 
is a well-known simple integrability (called smoothness) condition
\be
\int |f(t)|^\nu\,dt < \infty \quad {\rm for \ some} \ \nu \geq 1,
\en
which is equivalent to the stronger property that $Z_n$ have bounded
(and actually continuous) densities $p_n$ for all $n$ large enough,
cf. \cite{B-RR}. Moreover, (2.5) is equivalent to the strengthened
variant of (2.3)-(2.4) in the form of the uniform local limit theorem
of Gnedenko,
\be
\sup_x |p_n(x) - \varphi(x)| \rightarrow 0
\en
as $n \rightarrow \infty$,
as well as to the convergence of densities in any $L^s$-space
\be
\|p_n - \varphi\|_s = 
\Big(\int |p_n(x) - \varphi(x)|^s\,dx\Big)^{1/s} \rightarrow 0
\en
with an arbitrary fixed power $s>1$ 
(cf. \cite{G-K}, \cite{Pe1}, \cite{Pe2}, \cite{B3}).

\vskip7mm
\section{{\bf Relative Entropy with Respect to Normal Laws}}
\setcounter{equation}{0}

\noindent
For the order $\alpha=1$, the whole theory aimed at the central limit 
theorem for $D_\alpha$ has many new interesting features which
originate from Information Theory. It has deep connections with other
fields, including, for example, the theory of optimal transport. Therefore, 
in a few next sections, 
we separately discuss the questions which formally have nothing
to do with the convergence problems. As before, $Z$ denotes 
a standard normal random vector in $\R^d$.

If $X$ is a random vector in $\R^d$, for the relative entropy 
$D(X||Z)$ to be finite, it is necessary that $X$ has a density $p$ 
with respect to the Lebesgue measure $\mu_d$. In this case,
choosing $\mu = \mu_d$, the definition (1.4) with $q = \varphi$ becomes
\be
D(X||Z) = \int p(x)\, \log\frac{p(x)}{\varphi(x)}\, dx.
\en
This functional is finite, if and only if $X$ has a finite 
second moment $\E\,|X|^2$ and finite Shannon differential entropy 
$$
h(X) = -\int p \log p\,dx
$$
(the latter integral is well-defined in the Lebesgue sense as long as 
$\E\,|X|^2 < \infty$, although it may take the value $-\infty$). 
A similar description is valid for the relative entropy $D(X||Z')$ 
with respect to any normal random vector $Z'$ having a density 
on $\R^d$. 

In this connection, it is natural to ask about the best approximation 
in $D$ over all normal laws, that is, about the $D$-distance of $P$ 
from the class of all normal distributions on $\R^d$,
\be
D(X) = \inf D(X||Z')
\en
with infimum over all $Z'$ as above. This infimum is 
attained when the means $a = \E X = \E Z'$ and covariance matrices 
$R = {\rm cov}(X) = {\rm cov}(Z')$ of $X$ and $Z'$ coincide. 
In this case, we also have a description by means of entropy via
\be
D(X) = D(X||Z') = h(Z') - h(X).
\en
This follows from the simple algebraic identity
\bee
D(X||Z) 
 & = &
D(X||Z') + \frac{1}{2}\,|a|^2 + 
\frac{1}{2}\,\big({\rm Tr}(R) - \log\, {\rm det}(R) - d\big) \\
 & = &
D(X||Z') + \frac{1}{2}\,|a|^2 + \frac{1}{2}\, \sum_{i=1}^d
\big(\lambda_i - \log \lambda_i - 1\big),
\ene
where $\lambda_i$ denote the eigenvalues of $R$, and ${\rm Tr}(R)$ 
and ${\rm det}(R)$ denote respectively the trace and the determinant
of this matrix.

Thus, if $X$ has mean $a$ and covariance matrix $R$, then
$$
D(X||Z) = D(X) + \frac{1}{2}\,|a|^2 + \frac{1}{2}\, \sum_{i=1}^d
\big(\lambda_i - \log \lambda_i - 1\big).
$$
In particular, the smallness of $D(X||Z)$ forces
$X$ to have a small mean $a$, while the covariance matrix $R$ 
has to be close to the unit covariance matrix $I_d$ in the
Hilbert-Schmidt norm, for example.

By definition, $D(X)$ is invariant under all affine invertible 
transformations of the space. Hence, in many problems or
formulas, one may assume without loss of generality that $X$ 
has mean zero and identity covariance matrix. For example, in this case, 
(3.3) becomes
$$
D(X) = D(X||Z) = h(Z) - h(X).
$$

Another important representation is given by de Bruijn's formula
\be
D(X||Z) = \int_0^1 I(X_t||Z)\,\frac{dt}{2t},
\en
still assuming that $X$ has mean zero and a unit covariance matrix.
Here $X_t = \sqrt{t} X + \sqrt{1-t} Z$
with $Z$ being independent of $X$, and
\be
I(X||Z) = \int 
\Big|\frac{\nabla p}{p} - \frac{\nabla \varphi}{\varphi}\Big|^2\,p\,dx
\en
stands for the relative Fisher information hidden in the distribution
of a random vector $X$ in $\R^d$ with a smooth density $p$.
More generally, this important distance is well-defined as long as
$\sqrt{p}$ belongs to the Sobolev space $W_1^2(\R^d)$.

\vskip7mm
\section{{\bf Bounds for Relative Entropy via Other Distances}}
\setcounter{equation}{0}

\noindent
The relative entropy with respect to the standard normal law
may be connected with more popular and standard distances. 
Recall that $D$ dominates the total variation. Moreover, applying 
the inequality (1.5) with weight $w(x) = \frac{1+|x|}{2}$,
we obtain a lower bound
$$
D(X||Z) \geq \frac{c}{d}\,
\bigg(\int (1+|x|)\,|p(x) - \varphi(x)|\,dx\bigg)^2
$$
in terms of the weighted total variation distance
(up to some absolute constant $c>0$). It is therefore
natural to expect that $D$ can be used
to bound various metrics responsible for the convergence of
probability measures on $\R^d$ in the weak topology. 

One of the most natural such metrics (especially in high dimension) 
is the Kantorovich transport distance of power order $s \geq 1$, 
which for Borel probability measures $P$ and $Q$ on $\R^d$ 
is defined by
$$
W_s(P,Q) = \inf_\mu \Big(\int\!\!\!\int |x-y|^s\,d\mu(x,y)\Big)^{1/s}.
$$
Here, the infimum is running over all probability measures $\mu$
on the product space $\R^d \times \R^d$ with marginals $P$ and $Q$.

The value $W_s^s(P,Q)$ is interpreted as the minimal expense needed to
transport $P$ to $Q$, provided that it costs $|x-y|^s$ to move 
the ``particle'' $x$ to the ``particle'' $y$. 
As is well-known, $W_s$ represents a metric in the space of
all probability distributions on $\R^d$ with finite absolute moments
of order $s$ (\cite{V}).

One of the remarkable relations between the relative entropy and
the quadratic Kantorovich distance was obtained by 
Talagrand \cite{T}; it indicates that, for any random vector $X$ in 
$\R^d$ with distribution $P$,
\be
D(X||Z) \geq \frac{1}{2}\,W_2^2(P,Q), 
\en
where $Q$ denotes the standard normal law on $\R^d$.
The advantage of (4.1) is that $D(X||Z)$ is defined 
explicitly and may be easily computed or estimated 
in many practical situations, in contrast with $W_2$.

Upper bounds on the relative entropy are of large interest as well.
One classical bound involving the relative Fisher information
as defined in (3.5) indicates that
\be
D(X||Z) \leq \frac{1}{2}\, I(X||Z)
\en
for any random vector $X$ in $\R^d$ with density $p$ such that
$\sqrt{p} \in W_1^2$. In fact, this relation represents an 
information-theoretic reformulation of the logarithmic Sobolev 
inequality for the standard Gaussian measure on $\R^d$; 
it was discovered by Gross \cite{Gr}, but appeared earlier in
an equivalent form of the entropic isoperimetric inequality
in Stam \cite{St}. Let us refer an interested reader to \cite{B-G-R-S}
for the history and some refinements of (4.2) involving 
the transport distance $W_2$.

Another example in which the smoothness of the density is not
needed was shown in \cite{B-M}. If a random vector 
$X$ in $\R^d$ has a square integrable density $p$ and satisfies 
$\E\,|X|^2 = \E\,|Z|^2 = d$, then
\be
D(X||Z) \leq c_d\, \Delta_2 \log^{\frac{d}{4} + 1}(1/\Delta_2).
\en
Here 
$$
\Delta_2 = \|p - \varphi\|_2 = 
\Big(\int (p(x) - \varphi(x))^2\,dx\Big)^{1/2}
$$
is the $L^2$-distance between $p$ and $\varphi$ (assuming that 
$\Delta_2 \leq 1/e$), and $c_d$ is a positive constant depending 
on the dimension $d$ only. 

Consequently, one may also bound $D(X||Z)$ in
terms of the uniform or $L^\infty$-distance
$\Delta_\infty = {\rm ess\,sup}_x\, |p(x) - \varphi(x)|$, which
is finite when $p$ is bounded. Without referring to (4.3),
one can show by similar arguments that
\be
D(X||Z) \leq c_d\, \Delta_\infty \log^{\frac{d}{2} + 1}(1/\Delta_\infty),
\en
as long as $\Delta_\infty \leq 1/e$.

By Plancherel's theorem, the inequality (4.3) can be restated in terms of
the characteristic functions of $X$ and $Z$. In dimension $d=1$, 
such bounds were explored in \cite{B-C-G5} by involving Edgeworth 
corrections. Put
$$
f(t) = \E\,e^{itX}, \quad 
g_\gamma(t) = \Big(1 + \gamma\,\frac{(it)^3}{3!}\Big)\,e^{-t^2/2}
$$
with an arbitrary parameter $\gamma \in \R$. Assuming that $\E\,|X|^3 < \infty$
(in which case $f(t)$ has three continuous derivatives), it was proved that
\be
D(X||Z) \leq \gamma^2 + 
4\,\Big(\|f - g_\gamma\|_2 + \|f''' - g_\gamma'''\|_2\Big).
\en

\vskip7mm
\section{{\bf Convexity and Monotonicity along Convolutions}}
\setcounter{equation}{0}

\noindent
The $D$-distance from the class of all normal laws is convex 
under variance preserving transformations. This follows from 
the entropy power inequality which was discovered by Shannon
and rigorously proved by Stam \cite{St}, \cite{D-C-T} (in dimension one
and subject to minor refining comments). 

For a random vector
$X$ in $\R^d$ with density $p$, define the entropy power
$$
N(X) = \exp\Big\{\frac{2}{d}\,h(X)\Big\} = 
\exp\Big\{-\frac{2}{d}\,\int p \log p\,dx\Big\},
$$
assuming that the last integral exists in the Lebesgue sense.
Like the variance in dimension one, this functional is translation 
invariant and homogeneous of order $2$.

\vskip3mm
{\bf Theorem 5.1.} {\sl If the random vectors $X$ and $Y$
in $\R^d$ are independent and have densities, then
\be
N(X+Y) \geq N(X) + N(Y),
\en
provided that the entropies of $X$, $Y$, and $X+Y$ exist.
}

\vskip4mm
As was shown in \cite{B-C}, it may happen that the entropy of $X$ and
$Y$ exists and is finite, while it does not exist for the sum $X+Y$.
A more careful formulation of the entropy power inequality is that
we should make the convention that $N(X) = 0$ whenever the entropy of $X$
does not exist (including the case where the distribution of $X$ is
not absolutely continuous with respect to the Lebesgue measure).
With this convention, (5.1) holds true without any restriction.

An equivalent variant of (5.1) was proposed by Lieb \cite{Lieb2}:
$$
h(\sqrt{t}\,X+\sqrt{1-t}\,Y) \geq t h(X) + (1-t)\,h(Y).
$$
This inequality holds for all $0 < t < 1$, whenever
independent random vectors
$X$ and $Y$ in $\R^d$ have densities such that all entropies exist.
As a consequence,
$$
D(\sqrt{t}\,X+\sqrt{1-t}\,Y||Z) \leq t D(X||Z) + (1-t)\,D(Y||Z),
$$
provided that $X$ and $Y$ have mean zero. This may be
viewed as a convexity of the $D$-distance along convolutions.

In dimension one the latter relation implies that, given independent random variables
$X_1,\dots,X_n$ with variances $\sigma_k^2 = \Var(X_k)$
such that $\sigma_1^2 + \dots + \sigma_n^2 = 1$,
for the sum $S_n = X_1 + \dots + X_n$ we have
\be
D(S_n) \leq \sum_{k=1}^n \sigma_k^2\,D(X_k).
\en
This suggests that the distributions of $S_n$ have a non-increasing
$D$-distance from the class of all normal laws,
as long as $X_k$ are independent and identically distributed.
Although this is immediate along the powers $n = 2^m$,
the general case is more sophisticated; nevertheless we have the
following remarkable result proved by Artstein, Ball, Barthe and Naor
\cite{A-B-B-N1}.

\vskip5mm
{\bf Theorem 5.2.} {\sl Let $(X_k)_{k \geq 1}$ be i.i.d. random
vectors in $\R^d$ wih finite second moments. Then, for all $n \geq 2$,
\be
D(S_n) \leq D(S_{n-1}).
\en
}
\vskip2mm
Recall that $D(S_n) = D(Z_n)$ for the normalized sums
$Z_n = S_n/\sqrt{n}$. Hence, an equivalent formulation
of (5.3) is that the entropy $h(Z_n)$ represents a non-decreasing sequence.

In \cite{A-B-B-N1}, a more general property in the non-i.i.d. situation 
has been also established, cf. also Madiman and Barron \cite{M-B}.

\vskip5mm
{\bf Theorem 5.3.} {\sl Given independent random vectors
$X_1,\dots,X_n$ wih finite second moments, $n \geq 2$, we have
$$
h\bigg(\frac{1}{\sqrt{n}} \sum_{k=1}^n X_k\bigg) \, \geq \, \frac{1}{n}
\sum_{k=1}^n h\bigg(\frac{1}{\sqrt{n-1}} \sum_{j \neq k} X_j\bigg).
$$
}

\vskip7mm
\section{{\bf Entropic Central Limit Theorem and Orlicz Spaces}}
\setcounter{equation}{0}

\noindent
Let $(X_k)_{k \geq 1}$ be i.i.d. random vectors in $\R^d$ wih mean
zero and identity covariance matrix. Consider the normalized sums
$$
Z_n = \frac{X_1 + \dots + X_n}{\sqrt{n}},
$$
so that $Z_n \Rightarrow Z$ as $n\rightarrow \infty$ weakly in distribution
(where we recall that $Z$ denotes a standard normal random vector in $\R^d$).

Since the relative entropy $D(Z_n) = D(Z_n||Z)$ dominates $D_\alpha(Z_n||Z)$ 
for $0 < \alpha < 1$, it is natural to expect that the normal approximation
in $D$ requires additional hypotheses on the underlying distribution. 
A final conclusion is however similar to the CLT in the total variation norm.

\vskip2mm
{\bf Theorem 6.1.} {\sl For the convergence
\be
D(Z_n) \rightarrow 0 \quad {\sl as} \ n \rightarrow \infty,
\en
it is necessary and sufficient that $D(Z_n)$ be finite for some $n$.
}

\vskip5mm
Here the condition that $D(Z_n)$ is finite is the same as the
finiteness of entropy $h(Z_n)$.
By Theorem 5.2, the convergence in (6.1) is monotone and is
equivalent to the monotone convergence of entropies
$$
h(Z_n) \uparrow h(Z) \quad {\rm as} \ n \rightarrow \infty.
$$

Since the functional $D$ defined in (3.2) is affine invariant,
one may also state Theorem 6.1 without moment constraints:
Given an i.i.d. sequence $X_k$ with finite second moments, (6.1)
holds if and only if $D(Z_n)$ is finite for some $n$.

Theorem 6.1 is due to Barron \cite{Bar} who considered the one-dimensional
setting. His proof was based on the application of de Bruijn's identity
(3.4); this argument was simplified by Harremoës and Vignat \cite{H-V}.
Earlier, an information-theoretic approach to the weak CLT
was proposed by Linnik \cite{Li}, who explored basic properties and 
behavior on convolutions of the closely related functional
$$
L(X) =
-\int_{-\infty}^\infty p(x) \log p(x)\,dx - \frac{1}{2} \int_{-\infty}^\infty
x^2\,p(x)\,dx,
$$
assuming that a random variable $X$ has mean zero and a bounded
density. He emphasized that $L(X) = h(X) - h(Z') + {\rm const}$,
where $Z'$ is a normal random variable with the same variance as $X$.

A simple sufficient condition for the validity of the entropic convergence
in (6.1) is that $Z_n$ has a bounded density for some $n$, which is
described as the smoothness (integrability) condition (2.5) in terms of the
common characteristic function $f(t)$ of $X_k$. But in that case we have
more -- a uniform local limit theorem (2.6) and the convergence (2.7) 
of densities in $L^2$, which is stronger than the
convergence in $D$ according to the upper bounds (4.3)-(4.4)
for the relative entropy.

On the other hand, it may happen that (6.1) holds true, while all
densities $p_n$ remain unbounded. Generalizing the example in \cite{G-K},
Barron considered the symmetric, compactly supported densities
of the form
\[
w(x) = \bigg\{
\begin{array}{cc}
0, \ 
 & 
{\rm if } \ |x| > 1/e, \\
\frac{r}{2\,|x|\,\log^{r + 1}(1/|x|)}, \ 
 &
{\rm if} \ |x| < 1/e,
\end{array}
\]
with parameter $r > 0$. Define the common density of $X_k$ to be
$p(x) = \frac{1}{\lambda}\,w(x/\lambda)$, where the constant 
$\lambda>0$ is chosen so that $\E X_1^2 = 1$. Near the origin $x=0$ 
the $n$-th convolution power $p^{*n}(x)$ admits a lower bound
$$
p^{*n}(x) \geq \frac{c_n}{|x|\,\log^{r n+1}(1/|x|)}
$$
with some constant $c_n>0$. Hence, all densities $p_n$ of $Z_n$
are unbounded in any neighbourhood of zero and therefore do not
satisfy a uniform local limit theorem. But, it is easy to check that
the entropies $h(Z_n)$ are finite as long as $n > 1/r$. Hence,
$Z_n$ do satisfy the entropic CLT.

Using the decomposition of densities, it was shown in \cite{B3} that
the local limit theorems in the norms of the Lebesgue spaces $L^s(\R^d)$ 
by Gnedenko and Prokhorov and the entropic central limit
theorem by Barron may be united in a more general statement
on the convergence of densities in Orlicz spaces. 
Given a Young function
$\Psi$, that is, an even convex function on the real line such that
$\Psi(0) = 0$ and $\Psi(r)>0$ for $r>0$, the Orlicz norm
of a measurable function $u$ on $\R^d$ is defined by
$$
\|u\|_\Psi = \inf\Big\{\lambda > 0: \int \Psi(u(x)/\lambda)\,dx \leq 1\Big\}.
$$

For example, the choice of the power function $\Psi(r) = |r|^s$, 
$s \geq 1$, leads to the $L^s$-norm $\|u\|_s$. The limit case
$\|u\|_\infty$ is also included in this scheme as an Orlicz norm.
The extreme role of this norm is explained in particular by 
a simple observation that, under the normalization condition 
$\Psi(1)=1$, we always have
$$
\|u\|_\Psi \leq \max\{\|u\|_1,\|u\|_\infty\}.
$$
This implies in particular that $\|\varphi\|_\Psi$ is finite for
all Orlicz norms.

In the setting of Theorem 6.1 (the i.i.d. model on $\R^d$), we have
the following characterization. Let $\|\cdot\|$ be one of the Orlicz norms.

\vskip5mm
{\bf Theorem 6.2} (\cite{B3}). {\sl Suppose that $Z_n$ have densities $p_n$
for large $n$. For the convergence
\be
\|p_n - \varphi\| \rightarrow 0 \quad {\sl as} \ n \rightarrow \infty,
\en
it is necessary and sufficient that $\|p_n\|$ be finite for some $n$.
}

\vskip5mm
If $\Psi$ satisfies the $\Delta_2$-condition, that is,
$\Psi(2r) \leq C\Psi(r)$ for all $r$ with some constant $C$,
then (6.2) is equivalent to
$$
\int \Psi(p_n(x) - \varphi(x))\,dx \rightarrow 0,
$$
which holds true if and only if $\int \Psi(p_n(x))\,dx < \infty$
for some $n$. Thus, Theorem 6.2 unites local limit theorems
in all $L^s$-spaces. As for the entropic CLT, it corresponds
to Theorem 6.2 with a particular Young function
$$
\psi(r) = |r|\,\log(1+|r|)
$$
in view of the next general characterization of the convergence
in $D$, which was also established in \cite{B3}.

\vskip5mm
{\bf Theorem 6.3.} {\sl Given a sequence of random vectors
$\xi_n$ in $\R^d$ with densities $p_n$, the convergence
$D(\xi_n||Z) \rightarrow 0$ as $n \rightarrow \infty$ is 
equivalent to the following two conditions:

\vskip2mm
$a)$ \ $\E\,|\xi_n|^2  \rightarrow d$ as $n \rightarrow \infty$;

\vskip2mm
$b)$ \ $\|p_n - \varphi\|_\psi \rightarrow 0$ as $n \rightarrow \infty$,
or equivalently,

\vskip2mm
$b')$ $\int \psi(p_n(x) - \varphi(x))\,dx \rightarrow 0$ as $n \rightarrow \infty$.
}

\vskip7mm
\section{{\bf Rates of Convergence in the Entropic CLT}}
\setcounter{equation}{0}

\noindent
Here we describe some results from \cite{B-C-G4} about the rates of 
convergence in Theorem 6.1. As in the previous section, we continue 
to assume that the normalized sums $Z_n$ are defined for the sequence 
$X_k$ of i.i.d. random vectors in $\R^d$ treated as independent copies 
of a random vector $X$  with mean zero and identity covariance matrix.

The question about the rates may be attacked under proper moment 
assumptions. Otherwise, one cannot say anything definite. Indeed, in 
the one dimensional case, for any sequence of real numbers 
$\ep_n \downarrow 0$, the random variable $X$ may have a distribution 
$P$ such that
\be
D(Z_n) \geq \ep_n
\en
for all $n$ large enough. As was shown by Matskyavichyus \cite{Mat},
this is even true for the weaker Kolmogorov distance
$$
\rho_n = \sup_x\, |\P\{Z_n \leq x\} - \P\{Z \leq x\}|.
$$
The distribution $P$ with this property may be constructed as 
a convex mixture of centered Gaussian measures on the real line.
Since $\rho_n$ is dominated by the total variation distance between
the distributions of $Z_n$ and $Z_n$, while the latter is dominated
by the relative entropy (Pinsker's inequality), we get (7.1) as well.

In order to get an idea about the correct rate of decrease of the
relative entropy for growing $n$, one may note that, in the typical situation,
for suitably increasing values $T_n$,
$$
D(Z_n) \sim
\int_{|x| < T_n} \frac{(p_n(x) - \varphi(x))^2}{\varphi(x)}\,dx \, + \, 
{\rm small \ error \ term}.
$$
If $T_n$ is not too large, then the deviations $p_n(x) - \varphi(x)$ 
are of the order at most $1/\sqrt{n}$ for all points $x$ in the ball 
$|x| < T_n$.
Hence, under proper assumptions, one may expect that $D(Z_n||Z)$
will be of the order at most $1/n$ (this was already conjectured by
Johnson \cite{J}). A more precise assertion is given
in the following theorem; for simplicity we start with the one dimensional case,
thus assuming that $X$ has mean zero and variance one. 

Recall that the cumulants of $X$ are defined by
$$
\gamma_r = i^{-r}\,\frac{d}{dt} \log\, \E\,e^{it X}\Big|_{t=0} \quad
\big(\,\E\,|X|^r < \infty, \ r = 1,2,\dots\big).
$$ 
In particular, 
$\gamma_3 = \E X^3$ and $\gamma_4 = \E X^4 - 3$ in the case 
$\gamma_3 = 0$. Put
$$
\Delta_n(s) = (n \log n)^{-\frac{s-2}{2}}, \quad s \geq 2,
$$
with the convention that $\Delta_n(s) = 1$ for $s=2$.

\vskip5mm
{\bf Theorem 7.1} (\cite{B-C-G4}).
{\sl Suppose that $D(Z_n)$ is finite for some $n$, and $\E\,|X|^s < \infty$.

$a)$ \ In the case $2 \leq s < 4$, we have $D(Z_n) = o(\Delta_n(s))$
as $n \rightarrow \infty$.

$b)$ \ In the case $4 \leq s < 6$, 
$$
D(Z_n) = \frac{c_1}{n} + o(\Delta_n(s)), \quad 
c_1 = \frac{1}{12} \gamma_3^2.
$$

$c)$ \ In the case $6 \leq s < 8$, and if $\gamma_3 = 0$,
$$
D(Z_n) = \frac{c_2}{n^2} + o(\Delta_n(s)), \quad 
c_2 = \frac{1}{48} \gamma_4^2.
$$
}

Part $a)$ with $s=2$ corresponds to Theorem 6.1. As for other
values of $s$, the error term in (7.2) is nearly optimal up to 
a logarithmic factor, which can be seen from the next assertion.

\vskip5mm
{\bf Theorem 7.2} \cite{B-C-G4}. 
{\sl Let $\eta>0$ and $2 < s < 4$. There exists
an i.i.d. sequence $X_k$ with mean zero, variance one and
$\E\,|X|^s < \infty$, such that $D(X) < \infty$ and
\be
D(Z_n) \geq \frac{c}{(n \log n)^{\frac{s-2}{2}}\,(\log n)^\eta},
\quad n \geq n_0,
\en
where $n_0$ is determined by the distribution of $X$, and where
the constant $c>0$ depends on $s$ and $\eta$ only.
}

\vskip3mm
Choosing any $\eta$ in (7.2), we see that $D(Z_n||Z)$
decays at the rate which is worse than $1/n$.

The asymptotic expression in $c)$ holds true in particular, as long as 
the distribution of $X$ is symmetric about the origin, since then 
$\gamma_3 = 0$. It was also shown in \cite{B-C-G4} that without 
constraints on the cumulants, the right-hand sides in $b)-c)$ may be 
further generalized as an expansion in powers in $1/n$, namely
\be
D(Z_n) = \frac{c_1}{n} + \dots + \frac{c_r}{n^r} + o(\Delta_n(s)),
\en
where $r = [\frac{s-2}{2}]$ (the integer part) and where every $c_j$
represents a certain polynomial in the cumulants 
$\gamma_3,\dots,\gamma_{2j+1}$ (hence a polynomial in moments
of $X$ up to order $2j+1$). 
As a particular case, assume that $s \geq 4$ and $\gamma_j = 0$ for all 
$3 \leq j < k$ for some $k = 3,4,\dots,[s]$. Then (7.3) is simplified to
$$
D(Z_n) = \frac{\gamma_k^2}{2k!} \cdot \frac{1}{n^{k-2}} + 
O\Big( \frac{1}{n^{k-1}}\Big) + o(\Delta_n(s)).
$$
Previously, Johnson [J] had noticed (though in terms of the standardized 
Fisher information) that if $\gamma_k \neq 0$, 
$D(Z_n)$ cannot be of smaller order than $n^{k-2}$. A more precise lower bound
$$
\liminf_{n \rightarrow \infty} \Big[n^{k-2} D(Z_n)\Big] \geq \frac{\gamma_k^2}{2k!}
$$
was later derived by Harremo\"es \cite{H}.

In the multidimensional case, Theorem 7.1 is extended in a slightly
weaker form.

\vskip3mm
{\bf Theorem 7.3.} {\sl Let $d \geq 2$. Suppose that $D(Z_n)$ 
is finite for some $n$, and $\E\,|X|^s < \infty$ for an integer $s \geq 2$.
Then we have an expansion $(7.3)$ with the error term
$$
\Delta_n(s) = n^{-\frac{s-2}{2}}\, (\log n)^{-\frac{s-d}{2}}
$$
with the convention that $\Delta_n(s) = 1$ for $s=2$.
}

\vskip5mm
In particular, for $s=2$ we get a multidimensional variant of Theorem 6.1.

If $\E\,|X|^4 < \infty$, then
$D(Z_n) = O(1/n)$ for $d \leq 4$ and
$$
D(Z_n) = O\Big((\log n)^{\frac{d-4}{2}}/n\Big) \quad 
{\rm for} \ d \geq 5.
$$
However, if $\E\,|X|^5 < \infty$, then
$$
D(Z_n) = O(1/n)
$$ 
regardless of the dimension $d$. This slight difference between conclusions  
for different dimensions is due to the dimension-dependent asymptotic
$$
\int_{|x| > T} |x|^2\,\varphi(x)\,dx \, \sim \, c_d T^d\, \varphi(T) \ \
{\rm as} \ T \rightarrow \infty.
$$

The proof of (the one dimensional) Theorem 7.1 and a more precise 
expansion (7.3) is based on the non-uniform local limit theorem,
in which the density $p_n$ of $Z_n$ is approximated by the
Edgeworth correction of the normal density defined by
$$
\varphi_m(x) = \varphi(x) +  \varphi(x) \sum_{\nu=1}^{m-2}
\frac{q_\nu(x)}{n^{\nu/2}}, \quad m = [s].
$$
Here
\be
q_{\nu}(x) = \sum H_{\nu+2l}(x) \prod_{r=1}^{\nu}\frac 1{k_r!}
\Big(\frac{\gamma_{r+2}}{(r+2)!}\Big)^{k_r}, 
\en
where the summation runs over all non-negative integer solutions 
$(k_1,k_2,\dots,k_{\nu})$ to the equation 
$$
k_1+2k_2+\dots+\nu k_{\nu}=\nu \ \ {\rm with} \ \
l=k_1+k_2+\dots+k_{\nu}.
$$
As usual, $H_k$ denotes the Chebyshev-Hermite polynomial of
degree $k$ with the leading coefficient 1. Hence, the sum in (7.4) 
defines a polynomial in $x$ of degree at most $3(\nu-2)$. 
In particular, $\varphi_m(x) = \varphi(x)$ for the range $2 \leq s < 3$.

\vskip4mm
{\bf Theorem 7.4.}
{\sl Assume that $X$ has a finite absolute moment of a real order 
$s \ge 2$, and $Z_n$ admits a bounded density for some $n$. Then, for all 
$n$ large enough, $Z_n$ have continuous bounded densities $p_n$ satisfying
uniformly in $x \in \R$
\be
(1 + |x|^m)\,(p_n(x) - \varphi_m(x)) = o\big(n^{-\frac{s-2}{2}}\big)
\en
as $n \rightarrow \infty$. Moreover,
\begin{eqnarray}
(1 + |x|^s)\,(p_n(x) - \varphi_m(x)) 
 & = &
o\big(n^{-\frac{s-2}{2}}\big) + \nonumber \\
 & & \hskip-20mm
\big(1 + |x|^{s-m}\big)\,\big(O\big(n^{-\frac{m-1}{2}}\big) + o\big(n^{-(s-2)}\big).
\end{eqnarray}
}

If $s=m$ is integer, $m \geq 3$, Theorem 7.4 is well known; then (7.5) 
and (7.6) simplify to
\be
(1 + |x|^m)\,(p_n(x) - \varphi_m(x)) = o\big(n^{-\frac{m-2}{2}}\big).
\en
In this formulation the result is due to Petrov \cite{Pe1}; cf. \cite{Pe2}, p.\,211, 
or \cite{B-RR}, p.\,192. Without the term $1+|x|^m$, the relation (7.7) 
goes back to the results of Cram\'er and Gnedenko (cf. \cite{G-K}).

In the general (fractional) case, Theorem 7.4 has been obtained in \cite{B-C-G1,B-C-G2}
by using the technique of Liouville fractional integrals and derivatives. Assertion
(7.7) gives an improvement over (7.6) on relatively large intervals of the real axis,
and this is essential in the case of non-integer $s$.

\vskip7mm
\section{{\bf Berry-Esseen Bounds for Total Variation}}
\setcounter{equation}{0}

\noindent
We now consider a general scheme of random variables
which are not necessarily identically distributed, focusing on the
dimension $d=1$. Let $X_1,\dots,X_n$ be independent random variables
with mean zero and finite variances $\sigma_k^2 = \Var(X_k)$. Assuming
that $B_n = \sigma_1^2 + \dots + \sigma_n^2$ is positive, 
define the normalized sum
\be
Z_n = \frac{X_1 + \dots + X_n}{\sqrt{B_n}},
\en
so that $\E Z_n = 0$ and $\var(Z_n) = 1$.

It is well-known that $Z_n$ is nearly normal in the weak sense
under the Lindeberg condition. In order to quantify this property,
one usually uses the Lyapunov ratios (coefficients)
\be
L_s = \frac{1}{B_n^{s/2}} \sum_{k=1}^n \E\,|X_k|^s, \quad s>2,
\en
which are finite as long as all $X_k$ have finite absolute moments 
of a fixed order $s$. In a typical situation, these quantities
are getting smaller for growing values of $s$; for example, in the
i.i.d. case with $\E X_1^2 = 1$, we have
$$
L_s = n^{-\frac{s-2}{2}}\, \E\,|X_1|^s,
$$
which has a polynomial decay with respect to the number of
``observations" $n$. On the other hand, in general
the function $s \rightarrow L_s^{\frac{1}{s-2}}$ is non-decreasing, 
so that $L_3 \leq \sqrt{L_4}$, for example.

The classical Berry-Esseen bound indicates that
\be
\sup_x\, |\P\{Z_n \leq x\} - \P\{Z \leq x\}| \leq cL_3
\en
with some universal constant $c>0$, where $Z$ is a standard
normal random variable (cf. e.g. \cite{Pe2}). In the i.i.d. case
with $\E X_1^2 = 1$, it leads to the well-known estimate
$$
\sup_x\, |\P\{Z_n \leq x\} - \P\{Z \leq x\}| \leq 
\frac{c}{\sqrt{n}}\, \E\,|X_1|^3
$$
with a standard rate of normal approximation for the
Kolmogorov distance. Note that in general $L_3 \geq \frac{1}{\sqrt{n}}$.

An interesting question is how to extend the bound (8.3)
to strong distances such as the total variation and relative entropy.
Note, however, that these distances are useless, for example, 
when all summands have discrete distributions, in which case 
$\|P_{Z_n} - P_Z\|_{{\rm TV}} = 2$ and $D(Z_n||Z) = \infty$. 
Therefore, some assumptions are needed or desirable, such as an absolute
continuity of distributions $P_{X_k}$ of $X_k$. But even with this 
assumption we cannot exclude the case that our distances from $Z_n$ 
to the normal law may be growing when the $P_{X_k}$ are close to 
discrete distributions. To prevent such behaviour,
one may require that the densities of $X_k$ should be bounded on 
a reasonably large part of the real line. This can be guaranteed
quite naturally, by using the entropy functional
$h(X)$ or equivalently $D(X)$. If the latter is finite, then,
for example, the characteristic function $f(t) = \E\,e^{itX}$
is bounded away from 1 at infinity, and moreover
$$
|f(t)| \leq 1 - c\,e^{-4D(X)}, \quad \sigma |t| \geq \frac{\pi}{4},
$$
where $\sigma$ is the standard deviation of $X$, and $c>0$
is an absolute constant (cf. \cite{B-C-G3}). Thus, the finiteness 
of $D(X)$ guarantees that $P_X$ is separated from the class 
of discrete probability distributions, and if it is small, one may 
speak about the closeness of $P_X$ to normality in a rather strong 
sense. Using $D$ for both purposes, one can obtain refinements 
of Berry-Esseen's inequality (8.3) in terms of the total variation 
and the entropic distances to normality for the distributions of $Z_n$.
The following statement was proved in \cite{B-C-G5}.

\vskip5mm
{\bf Theorem 8.1.} {\sl Suppose that the random variables $X_k$
have finite absolute moments of the third order and satisfy
$D(X_k) \leq D$ for a number $D$. Then
\be
\|P_{Z_n} - P_Z\|_{\rm TV} \leq cL_3,
\en
where the constant $c$ depends on $D$ only.
}

\vskip5mm
In particular, in the i.i.d. case with $\E X_1^2 = 1$, we get
$$
\|P_{Z_n} - P_Z\|_{\rm TV} \leq \frac{c}{\sqrt{n}}\, \E\,|X_1|^3
$$
where the constant $c$ depends on $D(X_1)$ only. Related estimates in the 
i.i.d.-case were studied by many authors. For example, in the early 
1960's Sirazhdinov and Mamatov \cite{S-M} found an exact asymptotic 
$$
\|P_{Z_n} - P_Z\|_{\rm TV} = 
\frac{c_0}{\sqrt{n}}\, |\E X_1^3| + o\Big(\frac{1}{\sqrt{n}}\Big)
$$
with some universal constant $c_0$, which holds under 
the assumption that the distribution of $X_1$ has a non-trivial 
absolutely continuous component. Note that this statement refines
Prokhorov's theorem (2.2) under the 3-rd moment assumption.

Returning to Theorem 8.1, it was also shown in \cite{B-C-G5}
that if $L_3 \leq \frac{1}{64}$
and 
$$
D(X_k) \leq \frac{1}{24}\,\log \frac{1}{L_3},
$$ 
then (8.4) holds true with an absolute constant.

The condition in Theorem 8.1 may be stated in terms of maximum
of densities. If a random variable $X$ with finite standard 
deviation $\sigma$ has a density $p$ such that $p(x) \leq M$ 
for a number $M$, then $X$ has finite entropy, and moreover
\be
D(X) \leq \log(M\sigma\sqrt{2\pi e}).
\en
Indeed, the functional $X \rightarrow M\sigma$ with 
$M = \|p\|_\infty = {\rm ess\,sup}_x\, p(x)$ is affine invariant. Hence, (8.5) 
does not loose generality when $X$ has mean zero with $\sigma = 1$. 
But then (8.5) immediately follows from
$$
D(X) = h(Z) - h(X) = 
\int_{-\infty}^\infty p(x)\,\log\Big(p(x)\sqrt{2\pi e}\Big)\,dx.
$$

Thus, in the setting of Theorem 8.1, if the random variables
$X_k$ have densities $p_k \leq M_k$ such that $M_k \sigma_k \leq M$,
the inequality (8.5) holds true with a constant $c$ depending
$M$ only.

\vskip7mm
\section{{\bf Berry-Esseen Bounds for Relative Entropy}}
\setcounter{equation}{0}

\noindent
Theorem 8.1 has an analogue for the relative entropy, which was
derived in \cite{B-C-G5} in terms of the Lyapunov ratio 
$$
L_4 = \frac{1}{B_n^2} \sum_{k=1}^n \E X_k^4
$$
(cf. the definition (8.2) with $s=4$).
We keep the same setting and assumptions as in the previous section.

\vskip5mm
{\bf Theorem 9.1.} {\sl Suppose that the random variables $X_k$
have finite moments of the fourth order and satisfy
$D(X_k) \leq D$ for a number $D$. Then
\be
D(Z_n) \leq cL_4,
\en
where the constant $c$ depends on $D$ only. Moreover, if 
$L_4 \leq 2^{-12}$ and 
$$
D(X_k) \leq \frac{1}{48}\,\log \frac{1}{L_4},
$$ 
then $c$ may be chosen as an absolute constant.
}

\vskip5mm
In view of the bound (8.5), we obtain as a consequence that, 
if the random variables
$X_k$ have bounded densities $p_k$ such that $p_k(x) \leq M_k$
and $M_k \sigma_k \leq M$, the inequality (9.1) holds true 
with a constant $c$ depending on $M$ only.

One interesting feature of (9.1) is that it may be connected
with transport inequalities for the distributions $P_{Z_n}$
of $Z_n$ in terms of the quadratic Kantorovich distance $W_2$.
Indeed, applying Talagrand's entropy-transport inequality
(4.1), we conclude that
\be
W_2^2(P_{Z_n},P_Z) \leq cL_4,
\en
where $c$ depends on $D$. This relation, with an absolute
constant $c$, was discovered by Rio \cite{R}, who also studied 
more general Kantorovich distances $W_s$, by relating them 
to Zolotarev's ideal metrics (cf. also \cite{B2} for further
refinements and generalizations). It has also 
been noticed in \cite{R} that the 4-th moment condition is essential, 
so the Laypunov's ratio $L_4$ in (9.2) cannot be replaced with 
a function of $L_3$ including the i.i.d.-case.

In order to obtain the inequality (9.2) in full generality, that is,
without any constraints on $D(X_k)$ as in Theorem 9.1, the
entropic Berry-Esseen bound (9.1) has to be stated under
a different condition. 

\vskip5mm
{\bf Theorem 9.2.} {\sl If the characteristic function
$f_n(t) = \E\,e^{itZ_n}$ is vanishing outside the interval
$|t| \leq \frac{1}{4\sqrt{L_4}}$, then $(9.1)$ holds true
with an absolute constant~$c$.
}

\vskip5mm
This variant of Theorem 9.1 was proposed in \cite{B1}, with
an argument based on the application of the upper bound (4.5) 
for the relative entropy in terms of the corrected Fourier-Stieltjes
transforms.
Combining (9.1) with (4.1), we are led to the desired
relation (9.2), however under an additional hypothesis
on the support of $f_n(t)$. But, the latter may be removed
when applying (9.2) to the smoothed random variables
$$
Z_n(\tau) = \sqrt{1 - \tau^2}\,Z_n + \tau \xi, \quad 0 < \tau < 1,
$$
assuming that the random variable $\xi$ is independent of
$Z_n$ and has finite 4-th moment, with $\E \xi = 0$, $\E \xi^2 = 1$,
and with characteristic function vanishing on the interval of length 
of order 1. In that case 
\be
W_2^2(P_{Z_n(\tau)},P_{Z_n}) \leq 
\E\,(Z_n(\tau) - Z_n)^2 \leq 2\tau^2.
\en
Hence, if we choose $\tau \sim \sqrt{L_4}$, one may apply 
Theorem 9.2 to $Z_n(\tau)$, and then
the support assumption will be removed in view of (9.3).

Returning to Theorem 9.1, let us note that, in the i.i.d. case 
with $\E X_1^2 = 1$, we get
\be
D(Z_n) \leq \frac{c}{n}\, \E X_1^4,
\en
where the constant $c$ depends on $D(X_1)$ only. 
In fact, according to the second refining part of this theorem,
(9.4) holds true with an absolute constant, as long as
$n$ is sufficiently large, for example, if
$$
n \geq e^{12\,(1 + 4D(X_1))}\,\E X_1^4.
$$

Note also that the inequality (9.4) partly recovers Theorem 7.1 
for the power $s=4$ which yields a more
precise asymptotic expression
$$
D(Z_n) = \frac{1}{12\,n}\,| \E X_1^3|^2 + 
o\Big(\frac{1}{n\log n}\Big) \quad {\rm as} \ n \rightarrow \infty.
$$ 

In place of (9.4), one may also consider a more general 
scheme of weighted sums
\be
Z_n = a_1 X_1 + \dots + a_n X_n, \quad 
a_1^2 + \dots + a_n^2 = 1 \ \ (a_k \in \R),
\en
assuming that the random variables $X_k$ are independent
and identically distributed with mean zero, variance one, and
finite 4-th moment. Putting
$$
l_4(a) = a_1^4 + \dots + a_n^4, \quad a = (a_1,\dots,a_n),
$$
(9.1) yields
\be
D(Z_n) \leq c\,\E X_1^4\,l_4(a),
\en
where $c$ depends on $D(X_1)$. Berry-Esseen bounds for such 
weighted sums have been previously studied by Artstein, Ball, 
Barthe and Naor under the assumption that the distribution 
of $X_1$ satisfies a Poincar\'e-type inequality
$$
\lambda_1\,\Var(u(X_1)) \leq \E\,u'(X_1)^2.
$$
It is required to hold with some constant $\lambda_1 > 0$ 
(called a spectral gap) in the class of all bounded smooth 
functions $u$ on the real line (note that necessarily 
$\lambda_1 \leq 1$ due to the moment assumption $\E X_1^2 = 1$).
It was shown in \cite{A-B-B-N2} that
$$
D(Z_n)\, \leq\, \frac{2 l_4(a)}{\lambda_1 + (2 - \lambda_1)\, l_4(a)}\, D(X_1),
$$
or in a slightly modified form
$$
D(Z_n)\, \leq\, \frac{2D(X_1)}{\lambda_1}\,l_4(a).
$$
As well as in (9.6), here the right-hand side is proportional
to $l_4(a)$.

\vskip7mm
\section{{\bf R\'enyi and Tsallis Divergences with Respect to the Normal Law}}
\setcounter{equation}{0}

\noindent
We now turn to R\'enyi and Tsallis divergences of order $\alpha>1$ 
and describe in the next few sections some results taken mostly 
from \cite{B-C-G6}. As before, $Z$ denotes a standard normal 
random vector in $\R^d$.

If $X$ is a random vector in $\R^d$, for $D_\alpha(X||Z)$ to be 
finite it is necessary that $X$ have a density $p$ with respect to 
the Lebesgue measure $\mu_d$ on $\R^d$. Choosing 
$\mu = \mu_d$ in (1.1)-(1.2), these definitions become
\be
D_\alpha(X||Z) = \frac{1}{\alpha - 1} \log 
\int \Big(\frac{p(x)}{\varphi(x)}\Big)^\alpha\,\varphi(x)\, dx,
\en

\be
T_\alpha(X||Z) = \frac{1}{\alpha - 1} 
\bigg[\int \Big(\frac{p(x)}{\varphi(x)}\Big)^\alpha\,\varphi(x)\, dx - 1\bigg].
\en

This case is rather different compared to the case of the relative 
entropy ($\alpha = 1$), which can be seen as follows.
The finiteness of $D(X||Z)$ means that $X$ has finite second 
moment $\E\,|X|^2$ and finite entropy $h(X)$, which holds, 
for example, when the density $p$ is bounded.
But, for the finiteness of $D_\alpha(X||Z)$ with $\alpha>1$
it is necessary that $X$ be subgaussian, and moreover 
$\E\,e^{c|X|^2} < \infty$ for all $c < 1/(2\alpha^*)$, where 
$\alpha^* = \frac{\alpha}{\alpha-1}$
is the conjugate index. More precisely, putting
$$
T_\alpha =T_\alpha(X||Z), \quad
B = \big(1 + (\alpha - 1)T_\alpha\big)^{1/\alpha},
$$
we have
\be
\E\,e^{c|X|^2} \leq \frac{B}{(1 - 2\alpha^* c)^{\frac{d}{2\alpha^*}}}.
\en
It is however possible that $T_\alpha < \infty$, while 
$$
\E\,\exp\Big\{\frac{1}{2\alpha^*} |X|^2\Big\} = \infty.
$$

An alternative (although almost equivalent) variant of this property
may be given via the bound on the Laplace transform
\be
\E\,e^{\left<t,X\right>} \leq B\, e^{\alpha^* |t|^2/2}, \quad 
t \in \R^d.
\en
By Markov's inequality, this implies a subgaussian bound on tail 
probabilities
$$
\P\{\left<\theta,X\right> \geq r\} \leq 
B \exp\Big\{-\frac{r^2}{2\alpha^*}\Big\}, \quad r \geq 0,
$$
for any unit vector $\theta$.

Although the critical value $c = 1/(2\alpha^*)$ may not be 
included in (10.3), it may be included for sufficiently many 
convolutions of $p$ with itself. More precisely, consider 
the normalized sums
$$
Z_n = \frac{X_1 + \dots + X_n}{\sqrt{n}}
$$
of independent copies of $X$. If $n \geq \alpha$, then
\be
\E\,e^{|Z_n|^2/(2\alpha^*)} < \infty.
\en
Moreover, 
\be
\big|\E\,e^{|Z_n|^2/(2\alpha^*)} - \E\,e^{|Z|^2/(2\alpha^*)}\big|
\leq c_{n,d}\,\big((1 + \chi_\alpha^{1/\alpha})^n - 1\big),
\en
where $\chi_\alpha = \chi_\alpha(X,Z)$ is the Pearson-Vajda distance 
of order $\alpha$. According to the relations in (1.10), here the right-hand 
side may be further bounded in terms of $T_\alpha = T_\alpha(X||Z)$.

The proof of this interesting phenomenon is based upon
a careful application of the contractivity property of 
the Weierstrass transform.
One important consequence from it is that the function
$$
\psi(t) = \E\,e^{\left<t,X\right>} \, e^{-\alpha^* |t|^2/2}
$$
is vanishing at infinity and is integrable with any power 
$n \geq \alpha$. Moreover,
\be
\int \psi(t)^n\,dt \leq c_{n,d} B^n
\en
with some constants depending on $(n,d)$ only.

Similar conclusions can be made about the boundedness
of densities of $Z_n$. In section 6 we mentioned an example 
in which all $D(Z_n)$ are finite (for the parameter $r \geq 1$), 
while its densities $p_n$ remain unbounded.
This is no longer true for $D_\alpha$. 

Indeed, it follows from (10.1)-(10.2) that $p \in L^\alpha(\R^d)$ 
as long as $T_\alpha$ is finite. In that case, it belongs to all $L^\beta$,
$1 \leq \beta \leq \alpha$. Hence, in the case $\alpha \geq 2$
necessarily $p \in L^2$, so, the characteristic function $f$ of $X$
also belongs to $L^2$, which implies that
the density $p_2$ of $Z_2$ is bounded and continuous
(by the inverse Fourier formula). In the other case 
$\alpha < 2$, applying the Hausdorff-Young inequality, 
we obtain that $f$ belongs to the dual space $L^{\alpha^*}$.
Hence $f^n$ is integrable, whenever $n \geq \alpha^*$,
which implies that $Z_n$ has a bounded continuous density
$p_n$. Uniting both cases, we conclude that $Z_n$ have 
bounded continuous densities for all 
$n \geq n_\alpha = \max(2,\alpha^*)$.

This property may be considerably sharpened in terms of
pointwise subgaussian bounds on the densities.
Using contour integration, one can prove:

\vskip5mm
{\bf Theorem 10.1} (\cite{B-C-G6}). {\sl If $T_\alpha(X||Z) < \infty$, then 
for all $x \in \R^d$, the densities $p_n$ of $Z_n$ with 
$n \geq n_\alpha = \max(2,\alpha^*)$ are continuous and satisfy
\be
p_n(x) \, \le \, A_{\alpha,d}\, n^{d/2}\, e^{-|x|^2/(2\alpha^*)}\,
\psi\Big(\frac{x}{\alpha^*\sqrt n}\Big)^{n-n_\alpha},
\en
where $A_{\alpha,d}$ depends on $(\alpha,d)$ only. In particular, 
there exist constants $x_0 > 0$ and $\delta \in (0,1)$ depending 
on the density $p$ of $X$ such that for all $n$ large enough
\be
p_n(x) \, \le \, 
\delta^n  e^{-|x|^2/(2\alpha^*)}\,
\psi\Big(\frac{x}{\alpha^*\sqrt n}\Big)^{n/2} 
\quad 
{\sl whenever} \ \ |x| \geq x_0 \sqrt{n}.
\en
}

\vskip7mm
\section{{\bf Pearson's $\chi^2$-Distance to the Normal Law}}
\setcounter{equation}{0}

\noindent
As we have already mentioned, an interesting particular case
$\alpha=2$ leads to the Pearson's $\chi^2$-distance
$T_2 = \chi^2$ and the R\'eny divergence $D_2 = \log(1 + \chi^2)$.
For simplicity, let us consider the one dimensional situation. 
Thus, with respect to the standard normal law according to (10.2),
we have
\bee
\chi^2(X,Z)
 & = &
\int_{-\infty}^\infty \frac{p(x)^2}{\varphi(x)}\,dx - 1 \\
 & = &
\sqrt{2\pi}\, \sum_{k=1}^\infty \, \frac{1}{k!}
\int_{-\infty}^\infty x^{2k} p(x)^2\,dx, \quad Z \sim N(0,1),
\ene
where $X$ is a random variable with density $p$.

In this case, necessary and sufficient conditions for 
the finiteness of this distance may be given in terms of 
the characteristic function
$$
f(t) = \E\,e^{itX} = \int_{-\infty}^\infty e^{itx}\,p(x)\,dx, 
\quad t \in \R.
$$
The condition $\chi^2 = \chi^2(X,Z) < \infty$ ensures that 
$f(t)$ has square integrable derivatives $f^{(k)}(t)$ of any 
order $k$. Moreover, in that case, by Plancherel's theorem,
$$
\chi^2(X,Z) \, = \, \frac{1}{\sqrt{2\pi}} \,\sum_{k=1}^\infty \, 
\frac{1}{k!} \int_{-\infty}^\infty |f^{(k)}(t)|^2\,dt.
$$

According to (10.3), for all $c< \frac{1}{4}$,
\be
\E\,e^{cX^2} \leq \frac{B}{(1 - 4c)^{1/4}}, \quad 
B = \big(1 + \chi^2\big)^{1/2}, \quad 
\en
and it is possible that 
$\chi^2 < \infty$, while $\E\,e^{\frac{1}{4} X^2} = \infty$.
Nevertheless, 
$$
\E\,e^{\frac{1}{4} Z_n^2} < \infty \quad {\rm for \ all} \ n \geq 2,
$$
where $Z_n$ is the normalized sum of $n$ independent copies of $X$.

In fact, for $n=2$, the inequality (10.5) can be stated more precisely as
$$
\E\,e^{\frac{1}{4} Z_2^2} \leq 2\,(1+\chi^2).
$$
Equivalently, there is a corresponding refinement of the inequality 
(10.7) in the form without any convolution, namely
$$
\frac {1}{\sqrt{2\pi}} \int_{-\infty}^{\infty} 
f(iy)^2\,e^{-2y^2}\,dy \, \le \,1+\chi^2.
$$
The argument is based on the Plancherel formula
$$
\int_{-\infty}^{\infty} |f(iy)|^2\,e^{-2y^2}\,dy \, = \,
 \int_{-\infty}^{\infty} |\rho(t)|^2\,e^{-2t^2}\,dt,
$$
where $\rho$ is the Fourier transform of the function 
$g(x) = p(x)\,e^{x^2/4}$, assuming that it belongs to $L^2$
(that is, $\chi^2 < \infty$).

Let us also mention that, although the density $p$ does not need
be bounded in the case $\chi^2 < \infty$, the densities $p_n$ of all
normalized sums $Z_n$, $n \geq 2$, have to be bounded in this case.

\vskip7mm
\section{{\bf Exponential Series and Normal Moments}}
\setcounter{equation}{0}

\noindent
The $\chi^2$-distance from the standard normal law on the real line 
admits a nice description in terms of the so-called exponential series 
(following Cram\'er's terminology). Let us introduce basic notations and 
recall several well-known facts. Let
$$
H_k(x) =
(-1)^k\, \big(e^{-x^2/2}\big)^{(k)}\, e^{x^2/2}, \qquad 
k = 0,1,2,\dots \ \ (x \in \R),
$$
denote the $k$-th Chebyshev-Hermite polynomial. In particular, 
$$
H_0(x) = 1, \quad H_1(x) = x, \quad H_2(x) = x^2 - 1, \quad 
H_3(x) = x^3 - 3x.
$$
Each $H_k$ is a polynomial of degree $k$ with integer coefficients.
Depending on $k$ being even or odd, $H_k$ contains even respectively
odd powers only. It may be defined explicitly via
$$
H_k(x) = \E\,(x+iZ)^k, \quad Z \sim N(0,1).
$$

Being orthogonal to each other with weight function $\varphi(x)$, 
these polynomials form a complete orthogonal system 
in the Hilbert space $L^2(\R,\varphi(x)dx)$ with
$$
\E\, H_k(Z)^2 = \int_{-\infty}^\infty H_k(x)^2\, \varphi(x)\,dx = k!
$$
Equivalently, the Hermite functions $\varphi_k(x) = H_k(x) \varphi(x)$
form a complete orthogonal system in $L^2(\R,\frac{dx}{\varphi(x)})$ with
$$
\int_{-\infty}^\infty \varphi_k(x)^2\,\frac{dx}{\varphi(x)} = k!
$$
Hence, any complex valued function $u$ such that
$\int_{-\infty}^\infty |u(x)|^2\, e^{x^2/2} dx < \infty$
admits a unique representation in the form of the orthogonal series
\be
u(x) = \varphi(x) \sum_{k=0}^\infty \frac{c_k}{k!}\, H_k(x)
\en
which converges in $L^2(\R,\frac{dx}{\varphi(x)})$.
Here, the coefficients are given by
$$
c_k = \int_{-\infty}^\infty u(x)\,H_k(x)\, dx,
$$
and we have Parseval's identity
\be
\sum_{k=0}^\infty \frac{|c_k|^2}{k!} =
\int_{-\infty}^\infty \frac{|u(x)|^2}{\varphi(x)}\,dx.
\en

The functional series (12.1) representing $u$ is called an exponential series. 
The question of its pointwise convergence is rather delicate similarly to 
the pointwise convergence of ordinary Fourier series based on trigonometric 
functions. In particular, if $u(x)$ is vanishing at infinity and
has a continuous derivative such that the integral
$\int_{-\infty}^\infty |u'(x)|^2\, e^{x^2/2}\,dx$
is finite, it may be developed in an exponential series, which is 
absolutely and uniformly convergent on the real line, cf. Cram\'er \cite{Cr}. 
For example, for the Gaussian functions $u(x) = e^{-\lambda x^2}$ 
($\lambda > 0$), the corresponding exponential series can be explicitly 
computed. At $x=0$ it is absolutely convergent for $\lambda>\frac{1}{4}$, 
simply convergent for $\lambda = \frac{1}{4}$ and divergent for
$\lambda < \frac{1}{4}$.

Let $X$ be a random variable with density $p$, and let $Z$ be 
a standard normal random variable independent of $X$. 
Applying (12.1) to $p$, we obtain the following: If
\be
\int_{-\infty}^\infty p(x)^2\, e^{x^2/2}\,dx < \infty,
\en
then $p$ admits a unique representation in the form of the exponential series
\be
p(x) = \varphi(x) \sum_{k=0}^\infty \frac{c_k}{k!}\, H_k(x),
\en
which converges in $L^2(\R,\frac{dx}{\varphi(x)})$. Here, 
the coefficients are given by
$$
c_k = \int_{-\infty}^\infty H_k(x)\,p(x)\, dx = 
\E H_k(X)= \E\,(X + iZ)^k,
$$
which we call the normal moments of $X$. In particular, 
$c_0 = 1$, $c_1 = \E X$.

In general, $c_k$ exists as long as the $k$-th absolute moment 
of $X$ is finite. These moments are needed to develop 
the characteristic function of $X$ in a Taylor series around zero as follows:
\be
f(t) = \E\,e^{itX} = e^{-t^2/2}\, \sum_{k=0}^N 
\frac{c_k}{k!}\,(it)^k + o(|t|^N), \quad t \rightarrow 0.
\en
In particular, $c_k = 0$ for $k \geq 1$ when $X$ is standard normal,
similarly to the property of the cumulants 
$$
\gamma_k(X) = \frac{d^k}{i^k\,dt^k}\, \log f(t)|_{t=0}
$$
with $k \geq 3$ (using the branch of the logarithm determined by 
$\log f(0) = 0$).

Let us emphasize one simple algebraic property of normal moments.
Given a random variable $X$ with $\E X = 0$, $\E X^2 = 1$ and 
$\E\,|X|^k < \infty$ for some integer $k \geq 3$,
the following three properties are equivalent:

\vskip2mm
(i) \ $\gamma_r(X) = 0$\, for all\, $r = 3,\dots, k-1$;

\vskip1mm
(ii) \ $\E H_r(X) = 0$ for all\, $r = 3,\dots, k-1$;

\vskip1mm
(iii) \ $\E X^r = \E Z^r$ for all\, $r = 3,\dots, k-1$.

\vskip2mm
\noindent
In this case
\be
\gamma_k(X) \, = \, \E H_k(X) \, = \, \E X^k - \E Z^k.
\en

The moments of $X$ may be expressed in terms of the normal moments. 
Indeed, the Chebyshev-Hermite polynomials have the generating function
$$
\sum_{k=0}^\infty H_k(x)\, \frac{z^k}{k!} = e^{xz - z^2/2}, \quad
x,z \in \C,
$$
or equivalently,
$$
e^{xz} \, = \,
e^{z^2/2} \sum_{i=0}^\infty H_i(x)\, \frac{z^i}{i!} \, = \,
\sum_{i,j=0}^\infty H_i(x)\,\frac{z^{i+2j}}{i!j!\, 2^j}.
$$
Expanding $e^{xz}$ into the power series with $x=X$ and comparing
the coefficients, we get
$$
\E X^k = k! \sum_{j=0}^{[k/2]} \frac{1}{(k-2j)!\, j!\, 2^j}\, 
\E\, H_{k-2j}(X).
$$

Now, let us describe the connection between the normal moments and the 
$\chi^2$-distance. The series in (12.5) is absolutely convergent as 
$N \rightarrow \infty$, when $f$ is analytic in $\C$. 
Hence, assuming condition (12.3) so that to guarantee the finiteness
of a Gaussian moment according to (11.1), we have the expansion
$$
f(t) = e^{-t^2/2}\, \sum_{k=0}^\infty \frac{c_k}{k!}\,(it)^k, \quad t \in \C.
$$
Moreover, the Parseval identity (12.2) gives
$$
\sum_{k=0}^\infty \frac{c_k^2}{k!} =
\int_{-\infty}^\infty \frac{p(x)^2}{\varphi(x)}\,dx \, = \, 1 + \chi^2(X,Z),
$$
and we arrive at the following relation:

\vskip5mm
{\bf Theorem 12.1} (\cite{B-C-G6}). {\sl If $\chi^2(X,Z) < \infty$, then
\be
\chi^2(X,Z) \, = \, \sum_{k=1}^\infty \frac{1}{k!} \big(\E H_k(X)\big)^2.
\en
Conversely, if a random variable $X$ has finite
moments of any order, and the series in $(12.7)$ is convergent, then $X$ 
has an absolutely continuous distribution with finite distance $\chi^2(X,Z)$.
}

\vskip3mm
It looks surprising that a simple sufficient condition for the existence of 
a density $p$ of $X$ can be formulated in terms of moments of $X$ only. 
If $X$ is bounded, then it has finite moments of any order, and 
the property $\chi^2(X,Z) < \infty$ just means that $p$ is in $L^2$.
In that case we may conclude that $X$ has an absolutely
continuous distribution with a square integrable density, if and only if
the series in $(12.7)$ is convergent.

The identity (12.7) admits a natural generalization in terms of 
the random variables
$$
X_t = \sqrt{t}\, X + \sqrt{1-t}\, Z,
$$
where $Z \sim N(0,1)$ is independent of $X$. Namely, if
$\chi^2(X,Z) < \infty$, then, for all $t \in [0,1]$,
\be
\chi^2(X_t,Z) \, = \, \sum_{k=1}^\infty \frac{t^k}{k!}\,(\E H_k(X))^2.
\en
This yields another description of the normal moments via 
the derivatives of the $\chi^2$-distance:
$$
\big(\E H_k(X)\big)^2 \, = \, \frac{d^k t}{dt^k}\,\chi^2(X_t,Z)\big|_{t=0}, 
\quad k = 1,2,\dots
$$

\vskip7mm
\section{{\bf Behavior of R\'enyi Divergence under Convolutions}}
\setcounter{equation}{0}

\noindent
It is natural to raise the following obvious question, which appears when 
describing convergence in the CLT in the $D_\alpha$-distance with 
$\alpha>1$: Does it remain finite for sums of 
independent summands with finite $D_\alpha$-distances? The answer is 
affirmative and is made precise by virtue of the relation
\be
D_\alpha(aX+bY || Z) \leq D_\alpha(X||Z) + D_\alpha(Y||Z),
\en
where $Z \sim N(0,1)$. It holds true for all independent random 
variables $X,Y$ and for all $a,b \in \R$ such that $a^2 + b^2 = 1$.
Equivalently,
\begin{eqnarray}
1 + (\alpha - 1)\,T_\alpha(aX+bY||Z) 
 & \leq & \nonumber \\
 & & \hskip-40mm 
\big(1+(\alpha - 1)\,T_\alpha(X||Z)\big)\, 
\big(1 + (\alpha - 1)\,T_\alpha(Y||Z)\big).
\end{eqnarray}

The statement may be extended by induction to finitely many 
independent summands $X_1,\dots,X_n$ by the relation
\be
D_\alpha(a_1X_1+ \dots + a_n X_n || Z) \, \leq \,
D_\alpha(X_1||Z) + \dots + D_\alpha(X_n||Z),
\en
where $a_1^2 + \dots + a_n^2 = 1$. 

Let us note that for the relative entropy there is a stronger property
$$
D(a_1X_1+ \dots + a_n X_n || Z) \, \leq \, 
\max\{D(X_1||Z),\dots,D(X_n||Z)\},
$$
which follows from the convexity property (5.2).
However, this is no longer true for $D_\alpha$.
Nevertheless, for the normalized sums 
$$
Z_n = \frac{X_1 + \dots + X_n}{\sqrt{n}}
$$
with i.i.d. summands, (13.3) guarantees a sub-linear growth of the 
R\'enyi divergence with respect to $n$, i.e.,
\be
D_\alpha(Z_n||Z) \leq n D_\alpha(X_1||Z).
\en

The relation (13.1) follows from the contractivity property (1.8),
applied in the plane $\Omega = \R \times \R$ to  the random 
vectors $\widetilde X = (X,Y)$ and $\widetilde Z = (Z,Z')$, 
where $Z'$ is an independent copy of $Z$. Since
$$
D_\alpha(\widetilde X||\widetilde Z) = 
D_\alpha(X||Z) + D_\alpha(Y||Z'),
$$
we have
$$
D_\alpha\big(S(\widetilde X)||S(\widetilde Z)\big) \, \leq \, 
D_\alpha(X||Z) + D_\alpha(Y||Z')
$$
for any Borel measurable function $S:\R^2 \rightarrow \R$. 
It remains to apply this inequality with the linear function 
$S(x,y) =ax + by$.

In the case $\alpha = 2$, there is a simple alternative 
argument, which relies upon normal moments only and the 
representation (12.7) from Theorem 12.1. With this approach, 
one may use the binomial formula
for the Chebyshev-Hermite polynomials
\be
H_k(ax+by) = \sum_{i=0}^k 
C_k^i\, a^i b^{k-i}\, H_i(x) H_{k-i}(y), \quad x,y \in \R, 
\en
which holds true whenever $a^2 + b^2 = 1$ and implies
$$
\E H_k(aX+bY) = \sum_{i=0}^k C_k^i\, a^i b^{k-i}\, 
\E H_i(X)\, \E H_{k-i}(Y).
$$
A further application of Cauchy's inequality leads to 
$$
1 + \chi^2(aX+bY,Z) \leq 
\big(1+\chi^2(X,Z)\big)\, \big(1 + \chi^2(Y,Z)\big),
$$
which is exactly (13.2) for $\alpha=2$.

By the way, (13.5) yields 
$$
\E\, H_k(aX+bZ) = a^k\,\E H_k(X),
$$
which may be used in the formula (12.8) with $a = \sqrt{t}$ 
and $b = \sqrt{1-t}$.

One may also ask whether or not $\chi^2(aX + bY,Z)$ remains 
finite, when $\chi^2(X,Z)$ is finite, and $Y$ is ``small" enough. 
To this aim, one may derive a simple upper bound
$$
1 + \chi^2(aX+bY,Z) \leq
\frac{1}{|a|}\,\big(1 + \chi^2(X,Z)\big)\,\E\,e^{Y^2/2}
$$
under the same assumption $a^2 + b^2 = 1$ with $a \neq 0$.

\vskip7mm
\section{{\bf Examples of Convolutions}}
\setcounter{equation}{0}

\noindent
Let us now describe two examples of i.i.d. random variables 
$X,X_1,\dots,X_n$ such that for the normalized sums $Z_n$ 
and any prescribed integer $n_0 > 1$, we have
\be
\chi^2(Z_1,Z) = \dots = \chi^2(Z_{n_0 - 1},Z) = \infty, \ \ 
{\rm but} \ \ \chi^2(Z_{n_0},Z) < \infty.
\en

\vskip2mm
{\bf Example 14.1.} Suppose that $X$ has a symmetric density 
of the form
\be
p(x) = \int_0^\infty \frac{1}{\sigma \sqrt{2\pi}}\,e^{-x^2/2\sigma^2}
\,d\pi(\sigma^2),  \quad x \in \R,                                     
\en
where $\pi$ is a probability measure on the positive half-axis. 
It may be described as a density of the random variable $\sqrt{\xi} Z$,
where $\xi>0$ is independent of $Z$ and has distribution $\pi$.
The finiteness of $\chi^2(X,Z)$ implies that $\sigma^2 < 2$ for 
$\pi$-almost all $\sigma$, that is, $\P\{\xi<2\} = 1$.
Assuming this, introduce the distribution function
$F(\ep) = \P\{\xi \leq \ep\}$, $0 \leq \ep \leq 2$. 
It is easy to see that
$$
1 + \chi^2(X,Z) = \E\,\frac{1}{\sqrt{\xi + \eta - \xi \eta}},
$$
where $\eta$ is an independent copy of $\xi$. This implies that
$\chi^2(X,Z) < \infty$, if and only if 
\be
\int_0^1 \frac{F(\ep)^2}{\ep^{3/2}}\,d\ep < \infty \quad {\rm and} \quad
\int_1^2 \frac{(1 - F(\ep))^2}{(2 - \ep)^{3/2}}\,d\ep < \infty.
\en
One may note that $p$ is bounded, if and only if
$\E \frac{1}{\sqrt{\xi}} < \infty$, that is,
$$
\int_0^1 \frac{F(\ep)}{\ep^{3/2}}\,d\ep < \infty,
$$
which is a weaker condition when the support of the distribution
of $\xi$ is bounded away from the point $2$.

Based on this description, we now investigate convolutions of $p$
defined in (14.2). The normalized sum 
$Z_n = \frac{1}{\sqrt{n}}\,(X_1 + \dots + X_n)$ 
has density of a similar type
$$
p_n(x) = \int_0^\infty \frac{1}{\sigma \sqrt{2\pi}}\,e^{-x^2/2\sigma^2}
\,d\pi_n(\sigma^2).                                 
$$
More precisely, if $\xi_1,\dots,\xi_n$ are independent copies of
$\xi$, that are independent of independent copies
$\zeta_1,\dots,\zeta_n$ of $Z$, then 
$$
Z_n = \frac{1}{\sqrt{n}}\,
\big(\sqrt{\xi_1} \zeta_1 + \dots + \sqrt{\xi_n} \zeta_n\big) =
\sqrt{S_n} Z,
$$
where the last equality is understood in the sense of distributions
with $S_n = \frac{1}{n}\,(\xi_1 + \dots + \xi_n)$ being
independent of $Z$. Thus, the mixing measure $\pi_n$ can be 
recognized as the distribution of $S_n$.  Note that 
$\P\{S_n < 2\} = 1$ is equivalent to $\P\{\xi < 2\} = 1$ which is fulfilled.
Therefore, by (14.3), $\chi^2(Z_n,Z) < \infty$, if and only if
$$
\int_0^1 \frac{F_n(\ep)^2}{\ep^{3/2}}\,d\ep < \infty \quad {\rm and} \quad
\int_1^2 \frac{(1 - F_n(\ep))^2}{(2 - \ep)^{3/2}}\,d\ep < \infty,
$$
where $F_n$ is the distribution function of $S_n$. Since
$F(\ep)^n \leq F_n(\ep) \leq F(\ep n)^n$ and
$$
(1-F(2-\ep))^n \leq 1-F_n(2-\ep) \leq (1 - F(2 - \ep n))^n,
$$
which are needed near zero, these conditions may be simplified to
\be
\int_0^1 \frac{F(\ep)^{2n}}{\ep^{3/2}}\,d\ep < \infty, \quad
\int_1^2 \frac{(1 - F(\ep))^{2n}}{(2 - \ep)^{3/2}}\,d\ep < \infty.
\en

Now, suppose that $\pi$ is supported on $(0,2-\delta)$ for some 
$\delta>0$, so that the second integral in (14.4) is convergent.
Moreover, let $F(\ep) \sim \ep^\kappa$ as $\ep \rightarrow 0$ 
with parameter $\kappa > 0$, where the equivalence is 
understood up to a positive factor. Then, the first integral 
in (14.4) will be finite, if and only if $n > 1/(4\kappa)$. 
Choosing $\kappa = \frac{1}{4 (n_0 - 1)}$, we obtain 
the required property (14.1). In this example, one may 
additionally require that $\E X^2 = \E \xi = 1$.

\vskip5mm
{\bf Example 14.2.} Consider a density of the form
$$
p(x) = \frac{a_k}{1 + |x|^{1/2k}}\,e^{-x^2/4},  \quad x \in \R,
$$
where $a_k$ is a normalizing constant, $k = n_0 - 1$, and let 
$f_1$ denote its Fourier transform (the characteristic function). 
Define the distribution of $X$ via its characteristic function
$$
f(t) = q f_1(t) + (1 - q)\,\frac{\sin(\gamma t)}{\gamma t}
$$
with a sufficiently small $q > 0$ and
$\gamma = (3\,(1 + q f_1''(0))/(1 - q))^{1/2}$. It is easy 
to check that $f''(0) = -1$, which guarantees that $\E X= 0$, 
$\E X^2 = 1$. One can show that the densities $p_n$ of $Z_n$ 
admit the two-sided bounds
$$
\frac{b_n'}{1 + |x|^{n/2k}}\,e^{-x^2/4} \leq p_n(x) \leq
\frac{b_n''}{1 + |x|^{n/2k}}\,e^{-x^2/4} \quad (x \in \R),
$$
up to some $n$-dependent factors. Hence, again we arrive 
at the property (14.1).

\vskip7mm
\section{{\bf Super-additivity of $\chi^2$ with Respect to Marginals}}
\setcounter{equation}{0}

\noindent
A multidimensional CLT requires to involve some other properties of 
the $\chi^2$-distance in higher dimensions. The contractivity under 
mappings, 
\be
\chi^2(S(X),S(Z)) \leq \chi^2(X,Z), 
\en
has already been mentioned in (1.8); it holds in a general setting
and for all R\'enyi divergences. The inequality (15.1) may be considerably 
sharpened, when the distance is measured to the standard normal law in 
$\Omega = \R^d$. In order to compare the behavior of $\chi^2$-divergence 
with often used information-theoretic quantities, recall the definition 
of the Shannon entropy and the Fisher information,
$$
h(X) = -\int p(x)\,\log p(x)\,dx, \quad
I(X) = \int \frac{|\nabla p(x)|^2}{p(x)}\,dx,
$$
where $X$ is a random vector in $\R^d$ with density $p$ (assuming that 
the above integrals exist). These functionals are known to be subadditive and 
super-additive with respect to the components: Writing $X = (X',X'')$ with 
$X' \in \R^{d_1}$, $X'' \in \R^{d_2}$ ($d_1 + d_2 = d$), one always has
\be
h(X) \leq  h(X') + h(X''), \quad 
I(X) \geq I(X') + I(X'')
\en
cf. \cite{Lieb1}, \cite{C}. Both $h(X)$ and $I(X)$ themselves are not 
yet distances, so one also considers the relative entropy and the relative 
Fisher information with respect to other distributions. In particular, 
in the case of the standard normal random vector $Z \sim N(0,I_d)$ 
and random vectors $X$ with mean zero and identity covariance 
matrix $I_d$, they are given by
$$
D(X||Z) = h(Z) - h(X), \quad I(X||Z) = I(X) - I(Z). 
$$
Hence, by (15.2), these distances are both super-additive, i.e.,
\bee
D(X||Z) 
 & \geq & 
D(X'||Z') + D(X''||Z''), \\
I(X||Z) 
 & \geq &
I(X'||Z') + I(X''||Z''),
\ene
where $Z'$ and $Z''$ are standard normal in $\R^{d_1}$ and 
$\R^{d_2}$ respectively (both inequalities become equalities, 
when $X'$ and $X''$ are independent).

One can establish a similar property for the $\chi^2$-distance, 
which can be more conveniently stated in the setting of a Euclidean 
space $H$, say of dimension $d$, with norm $|\cdot|$ and inner 
product $\left<\cdot,\cdot\right>$. If $X$ is a random vector in 
$H$ with density $p$, and $Z$ is a normal random vector with 
mean zero and an identity covariance operator
$I_d$, then (according to the abstract definition),
$$
\chi^2(X,Z) = \int_H \, \frac{p(x)^2}{\varphi(x)}\, dx - 1 = 
\int_H \frac{(p(x) - \varphi(x))^2}{\varphi(x)}\, dx,
$$
where $\varphi(x) = (2\pi)^{-d/2}\,e^{-|x|^2/2}$ ($x \in H$)
is the density of $Z$.

\vskip5mm
{\bf Theorem 15.1.} {\sl Given a random vector $X$ in $H$ and 
an orthogonal decomposition $H = H' \oplus H''$ into two linear 
subspaces $H',H'' \subset H$ of dimensions $d_1,d_2 \geq 1$, 
for the orthogonal projections 
$X' = {\rm Proj}_{H'}(X)$, $X'' = {\rm Proj}_{H''}(X)$, we have
\be
\chi^2(X,Z) \, \geq \, \chi^2(X',Z') + \chi^2(X'',Z''),
\en
where $Z,Z',Z''$ are standard normal random vectors in $H,H',H''$, 
respectively.
}

\vskip5mm
Note, however, that (15.3) will not become an equality for independent 
components $X',X''$.

Let us explain this inequality in the simple case $H = \R^2$ with
$d_1 = d_2 = 1$. The finiteness of $\chi^2(X,Z)$ means 
that the random vector $X = (\xi_1,\xi_2)$ has density 
$p = p(x_1,x_2)$ such that
$$
\int_{-\infty}^\infty \int_{-\infty}^\infty 
p(x_1,x_2)^2\, e^{(x_1^2 + x_2^2)/2}\,dx_1dx_2 < \infty.
$$
The Hermite functions 
$$
\varphi_{k_1,k_2}(x_1,x_2) = 
\varphi(x_1)\varphi(x_2)\, H_{k_1}(x_1) H_{k_2}(x_2)
$$
form a complete orthogonal system in $L^2(\R^2)$, where now 
$\varphi$ denotes the one dimensional standard normal density. 
Hence, the density $p$
admits a unique representation in the form of the exponential series
\be
p(x_1,x_2) \, = \, 
\varphi(x_1)\varphi(x_2) \sum_{k_1=0}^\infty \sum_{k_2=0}^\infty \,
\frac{c_{k_1,k_2}}{k_1! k_2!}\, H_{k_1}(x_1) H_{k_2}(x_2),
\en
converging in $L^2(\R,\frac{dx_1 dx_2}{\varphi(x_1)\varphi(x_2)})$ 
with coefficients (mutual normal moments)
$$
c_{k_1,k_2} \, = \, \int_{-\infty}^\infty \int_{-\infty}^\infty 
H_{k_1}(x_1) H_{k_2}(x_2)\,p(x_1,x_2)\,dx_1dx_2 \, = \,
\E\, H_{k_1}(\xi_1) H_{k_2}(\xi_2).
$$
Moreover, we have Parseval's equality
\be
1 + \chi^2(X,Z) \, = \, \int_{-\infty}^\infty \!\int_{-\infty}^\infty
\frac{p(x_1,x_2)^2}{\varphi(x_1)\varphi(x_2)}\,dx_1dx_2 \ = 
\sum_{k_1=0}^\infty \sum_{k_2=0}^\infty\,\frac{c_{k_1,k_2}^2}{k_1! k_2!}.
\en

Now, integrating (15.4) over $x_2$ and separately over $x_1$, 
we obtain similar representations for the marginal densities
\bee
p_1(x_1) 
 & = & 
\varphi(x_1) \sum_{k_1=0}^\infty\,
\frac{c_{k_1,0}}{k_1!}\, H_{k_1}(x_1), \\
p_2(x_2) 
 & = & 
\varphi(x_2) \sum_{k_2=0}^\infty\,
\frac{c_{0,k_2}}{k_2!}\, H_{k_2}(x_2).
\ene
Hence, by Theorem 12.1,
$$
\chi^2(\xi_1,\xi) \, = \, \sum_{k_1 = 1}^\infty \frac{c_{k_1,0}^2}{k_1!}, \quad
\chi^2(\xi_2,\xi) \, = \, \sum_{k_2 = 1}^\infty \frac{c_{0,k_2}^2}{k_2!} \qquad
(\xi \sim N(0,1)).
$$
But, these quantities appear as summands in (15.5).

\vskip7mm
\section{{\bf Edgeworth Expansion for Densities and Truncated Distances}}
\setcounter{equation}{0}

\noindent
The study of the central limit theorem for $T_\alpha$-distances including
the entropic CLT involves the Edgeworth expansion for densities under 
moment assumptions, which we briefly discussed in Section 7, cf. Theorem 7.4.
Let us state once more its particular case (7.7).

Suppose that we have independent copies $(X_n)_{n \geq 1}$ of 
a random variable $X$ with mean zero and variance one, and let 
$$
Z_n = \frac{X_1 + \dots + X_n}{\sqrt{n}}.
$$ 

\vskip2mm
{\bf Theorem 16.1.}
{\sl Assume that $X$ has a finite absolute moment of an integer order 
$k\ge 3$, and $Z_n$ admits a bounded density for some $n$. Then, for all 
$n$ large enough, $Z_n$ have continuous bounded densities $p_n$ satisfying
uniformly in $x \in \R$
\be
p_n(x)=\varphi(x) +  \varphi(x) \sum_{\nu=1}^{k-2}
\frac{q_{\nu}(x)}{n^{\nu/2}} + o\Big(\frac 1{n^{(k-2)/2}}\Big)\frac 1{1+|x|^k}.
\en
}

Recall that in this formula
\be
q_{\nu}(x) = \sum H_{\nu+2l}(x) \prod_{m=1}^{\nu}\frac 1{k_m!}
\Big(\frac{\gamma_{m+2}}{(m+2)!}\Big)^{k_m}, 
\en
where $\gamma_r$ denotes the $r$-th cumulant of $X$, and
the summation runs over all non-negative integer solutions 
$(k_1,\dots,k_{\nu})$ to the equation $k_1+2k_2+\dots+\nu k_{\nu}=\nu$ 
with $l=k_1+k_2+\dots+k_{\nu}$.
The sum in (16.2) defines a polynomial in $x$ of degree at most $3(k-2)$. 

For example, for $k=3$ (16.1) yields
$$
p_n(x) = \varphi(x) + \frac{\gamma_3}{3!\sqrt{n}}\,H_3(x) \varphi(x)
 + o\Big(\frac{1}{\sqrt{n}}\Big)\frac 1{1+|x|^3}, \quad \gamma_3 = \E X^3.
$$
More generally, if $\gamma_3 = \dots = \gamma_{k-1} = 0$,
that is, the first $k-1$ moments of $X$ coincide with those of 
$Z \sim N(0,1)$, then (16.1) is simplified to
$$
p_n(x)=\varphi(x) + \frac{\gamma_k}{k!}\,H_k(x) \varphi(x) \, n^{-\frac{k-2}{2}}
 + o\big(n^{-\frac{k-2}{2}}\big)\frac 1{1+|x|^k}
$$
with $\gamma_k = \E X^k - \E Z^k$ (cf. (12.6)).

The condition on the boundedness of $p_n$ for some 
$n = n_0$ (and then for all $n \geq n_0$) may be described in terms of the 
characteristic function of $X$ as the smoothness property (2.5). It appears 
as a necessary and sufficient condition for the uniform local limit theorem (2.6).

Theorem 16.1 allows one to investigate an asymptotic behaviour
of the ``truncated" $T_\alpha$-distances, that is, the integrals of the form
$$
I_\alpha(M) = 
\int_{|x|\le M}\Big(\frac{p_n(x)}{\varphi(x)}\Big)^\alpha\,\varphi(x)\,dx - 1.
$$
Choosing
$$
M = M_n(s) = \sqrt{2(s-1)\log n}
$$ 
with a fixed integer $s \geq 2$ and applying (16.1) with
$k = 2s$, we get an expansion
\be
I_\alpha(M_n(s)) = \sum_{j=1}^{s-1} b_j\,n^{-j} + o\big(n^{-(s-1)}\big)
\en
with coefficients
$$
b_j = \sum \frac{(\alpha)_{m_1 + \dots + m_{2j-1}}}{m_1! \dots m_{2j-1}!}\,
\int_{-\infty}^\infty q_1(x)^{m_1} \dots q_{2j-1}(x)^{m_{2j-1}}\,\varphi(x)\,dx.
$$
Here we use the standard notation 
$(\alpha)_m = \alpha (\alpha - 1) \dots (\alpha - m + 1)$, while
the sum extends over all integers $m_1, \dots, m_{2j-1} \geq 0$ 
such that 
$$
m_1 + 2m_2 + \dots + (2j-1)\,m_{2j-1} = 2j.
$$ 

In particular,
when $\gamma_j=0$ for $j=3,\dots,s-1$ $(s \geq 3)$, we have
$$
I_\alpha(M_n(s)) = \alpha (\alpha-1)\, 
\frac {\gamma_s^2}{2s!}\,\frac 1{n^{s-2}} + O\big(n^{-(s-1)}\big).
$$

Theorem 16.1 may also be used to control the truncated
$T_\infty$-distance. In particular, if $k \geq 4$ and $\E X^3 = 0$, 
we have $\gamma_3 = 0$, so that (16.1) takes the form
$$
p_n(x) =
\varphi(x) + \frac{\gamma_4}{24\, n} H_4(x) \varphi(x) +
\varphi(x) \sum_{\nu=3}^{k-2}
\frac{q_\nu(x)}{n^{\nu/2}} + o\Big(\frac{1}{n^{(k-2)/2}}\Big)\,
\frac{1}{1 + |x|^k},
$$
where the sum is empty in the case $k=4$. Let us rewrite this
representation as
$$
\frac{p_n(x) - \varphi(x)}{\varphi(x)} =
\frac{\gamma_4}{24\, n} H_4(x) + R_n(x) +
o\Big(\frac{1}{n^{(k-2)/2}}\Big)\,\frac{e^{x^2/2}}{1 + |x|^k}.
$$
If $|x| \leq M_n(s)$
with a fixed $s \geq 1$, then, using the property that the degree
of every polynomial $q_\nu(x)$ does exceed $3(k-2)$, we get
$$
|R_n(x)| \leq \sum_{\nu=3}^{k-2} \frac{|q_\nu(x)|}{n^{\nu/2}}
\leq C \frac{(\log n)^{\frac{3(k-2)}{2}}}{n^{3/2}} \leq \frac{C'}{n},
$$
while $|H_4(x)| \leq C(\log n)^2$,
where all constants do not depend on $x$. In addition,
$$
\frac{e^{x^2/2}}{1 + |x|^k} \leq n^{s-1}.
$$
In order to get the rate in the remainder term of order $1/n$, we
therefore need to assume that $\frac{k-2}{2} \geq s$, that is,
$k \geq 2s+2$. As a result, in this case
\be
\sup_{|x| \leq M_n(s)} \frac{|p_n(x) - \varphi(x)|}{\varphi(x)} =
O\Big(\frac{(\log n)^2}{n}\Big).
\en

Let us also mention that
Theorem 16.1 may be extended to the multidimensional case,
cf. \cite{B-RR}, Theorem 19.2. In that case, each $q_\nu$ represents
a polynomial whose coefficients involve mixed cumulants of the
components of $X$ up to order $\nu+2$. Correspondingly, we
obtain an expansion (16.3) and the asymptotic formula (16.4)
for the truncated Tsallis distances.

\vskip7mm
\section{{\bf Edgeworth Expansion and CLT for R\'enyi Divergences}}
\setcounter{equation}{0}

\noindent
Let us now state the central limit theorem with respect to the
$\chi^2$-distance, together with an expansion similarly to Theorem 7.1
about the rates of convergence in the entropic CLT. The main difference
is now the property that the finiteness of the $\chi^2$ distance
guarantees existence of all moments.

Thus, suppose that we have independent copies $X_n$ of a random variable
$X$ with mean zero and variance one, and let 
$$
Z_n = \frac{X_1 + \dots + X_n}{\sqrt{n}}
$$ 
denote the normalized sum of the first $n$ summands.

\vskip5mm
{\bf Theorem 17.1} (\cite{B-C-G6}). {\sl $\chi^2(Z_n,Z)\to 0$ 
as $n\to\infty$ if and only if $\chi^2(Z_n,Z)$ is finite for some $n=n_0$, and
\be
\E\,e^{tX} < e^{t^2} \quad \text{for all} \ t\ne 0.
\en
In this case, the $\chi^2$-divergence admits an Edgeworth-type expansion
\be
\chi^2(Z_n,Z) \, = \,
\sum_{j=1}^{s-2} \frac{c_j}{n^j} + O\Big(\frac{1}{n^{s-1}}\Big) \quad as
\ n \to\infty,
\en
which is valid for every $s=3,4,\dots$ with coefficients $c_j$ representing 
certain polynomials in the moments $\alpha_k = \E X^k$, $k = 3,\dots,j+2$.
}

\vskip3mm
For $s=3$ (17.2) becomes
$$
\chi^2(Z_n,Z)\,=\,\frac{\alpha_3^2}{6n} + O\Big(\frac{1}{n^2}\Big),
$$
and if $\alpha_3 = 0$ (as in the case of symmetric distributions), 
one may turn to the next moment of order $s=4$, for which (17.2) yields
\be
\chi^2(Z_n,Z)\,=\,\frac{(\alpha_4 - 3)^2}{24\,n^2} + 
O\Big(\frac{1}{n^3}\Big).
\en

The property $\chi^2(Z_n,Z) < \infty$ is rather close to the subgaussian 
condition (17.1). As we know, it implies that (17.1) is fulfilled for all 
$t$ large enough, as well as near zero due to the variance assumption. 
It may happen, however, that (17.1) is fulfilled for all $t \neq 0$ except 
just one value $t = t_0$, and then there is no CLT for the 
$\chi^2$-distance. 

The convergence to zero, and even the verification of the boundedness 
of $\chi^2(Z_n,Z)$ in $n$ is rather delicate. This problem was first 
studied in the early 1980's by Fomin \cite{F} in terms of the exponential series
for the density of $X$, 
$$
p(x) = \varphi(x)\, \sum_{k=1}^\infty\, 
\frac{\sigma_k}{2^k k!}\,H_{2k}(x).
$$
As a main result, he proved that $\chi^2(Z_n,Z) = O(\frac{1}{n})$ as $
n \rightarrow \infty$,  assuming that $p$ is compactly supported, 
symmetric, piecewise differentiable, 
such that the series coefficients satisfy $\sup_{k \geq 2} \sigma_k < 1$. 
This sufficient condition was verified for the uniform distribution on 
the interval $(-\sqrt{3},\sqrt{3})$ (this length is caused by 
the assumption $\E X^2 = 1$). 

A similar characterization as in Theorem 17.1 continues to hold in 
the multidimensional case for mean zero i.i.d. random vectors 
$X,X_1,X_2,\dots$ in $\R^d$ normalized  to have an identity covariance
matrix. Moreover, one may extend these results to the range of indexes 
$\alpha>1$, arriving at the following statement proved in \cite{B-C-G6}. 
As before, we denote by $\alpha^* = \frac{\alpha}{\alpha - 1}$ 
the conjugate index, and by $Z$ a random vector in $\R^d$ having 
a standard normal distribution. 

\vskip4mm
{\bf Theorem 17.2.} {\sl $D_\alpha(Z_n||Z)\to 0$ as $n\to\infty$, 
if and only if $D_\alpha(Z_n||Z)$ is finite for some $n$, and
\be
\E\,e^{\left<t,X\right>} < e^{\alpha^* |t|^2/2} \quad \text{for all} \ \, 
t \in \R^d, \ t\ne 0.
\en
In this case, $D_\alpha(Z_n||Z) = O(1/n)$, and even 
$$
D_\alpha(Z_n||Z) = O(1/n^2), 
$$
provided that the distribution of $X$ is symmetric about the origin.
}

\vskip3mm
Thus, in addition to the strength of normal approximation,
the convergence in the R\'enyi distance says a lot about the character 
of the underlying distributions. Thanks to the existence of all moments of 
$X$, an Edgeworth-type expansion for $D_\alpha$ and $T_\alpha$ also holds 
similarly to (17.2), involving the mixed cumulants of the components of $X$. 
Such expansion shows in particular an equivalence
$$
D_\alpha(Z_n||Z) \sim T_\alpha(Z_n||Z) \sim \frac{\alpha}{2}\,\chi^2(Z_n,Z),
$$
once these distances tend to zero. Moreover, an Edgeworth-type 
expansion allows to establish the monotonicity property of $D_\alpha(Z_n||Z)$ 
with respect to (large) $n$, in analogy with the known property of 
the relative entropy.

\vskip7mm
\section{{\bf Non-Uniform Local Limit Theorems}}
\setcounter{equation}{0}

\noindent
As a closely related issue, and in fact, as an effective application,
the Renyi divergence appears naturally in the study of normal approximation 
for densities $p_n$ of $Z_n$ in the form of non-uniform local limit theorems. 
In the setting of Theorem 17.2 we have:

\vskip4mm
{\bf Theorem 18.1} (\cite{B-C-G6}). {\sl Suppose that $D_\alpha(Z_n||Z)$ 
is finite for some $n$, and let the property $(17.4)$ be fulfilled. 
Then, for all $n$ large enough and for all $x \in \R^d$,
\be
|p_n(x) - \varphi(x)| \, \leq \, \frac{c}{\sqrt{n}}\, e^{-|x|^2/(2\alpha^*)},
\en
where the constant $c$ does not depend on $n$. Moreover, the rate 
$1/\sqrt{n}$ may be improved to $1/n$, if
the distribution of $X$ is symmetric about the origin.
}

\vskip5mm
Let us recall that $\alpha>1$ and $\alpha^* = \frac{\alpha}{\alpha - 1}$
denotes the conjugate index.

In dimension $d=1$ one can give a more precise statement, using the 
cumulants $\gamma_k$ of $X$, cf. \cite{B-C-G6}. In this case,
the basic moment assumption is that $\E X = 0$ and $\E X^2 = 1$.

\vskip3mm
{\bf Theorem 18.2.} {\sl Suppose that $D_\alpha(Z_n||Z)$ 
is finite for some $n$, and let the condition $(17.4)$ hold. If 
$\gamma_3 = \dots = \gamma_{s-1} = 0$ for some integer $s \geq 3$, then
\be
\sup_{x \in \R} \, \frac{|p_n(x) - \varphi(x)|}{\varphi(x)^{1/\alpha^*}} \, = \, 
\frac{a_s\,|\gamma_s|}{s!}\, n^{-\frac{s-2}{2}} + O\big(n^{-\frac{s-1}{2}}\big),
\en
where 
$$
a_s = \sup_{x \in \R}\, \big[\,\varphi(x)^{1/\alpha}\, |H_s(x)|\big].
$$
}

In the case $s=3$, there is no restriction on the cumulants, and 
we obtain the inequality (18.1). If $\E X^3 = 0$, then $\gamma_3 = 0$, 
and one may turn to the next moment of order $s = 4$, which yields
the rate $1/n$ in (18.1). As for the cumulant coefficient in (18.2), 
let us recall that
$$
\gamma_s = \E H_s(X) = \E X^s - \E Z^s
$$

To compare this result with the local limit theorem (16.1), note that, 
assuming the existence of moments of order $s$ and that $Z_n$ has 
a bounded continuous density $p_n$ for large~$n$, the Edgeworth 
expansion in Theorem 16.1 with $k = s$ allows to derive a much 
weaker statement, namely
$$
\sup_{x \in \R} \ (1 + |x|^s)\,|p_n(x) - \varphi(x)| \, = \, 
\frac{a_s'\,|\gamma_s|}{s!}\, n^{-\frac{s-2}{2}} + o\big(n^{-\frac{s-2}{2}}\big)
$$
with $s$-dependent factors
$$
a_s' = \sup_{x \in \R}\ (1 + |x|^s)\, |H_s(x)|\,\varphi(x).
$$

The condition (17.4) is almost necessary for 
the conclusion such as (18.2) and even for a weaker one. Indeed, 
arguing in the multidimensional setting, suppose that
\be
\liminf_{n \rightarrow \infty} \, 
\sup_{x \in \R^d} \ \frac{p_n(x) - \varphi(x)}{\varphi(x)^{1/\alpha^*}} 
\, < \, \infty,
\en
so that 
$$
p_n(x) \leq \varphi(x) + C \varphi(x)^{1/\alpha^*}
$$
for infinitely many $n$ with some constant $C$.
Multiplying this inequality by $e^{\left<t,x\right>}$ with $t \in \R^d$
and integrating over the variable $x$, we get
$$
\big(\E\,e^{\left<t,X\right>/\sqrt{n}}\big)^n = 
\E\,e^{\left<t,Z_n\right>} \leq e^{|t|^2/2} + 
A C\,e^{\alpha^* |t|^2/2}
$$
with constant $A = (2\pi)^{d/(2\alpha)}\,(\alpha^*)^{d/2}$.
Now substitute $t$ with $t\sqrt{n}$ and raise this inequality 
to the power $1/n$. Letting $n \rightarrow \infty$ along a suitable 
subsequence, we arrive in the limit at
\be
\E\,e^{\left<t,X\right>} \leq e^{\alpha^* |t|^2/2}
\en
for all $t \in \R^d$.
Thus, this subgaussian property is indeed implied by the local limit 
theorem in the form (18.3).

\vskip7mm
\section{{\bf Comments on the Proofs}}
\setcounter{equation}{0}

\noindent
Let us comment on some steps needed for
the proof of Theorems 17.1--17.2 and 18.1--18.2.

As we have already explained, for the non-uniform local limit theorem (18.1),
the condition (18.3) is necessary (which is a slightly weakened
form of (17.4)). A similar conclusion can be made about Theorem 17.2,
and it is sufficient to require in analogue with (18.3) that
\be
\liminf_{n \rightarrow \infty} \, 
\Big[\,\frac{1}{n}\,D_\alpha(Z_n||Z)\Big] = 0.
\en
For this aim, let us return to the bound (10.4) on the Laplace transform 
which for the random vector $Z_n$ in place of $X$ gives
$$
\E\,e^{\left<t,Z_n\right>} \leq B_n\, e^{\alpha^* |t|^2/2}, \quad 
t \in \R^d,
$$
where
$$
B_n = \Big(1 + (\alpha - 1)T_\alpha(Z_n||Z)\Big)^{1/\alpha} = \,
\exp\Big\{\frac{1}{\alpha^*}\,D_\alpha(Z_n||Z)\Big\}.
$$
Replacing $t$ with $t\sqrt{n}$, one may rewrite this inequality as
\be
\E\,e^{\left<t,X\right>} \leq B_n^{1/n}\, e^{\alpha^* |t|^2/2}, \quad 
t \in \R^d.
\en
Assuming that (19.1) holds, we get that $B_n^{1/n} \rightarrow 1$
along a suitable subsequence, and then
(19.2) yields in the limit the inequality (18.4).

Thus, if 
$$
\E\,e^{\left<t,X\right>} > e^{\alpha^* |t|^2/2}
$$ 
for some $t \in \R^d$, then (19.1) does not hold, that is,
$D_\alpha(Z_n||Z)\ge cn$ for all $n$ with some constant 
$c>0$. In this case $D_\alpha(Z_n||Z)$ has a maximal growth rate, 
in view of the sub-linear upper bound (13.4).

In order to obtain the strict inequality in (17.4), a more delicate analysis
is required in dimension $d=1$ about the asymptotic behavior of the integrals
\be
I_{nk} = \int_{-\infty}^\infty (\E\,e^{tZ_{nk}})^2\,e^{-\alpha^* t^2}\,dt =
\sqrt{\frac{\pi}{\alpha^*}}\ \E\,e^{\frac{1}{2\alpha^*} Z_{nk}^2}.
\en
Assuming that $D_\alpha(Z_n||Z) \rightarrow 0$,
or equivalently $\chi_\alpha(Z_n,Z) \rightarrow 0$, we have (18.4), that is,
\be
\psi(t) \, \equiv \, \E\,e^{t X} \, e^{-\alpha^* t^2/2} \leq 1, \quad t \in \R.
\en
Moreover, from the upper bound (10.6) it follows that, for any integer 
$k \geq \alpha/2$,
\be
\lim_{n \rightarrow \infty} I_{nk} = 
\sqrt{\frac{\pi}{\alpha^*}}\, \E\,e^{\frac{1}{2\alpha^*} Z^2} =
\sqrt{\pi\,(\alpha - 1)}.
\en

The function $\psi(t)$ is extended to the complex plane as an entire
function. Using the Taylor expansion of $\psi$ near zero, one can show that,
for any sufficiently small $\delta>0$, the integral in (19.3) when it is
restricted to the interval $|t| \leq \delta \sqrt{nk}$ behaves like
$\sqrt{\pi\,(\alpha - 1)} + o(1)$. Therefore, by (19.5), it is necessary that
\be
\int_{|t| > \delta} \psi(t)^{2nk}\,dt = o\Big(\frac{1}{\sqrt{n}}\Big).
\en
But, if we assume that $\psi(t_0) = 1$ for some $t_0 \neq 0$ and
recall (18.4), the above integral being restricted to a neighborhood
of $t_0$ will be at least $cn^{-1/2m}$ for some real $c>0$ and 
an integer $m \geq 1$ (using the Taylor expansion of $\psi$ near
the point $t_0$). This would contradict to (19.6), and as
a result, the inequality in (19.4) must be strict for any $t \neq 0$.

The necessity part in Theorem 17.2 in the multidimensional
situation can be reduced to dimension one, by applying
the contractivity property (1.8) of the functional $D_\alpha$.
Indeed, consider the i.i.d. sequence $\left<X_i,\theta\right>$ with 
unit vectors $\theta$. Then, assuming that $D_\alpha(Z_n||Z)\to 0$ 
as $n \to \infty$, we get
$$
D_\alpha(\left<Z_n,\theta\right>||\left<Z,\theta\right>) \, \leq \, 
D_\alpha(Z_n||Z)\to 0.
$$
Since $\E \left<X_i,\theta\right> = 0$, $\E \left<X_i,\theta\right>^2 = 1$, 
and $\left<Z,\theta\right> \sim N(0,1)$,
we may apply the one dimensional variant of this theorem which gives
$$
\E\,e^{r\left<X,\theta\right>} < e^{\alpha^* r^2/2} \quad \text{for all} \ \ 
r \ne 0,
$$
thus proving the necessity part in Theorem 17.2.

For the sufficiency part, one needs to explore the asymptotic behavior 
of the integrals
$$
(\alpha - 1)\, T_\alpha(Z_n||Z) = \int_{\R^d} w_n^\alpha(x)\, dx - 1, \quad
w_n(x) = p_n(x)\,\varphi(x)^{-1/\alpha^*}.
$$
For this aim, we split the integration into the four shell-type regions. 
The behavior of the integrals
$$
I_0 = \int_{|x| < M_n} w_n^\alpha(x)\, dx, \quad M_n = \sqrt{2\,(l-1)\log n},
$$
may be studied by virtue of the Edgeworth 
expansion for $p_n(x)$ on the balls $|x| < M_n$ with a non-uniform error term,
which we discussed in Section 16. Note that $I_0$ represents a truncated Tsallis
distance, which admits an Edgeworth-type expansion (16.3). It leads
to the required expansion (17.2) in Theorem 17.1 in dimension one.
In the multidimensional case, Theorem 16.1 is stated similarly as an expansion
\be
p_n(x) = \varphi_s(x) + \frac{o(n^{-(s-2)/2})}{1+|x|^s}, \quad
\varphi_s(x) = \varphi(x) +  \varphi(x) \sum_{k=1}^{s-2}
\frac{q_k(x)}{n^{k/2}},
\en
where each $q_k$ represents a polynomial whose coefficients involve 
mixed cumulant of the components of $X$ of order up to $k+2$ 
(cf. \cite{B-RR}, Theorem 19.2). In particular, if the distribution 
of $X$ is symmetric about the origin, then $q_1(x) = 0$ and there is no 
$1/\sqrt{n}$ term in (19.7). In this way, we will arrive at the Edgeworth-type 
expansion for $I_0$ similarly to dimension one, which implies that 
$I_0 - 1 = O(1/n)$ in general, and $I_0 - 1 = O(1/n^2)$ when 
the distribution of $X$ is symmetric.

It remains to establish a polynomial smallness of the integrals
\bee
I_1 = \int_{|x| > x_0 \sqrt{n}}\, w_n^\alpha(x)\, dx, \ \ \ \
 &
I_2 = \int_{x_1 \sqrt{n} < |x| < x_0 \sqrt{n}}\, w_n^\alpha(x)\, dx, \\
I_3 = \int_{M_n < |x| < x_1 \sqrt{n}}\, w_n^\alpha(x)\, dx 
 &
\ene
with $x_1>0$ being any fixed small number, and $x_0>x_1$ depending 
on the density $p$. For simplicity, let us assume that $D(X||Z)$ is finite, 
and rewrite the condition (17.4) as
\be
\psi(t)  \equiv \E\,e^{\left<t,X\right>} \, e^{-\alpha^* |t|^2/2}< 1 \quad 
{\rm for \ all} \ t \neq 0.
\en

To bound these integrals, one should involve Theorem 10.1.
By the pointwise bound (10.9), we have
\be
w_n^\alpha(x) \, \le \, c_{\alpha,d}\,
\delta^{\alpha n} \,\psi\Big(-\frac{x}{\alpha^*\sqrt n}\Big)^{\alpha n/2} 
\en
in the region $|x| \geq x_0 \sqrt{n}$ for some $x_0 > 0$, while, by (10.8),
for all $x \in \R^d$,
\be
w_n^\alpha(x) \, \le \, c_{\alpha,d}\, n^{\alpha d/2}\, 
\psi\Big(-\frac{x}{\alpha^*\sqrt n}\Big)^{\alpha(n - n_\alpha)} 
\en
with some $(\alpha,d)$-depending constants, where
$n_\alpha = \max(2,\alpha^*)$ and $n \geq n_\alpha$.
Hence, by (19.9), $I_1$ has an exponential decay with respect to $n$ 
due to (19.8) and the integrability of $\psi$ with any power $k \geq \alpha$,
cf. (10.7). For the region of $I_2$, thanks to (19.8), we have
$\delta = \max_{x_0 \leq |t| \leq x_1} \psi(\frac{t}{\alpha^*}) < 1$. 
Hence, by (19.10), $I_2$ has an exponential decay as well.

Finally, using the analyticity of the characteristic function $f$ of $X$
near zero, we have
$$
\psi(u) \leq e^{-(\alpha^* - 1) |u|^2/4}
$$ 
in a sufficiently small ball $|u|<r$. As a consequence,
$I_3 = o(n^{-\beta})$ for any prescribed value $\beta > 2$ by
choosing a sufficiently large value of the parameter $l$ 
in the definition of $M_n$.

Theorems 18.1-18.2 are proved with similar arguments.

\vskip7mm
\section{{\bf Some Examples and Counter-Examples}}
\setcounter{equation}{0}

\noindent
Given a random variable $X$ with $\E X=0$, $\E X^2=1$, consider the function 
$\psi(t)=e^{-t^2}\, \E\, e^{tX}$ ($t \in \R$). As before, put
$Z_n = (X_1 + \dots + X_n)/\sqrt{n}$,
where $X_k$ are independent copies of $X$. One immediate consequence of 
Theorem 17.1 with $n_0 = 1$ is the following characterization. As usual,
$Z$ denotes a standard normal random variable.

\vskip3mm
{\bf Theorem 20.1.} {\sl Let the random variable $X$ have a density $p$
such that
\be
\int_{-\infty}^\infty p(x)^2\,e^{x^2/2}\,dx < \infty.
\en
Then $\chi^2(Z_n,Z) \rightarrow 0$ as $n \rightarrow \infty$, 
if and only if
\be
\psi(t) < 1 \quad for \ all \ \ t \neq 0.
\en
}
\vskip2mm
The assumption (20.1) is fulfilled, for example, when $X$ is bounded and has 
a square integrable density. 

We now illustrate Theorem 20.1 and 
the more general Theorem 17.2 with a few examples (mostly in dimension one).

\vskip2mm
{\bf Uniform distribution.} If $X$ is uniformly distributed on the segment 
$[-\sqrt 3,\sqrt 3]$, then
\be
\E\,e^{tX} =\frac{\sinh(t\sqrt 3)}{t\sqrt 3} < e^{t^2/2}, \quad t \in \R \ \ 
(t \neq 0),
\en
so that (20.2) holds. In this case the first moments are given by 
$\alpha_2 = 1$, $\alpha_3 = 0$, $\alpha_4 = \frac{9}{5}$, and
by Theorem 20.1, $\chi^2(Z_n,Z) \rightarrow 0$ as $n \rightarrow \infty$.
Moreover, Theorem 17.1 provides an asymptotic expansion 
$$
\chi^2(Z_n,Z) \, = \, \frac{3}{50\,n^2} + O\Big(\frac{1}{n^3}\Big).
$$

In fact, the property (20.3) means that the condition (17.4) of a more general 
Theorem 17.2 is fulfilled for all $\alpha > 1$, and
we obtain a stronger assertion 
$$
D_\alpha(Z_n||Z) = \frac{\alpha}{2}\,\chi^2(Z_n,Z) + O\Big(\frac{1}{n^3}\Big).
$$

{\bf Convex mixtures of centered Gaussian measures.} 
Following Example 14.1, consider the densities of the form
$$
p(x) = \int_0^2 \frac{1}{\sigma \sqrt{2\pi}}\,e^{-x^2/2\sigma^2}
\,d\pi(\sigma^2),  \quad x \in \R,                                     
$$
where $\pi$ is a probability measure on the interval $(0,2)$ with
$\int_0^2 \sigma^2 d\pi(\sigma^2) = 1$. The random variable $X$ 
with this density has mean zero and variance one. In addition, 
the Laplace transform
$$
\E\,e^{tX} = \int_0^2 e^{\sigma^2 t^2/2} \,d\pi(\sigma^2)
$$
satisfies (20.2). Recall that
$\chi^2(Z_n,Z) < \infty$ for some $n$, if and only if the distribution 
function $F(\ep) = \pi(0,\ep]$ satisfies the condition (14.4).
Thus, $\chi^2(Z_n,Z) \rightarrow 0$ as 
$n \rightarrow \infty$, if and only if the measure $\pi$ satisfies (14.4) for some $n$.
In this case, we obtain the expansion (17.3) which reads
$$
\chi^2(Z_n,Z) \, = \, \frac{3\,(m-1)^2}{8 n^2} + O\Big(\frac{1}{n^3}\Big), 
\quad m = \int_0^\infty \sigma^4 \,d\pi(\sigma^2).
$$

\vskip2mm
{\bf Distributions with Gaussian component.}
Consider random variables
$$
X = a\xi + bZ \quad (a^2 + b^2 = 1, \ a,b>0),
$$
assuming that $\E \xi = 0$, $\E\xi^2 = 1$, with $Z \sim N(0,1)$ 
being independent of $\xi$. The distribution of $X$ is a convex mixture 
of shifted Gaussian measures with variance $b^2$. It admits a density
$$
p(x) = \frac{1}{b}\,\E\,\varphi\Big(\frac{x - a\xi}{b}\Big), \quad x \in \R.
$$
To ensure finiteness of $\chi^2(Z_n,Z)$ with some $n$, 
the Laplace transform of the distribution of $\xi$ should 
admit a subgaussian bound
\be
\E\,e^{t \xi} \leq e^{\sigma^2 t^2/2}, \quad t \in \R,
\en
with some $\sigma>0$, in which case $\E\,e^{c \xi^2} < \infty$ whenever 
$c < 1/(2\sigma^2)$ (due to the moment assumptions on $\xi$, necessarily 
$\sigma \geq 1$). Let $\eta$ be an independent copy of $\xi$.
Squaring $p(x)$ and using $2\xi \eta \leq \xi^2 + \eta^2$, we get
\bee
1 + \chi^2(X,Z) 
 & = &
\frac{1}{\sqrt{1 - a^4}}\,\E\,e^{\frac{a^2}{2b^2(1 + a^2)}\,
(2\xi \eta - a^2(\xi^2 + \eta^2))} \\
 & \leq &
\frac{1}{\sqrt{1 - a^4}}\,\Big(\E\,e^{\frac{a^2}{2(1 + a^2)}\,\xi^2}\Big)^2.
\ene
Hence, $\chi^2(X,Z) < \infty$ whenever 
$a < a_\sigma = \frac{1}{\sqrt{\sigma^2 - 1}}$,
which is automatically fulfilled in the case $\sigma^2 \leq 2$. 
Moreover, by (20.4), the condition (20.2) is fulfilled as well. Thus,
$\chi^2(Z_n,Z) \rightarrow 0$ as $n \rightarrow \infty$, if 
$a < \frac{1}{\sqrt{\sigma^2 - 1}}$. In the case $\sigma^2 \leq 2$, 
this convergence holds for all admissible $(a,b)$.

\vskip4mm
{\bf Distributions with finite Gaussian moment.} If a random variable 
$X$ with mean zero and variance one has finite $\E\,e^{cX^2}$ 
($c>0$), then (20.4) is fulfilled for some $\sigma \geq 1$.
This means that (17.4) is fulfilled for any $\alpha>1$ such that
$\alpha^* < \sigma^2$. Therefore, if $D_\alpha(X||Z) < \infty$, then
$D_\alpha(Z_n||Z) \rightarrow 0$ with any 
$\alpha < \frac{\sigma^2}{\sigma^2 - 1}$.

\vskip2mm
{\bf Conditions in terms of exponential series.}
Consider a symmetric density of the form
$$
p(x) = \varphi(x) + \varphi(x)  \sum_{k=2}^\infty \frac{\sigma_k}{2^k k!}\, H_{2k}(x),
\quad x\in\R,
$$
so that $\E X=0$ and $\E X^2 = 1$ for the random variable $X$ with density $p$. 
As we discussed in
Section 12, the condition (20.1) is fulfilled, if and only if the series
$$
\chi^2(X,Z) \, = \, \sum_{k=2}^\infty \frac{(2k)!}{4^k\,k!^2}\,\sigma_k^2 
\, \sim \, \sum_{k=2}^\infty \frac{1}{\sqrt{k}}\,\sigma_k^2
$$
is convergent (which is fulfilled automatically, when $p$ is compactly 
supported and bounded). Using the identity
$$
\int_{-\infty}^\infty e^{tx} H_{2k}(x) \varphi(x)\,dx =
t^{2k}\,e^{t^2/2}
$$
and assuming additionally that 
$\sup_{k\ge 2} \sigma_k \le 1$, we also have
$$
\E\,e^{tX} \, = \, e^{t^2/2}\,
\bigg[1+\sum_{k=2}^{\infty}
\frac{\sigma_k}{k!}\Big(\frac{t^2}2\Big)^k\bigg]
 \, \le \, e^{t^2/2}\Big(e^{t^2/2}-\frac{t^2}{2}\Big) \, < \, e^{t^2}, 
\quad t\ne 0.
$$
Hence, in this case, by Theorem 20.1, $\chi^2(Z_n,Z) \rightarrow 0$ as 
$n \rightarrow \infty$. Moreover,  according to the expansion (17.3),
$\chi^2(Z_n,Z) = O(1/n^2)$.
This assertion strengthens the result of \cite{F}.

\vskip2mm
{\bf Log-concave probability distributions.} More examples including those 
in higher dimensions illustrate the multidimensional Theorem 17.2 within 
the class of densities $p(x) = e^{-V(x)}$ supported on some open convex 
region $\Omega \subset \R^d$. Let $V$ be a $C^2$-convex function with Hessian 
satisfying $V''(x) \geq c\,I_d$ in the sense of positive definite 
matrices ($c > 0$). The probability measures with such densities are known 
to admit logarithmic Sobolev inequalities (via the Bakry-Emery criterion).
In particular, they satisfy transport-entropy inequalities which in turn
can be used to get a subgaussian bound
$$
\E\,e^{\lambda g(X)} \leq e^{\lambda^2/(2c)}, \quad \lambda \in \R.
$$
Here, $g$ may be any function on $\R^d$ with a Lipschitz semi-norm
$\|g\|_{\rm Lip} \leq 1$, such that $\E\,g(X) = 0$ (cf. \cite{Led}, \cite{B-G1}).
In particular, if $\E X = 0$, one may choose linear functions 
$g(x) = \left<x,\theta\right>$, $|\theta| = 1$. Hence, the condition (17.4)  
is fulfilled for $c > \frac{1}{\alpha^*}$. Moreover, the property 
$D_\alpha(X||Z) < \infty$ will also hold in this case. Indeed, by the convexity
of $V$, we have
$$
V(x) \geq V(x_0) + \left<V'(x_0), x-x_0\right> + \frac{c}{2}\,|x-x_0|^2
$$
for all $x,x_0 \in \Omega$, which gives an upper pointwise bound
$p(x) \leq c_0\,e^{\left<v,x\right> - \frac{c}{2} |x|^2}$, 
$x \in \Omega$,
with some $c_0>0$ and $v \in \R^d$. Applying Theorem 17.2, we get:

\vskip4mm
{\bf Corollary 20.2.} {\sl If a random vector $X$ in $\R^d$ with mean zero 
and identity covariance matrix has density $p = e^{-V}$ such that 
$V'' \geq c\,I_d$ $(0 < c \leq 1)$ on the supporting open convex region,
then $D_\alpha(Z_n||Z) \rightarrow 0$ as $n \rightarrow \infty$ for all
$\alpha < \frac{1}{1-c}$.
}

\vskip5mm
{\bf Convolution of Bernoulli with Gaussian.}
One might wonder whether or not it is possible to replace the condition 
(17.1) in Theorem 17.1 with a slightly weaker requirement 
$\E\,e^{tX} \leq e^{t^2}$ (hoping that the strict inequality 
would automatically hold in view of the assumption $\E X^2 = 1$). 
The answer is negative, including the $D_\alpha$-case
as in Theorem 17.2 with its condition (17.4). Put 
$\beta = \frac{\alpha}{\alpha - 1}$ for a fixed $\alpha>1$.

\vskip5mm
{\bf Theorem 20.3.} {\sl There exists a random variable $X$ with 
$\E X = 0$, $\E X^2 = 1$, and $D_\alpha(X||Z) < \infty$ for $Z \sim N(0,1)$,
such that the inequality
\be
\E\,e^{tX} < e^{\beta t^2/2}
\en
is fulfilled for all $t \neq 0$ except for exactly one point $t_0 \neq 0$.
}

\vskip4mm
Since (20.5) is violated (although at one point only), Theorem 17.2 
implies that convergence $D_\alpha(Z_n||Z) \rightarrow 0$ does not hold.
Let us describe explicitly one family of distributions satisfying the assertion 
of this theorem. Returning to one of the previous examples, consider 
random variables of the form $X = a\xi + bZ$ $(a,b>0)$, assuming that 
$\xi$ takes two values $q$ and $-p$ with probabilities $p$ and $q$,
respectively ($p,q > 0$, $p + q = 1$), while $Z \sim N(0,1)$ is independent 
of $\xi$. Then $\E X = 0$, and we have the constraint
\be
\E X^2 = pq\, a^2 + b^2 = 1.
\en
The condition $D_\alpha(X||Z) < \infty$ obviously holds since $b<1$. 

It is known that the smallest positive
constant $\sigma^2 = \sigma^2(p,q)$ in the inequality
\be
\E\,e^{t\xi} = p e^{qt} + q e^{-pt} \leq e^{\sigma^2 t^2/2}, 
\quad t \in \R,
\en
is given by 
$$
\sigma^2 = \frac{p-q}{2\,(\log p - \log q)}
$$
(called the subgaussian constant for the Bernoulli distribution, 
cf \cite{B-H-T}, Proposition 2.3). Hence
$$
\E\,e^{tX} \leq e^{(\sigma^2 a^2 + b^2)\, t^2/2}, \quad t \in \R,
$$
with an optimal constant in the exponent on 
the right-hand side. Thus, according to (20.5), we get another constraint 
$\sigma^2 a^2 + b^2 = \beta$. Combining it with (20.6), we find that
$$
a^2 = \frac{\beta - 1}{\sigma^2 - pq}, \quad
b^2 = \frac{\sigma^2 - \beta pq}{\sigma^2 - pq},
$$
which makes sense if $\sigma^2 > \beta pq$. 
It is easy to see that (20.7) becomes an equality for 
$t_0 = -2\,(\log p - \log q)$, which is a unique non-zero point with 
such property, as long as $p \neq q$. Therefore, the random variable 
$X$ satisfies the assertion of Theorem 20.3, if and only if 
$$
\frac{p-q}{2\,(\log p - \log q)} > \beta pq.
$$
This inequality does hold, provided that $p$ is sufficiently close to 0 or 1,
although it is not true for a neighborhood of $1/2$ (since at this point
the inequality becomes $1 > \beta$).

\vskip7mm
\section{{\bf Sufficient Conditions for Convergence in $D_\alpha$}}
\setcounter{equation}{0}

\noindent
Following Kahane \cite{K}, a random vector $X$ in $\R^d$ is called 
subgaussian, or its distribution is called subgaussian, if 
$\E\,e^{cX^2} < \infty$ for some
$c>0$. Equivalently,  it has subgaussian tails, i.e.
$$
\P\{|X| \geq r\} \leq C \,e^{-cr^2/2}, \quad r > 0,
$$
for some positive constants $C$ and $c$ which do not depend on $r$
(here one may choose $C=2$ at the expense of a smaller value of $c$).
If $X$ has mean zero, this property may also be stated in terms of the
Laplace transform via the relation
\be
\E\,e^{\left<t,X\right>} \leq e^{\sigma^2 |t|^2/2}, \quad t \in \R^d,
\en
which should hold with some $\sigma^2$. Then, the optimal value 
$\sigma^2 = \sigma^2(X)$ is often called the subgaussian constant
of the distribution of $X$.

Let $(X_n)_{n \geq 1}$ be independent copies of a random vector $X$ 
in $\R^d$ with mean zero and identity covariance matrix, and let 
\be
Z_n = \frac{1}{\sqrt{n}}\,(X_1 + \dots + X_n).
\en
As before, denote by $Z$ a standard normal random vector in $\R^d$.
As we know from Theorem 17.2, a necessary condition for the
convergence $D_\alpha(Z_n||Z) \rightarrow 0$ as $n \rightarrow \infty$
for an index $\alpha>1$ is that $X$ is subgaussian with a subgaussian constant
satisfying $\sigma^2(X) \leq \alpha^*$.

Let us now comment on the other necessary condition in Theorem 17.2,
$D_\alpha(Z_n||Z) <\infty$ for some $n = n_\alpha$ (note that in
the previous practical examples we assume that it holds with $n_0 = 1$).
As we know from Part II, it is stronger than the boundedness of
density $p_n$ of $Z_n$. However, together with subgaussianity, the
boundedness of densities turns out to be sufficient for the convergence in
$D_\alpha$ within a corresponding range of indices.

\vskip5mm
{\bf Theorem 21.1.} {\sl 
Suppose that $Z_n$ have bounded densities for large $n$. Then, under
the condition $(21.1)$, we have $D_\alpha(Z_n||Z) \rightarrow 0$ as 
$n \rightarrow \infty$ for any $\alpha < \frac{\sigma^2}{\sigma^2 - 1}$ 
(that is, if $\alpha^* > \sigma^2$).
}

\vskip2mm
Note that necessarily $\sigma^2 \geq 1$ due to the assumption that $X$ has 
mean zero and unit covariance matrix (by comparing both sides of (21.1)
with small $t$). The value $\sigma^2 = 1$ is quite possible, like in the example 
of the uniform distribution from the previous section. This case, which we discuss
in details in the next sections, corresponds to the convergence
of $Z_n$ in all $D_\alpha$ simultaneously.

\vskip4mm
{\bf Corollary 21.2.} {\sl  The convergence $D_\alpha(Z_n||Z) \rightarrow 0$ 
as $n \rightarrow \infty$ holds true for any $\alpha$, if and only if
$Z_n$ have bounded densities for large $n$, and $\sigma^2(X) = 1$.
}

\vskip5mm
Let us recall that the boundedness of densities $p_n$ of $Z_n$ for some
(and then for all large) $n$ may be related to the integrability property of the Laplace 
transform along the imaginary axis in the complex plane as the smoothness condition
$$
\int |f(t)|^\nu\,dt < \infty \quad {\rm for \ some} \ \nu \geq 1,
$$
where $f(t) = \E\,e^{i\left<t,X\right>}$, $t \in \R^d$.

Theorem 21.1 follows from Theorem 17.2 and the following general observation
from \cite{B4}. If a subgaussian random vector $X$ with $\sigma^2(X) \leq \sigma^2$
has a density bounded by $M$, then the densities $p_n$ of $Z_n$, $n \geq 2$,
admit an upper pointwise bound
$$
p_n(x) \leq e^{d/2} M\,\exp\Big\{-\frac{n-1}{2n \sigma^2}\,|x|^2\Big\}, \quad
x \in \R^d.
$$
This implies
$$
\int \Big(\frac{p_n(x)}{\varphi(x)}\Big)^\alpha \varphi(x)\,dx \leq c
\int \exp\Big\{-\frac{1}{2}\,
\Big(\alpha\, \frac{n-1}{n \sigma^2} - (\alpha-1)\Big)\,|x|^2\Big\}\, dx,
$$
where the constant $c$ does not depend on $x$. The last integral is
convergent for sufficiently large $n$, as long as $\alpha^* > \sigma^2$, and
then we conclude that $D(Z_n||Z)$ is finite.

\vskip7mm
\section{{\bf Strictly Subgaussian Distributions}}
\setcounter{equation}{0}

\noindent
In Corollary 21.2 we obtain a rather interesting class of subgaussian probability 
distributions. In what follows we restrict ourselves to dimension $d=1$.

\vskip4mm
{\bf Definition 22.1.} We say that a subgaussian random variable $X$ is strictly 
subgaussian, or the distribution of $X$ is strictly subgaussian, 
if the inequality
\be
\E\,e^{tX} \leq e^{\sigma^2 t^2/2}, \quad t \in \R,
\en
holds with (best possible) constant $\sigma^2 = \Var(X)$. 

\vskip4mm
Thus, if the random variable $X$ has mean zero and variance one,
the normalized sums (21.2) satisfy $D_\alpha(Z_n||Z) \rightarrow 0$, if and only if
$Z_n$ have bounded densities for large $n$, and if $X$ is strictly subgaussian.

This class was apparently first introduced in an explicit form by Buldygin and 
Kozachenko in \cite{B-K1} under the name ``strongly subgaussian" and then
analysed in more details in their book \cite{B-K2}. Recent investigations include the
work by Arbel, Marchal and Nguyen \cite{A-M-N} providing some examples and 
properties and by Guionnet and Husson \cite{G-H}. In the latter paper, (22.1) appears
as a condition for the validity of large deviation principles for the largest
eigenvalue of Wigner matrices with the same rate function as in the case of
Gaussian entries.

A simple sufficient condition for the strict subgaussianity was given by 
Newman in terms of location of zeros of the characteristic function
$$
f(z) = \E\,e^{izX}, \quad z \in \C,
$$
which is extended, by the subgaussian property, from the real line to the complex 
plane as an entire function of order at most 2. As was stated in \cite{N1}, $X$ 
is strictly subgaussian, as long as $f(z)$ has only real zeros in $\C$ 
(a detailed proof was later given in \cite{B-K2}). Such probability distributions 
form an important class denoted by $\mathfrak L$, introduced and studied 
by Newman in the mid 1970's in connection with the Lee-Yang property which 
naturally arises in the context of ferromagnetic Ising models, 
cf. \cite{N1,N2,N3,N-W}. This class is rather rich; it is closed under 
infinite convergent convolutions and under weak limits. For example, it 
includes Bernoulli convolutions and hence convolutions of uniform 
distributions on bounded symmetric intervals.
Some classes of strictly subgaussian distributions including those
outside $\mathfrak L$ have been recently discussed in \cite{B-C-G8}.

Let us turn to the basic properties of strictly subgaussian distributions.
Immediate consequences of the inequality (22.1) are the finiteness of 
moments of all orders of $X$ and in particular the relations 
$$
\E X =0 \quad   \text{ and} \quad \E X^2 \leq \sigma^2,
$$
which follow by an expansion of both sides of (22.1) around 
the point $t=0$. Thus, the word ``strictly" in Definition 22.1 reflects 
the requirement that the variance of $X$ is exactly $\sigma^2$ 
in contrast with the usual subgaussianity, when (22.1) is required 
to hold for all $t$ with some constant $\sigma^2$.
In addition to the properties $\E X = 0$ and $\E X^2 = \sigma^2$, 
the Taylor expansion of the exponential function in (22.1) 
around zero implies as well that necessarily
$$
\E X^3 = 0, \quad \E X^4 \leq 3\sigma^4.
$$
Here an equality is attained for symmetric normal distributions
(but not exclusively so).

The next statements are elementary.

\vskip4mm
{\bf Proposition 22.2.} {\sl If the random variables 
$X_1,\dots,X_n$ are independent and strictly subgaussian, 
then their sum $X = X_1 + \dots + X_n$ is strictly subgaussian.
}

\vskip5mm
{\bf Proposition 22.3.} {\sl If a sequence of strictly subgaussian 
random variables $(X_n)_{n \geq 1}$ converges weakly in 
distribution to a random variable $X$ with finite second moment, 
and $\Var(X_n) \rightarrow \Var(X)$ as $n \rightarrow \infty$,
then $X$ is strictly subgaussian.
}

\vskip2mm
Combining Proposition 22.2 with Proposition 22.3, we obtain:

\vskip5mm
{\bf Corollary 22.4.} {\sl Suppose that independent, strictly 
subgaussian random variables $(X_n)_{n \geq 1}$ have variances 
satisfying $\sum_{n=1}^\infty \Var(X_n) < \infty$. Then the series
$$
X = \sum_{n=1}^\infty X_n
$$ 
represents a strictly subgaussian random variable.
}

\vskip5mm
Here, the assumption that  $\sum_{n=1}^\infty \Var(X_n) < \infty$
ensures that the series $\sum_{n=1}^\infty X_n$ is convergent with
probability one (by the Kolmogorov theorem), so that the partial sums
of the series are weakly convergent to the distribution of $X$.

Thus, the class of strictly subgaussian distributions is closed in 
the weak topology under infinite convolutions. Obviously, it is also 
closed when taking convex mixtures.

\vskip3mm
{\bf Proposition 22.5.} {\sl If $X_n$ are strictly subgaussian random 
variables with $\Var(X_n) = \sigma^2$, and $\mu_n$ are distributions 
of $X_n$, then for any sequence $p_n \geq 0$ such that 
$\sum_{n=1}^\infty p_n = 1$, the random variable with distribution
$$
\mu = \sum_{n=1}^\infty p_n \mu_n
$$
is strictly subgaussian and has variance $\Var(X) = \sigma^2$.
}

\vskip5mm
One should also mention that, if $X$ is strictly subgaussian, then
$\lambda X$ is strictly subgaussian for any $\lambda \in \R$.

Finally, let us give a simple sufficient condition for a stronger property
in comparison with (22.2). Introduce the log-Laplace transform
$$
K(t) = \log \E\,e^{tX}, \quad t \in \R.
$$

\vskip3mm
{\bf Proposition 22.6} (\cite{B-C-G8}). {\sl Let $X$ be a non-normal strictly 
subgaussian random variable. If the function
$
t \rightarrow K(\sqrt{|t|})
$
is concave on the half-axis $t>0$ and concave on the half-axis $t<0$,
then, for any $t_0 > 0$, there exists $c = c(t_0)$,
$0 < c < \sigma^2 = \Var(X)$, such that
\be
\E\,e^{tX} \leq e^{c t^2/2}, \quad |t| \geq t_0.
\en
}

In a more compact form, for any $t_0>0$,
\be
\sup_{|t| \geq t_0} \Big[ \frac{1}{t^2} \log \E\,e^{tX}\Big] < 
\frac{1}{2}\,\Var(X).
\en

\vskip5mm
\section{{\bf Zeros of Characteristic Functions}}
\setcounter{equation}{0}

\noindent
One may try to describe the class of all strictly subgaussian 
distributions, for example, in terms of the characteristic function
\be
f(z) = \E\,e^{izX}, \quad z \in \R.
\en
The subgaussian property (22.1), being required with some $\sigma>0$, 
ensures that $f$ has an analytic extension to the whole complex plane 
$\C$ as an entire function of order at most 2, extending the definition 
(23.1) to arbitrary complex values of $z$. Note that if the characteristic 
function $f(z)$ of a subgaussian distribution does not have any real or 
complex zeros, a well-known theorem due to Marcinkiewicz implies 
that the distribution of $X$ is already normal, cf. \cite{M}.
Thus, richer classes of subgaussian distributions like the strictly 
subgaussian distributions need to have zeros. Interesting questions 
in this context are ``what locations of a single zero of $f(z)$ would
be compatible with the strict subgaussian property and
the assumption that $f(z)$ is a characteristic function" and 
``to what extent does the Hadamard product representation of $f(z)$
in terms of zeros correspond to a stochastic decomposition of $X$ 
as a sum of independent random variables?"

In particular, an application of Goldberg-Ostrovskiĭ's refinement of 
Hadam\-ard's factorization theorem leads to the 
following simple sufficient condition for strictly subgaussian distributions
(due to Newman \cite{N1}, as we mentioned before).

\vskip5mm
{\bf Theorem 23.1.} {\sl Let $X$ be a subgaussian random variable
with mean zero. If all zeros of $f(z)$ are real, then $X$ is strictly 
subgaussian. 
}

\vskip5mm
Let us recall Goldberg-Ostrovskiĭ's theorem \cite{G-O}: 
If an entire ridge function $f(z)$ of a finite order has only 
real roots, then it can be represented as the product
\be
f(z) = c\,e^{i\beta z - \gamma z^2/2} \prod_{n \geq 1} 
\Big(1 - \frac{z^2}{z_n^2}\Big), \quad z \in \C,
\en
for some $c \in \C$, $\beta \in \R$, $\gamma \geq 0$, and $z_n > 0$
such that $\sum_{n \geq 1} z_n^{-2} < \infty$.

In the case where $f(z)$ is the characteristic function of a subgaussian 
random variable $X$ with mean zero and variance $\sigma^2 = \Var(X)$,
it has to be a ridge entire function of order $\rho \leq 2$. 
If $f$ has only real zeros, the distribution of $X$ 
must be symmetric about the origin, and the representation 
(23.2) is applicable. Here, since $f(0) = 1$, $f'(0) = 0$ and 
$f''(0) = -\sigma^2$, we necessarily have $c = 1$ and 
$\beta = 0$. Hence, this representation is simplified to
\be
f(z) = e^{-\gamma z^2/2} \prod_{n \geq 1} 
\Big(1 - \frac{z^2}{z_n^2}\Big)
\en
with
\be
\frac{1}{2} \sigma^2 = 
\frac{1}{2} \gamma + \sum_{n \geq 1} \frac{1}{z_n^2},
\en
so that $\gamma \leq \sigma^2$. Applying (23.3) with 
$z = -it$, $t \in \R$, we get a similar representation for 
the Laplace transform
$$
\E\,e^{tX} = e^{\gamma t^2/2} \prod_{n \geq 1} 
\Big(1 + \frac{t^2}{z_n^2}\Big),
$$
which implies the desired bound (22.1) by applying
the inequality $1+x \leq e^x$ together with (23.4).
If this product is non-empty (that is, $X$ is non-normal),
we actually obtain a stronger property such as (22.2)-(22.3)
according to Proposition 22.6.

Let us rewrite (23.3) in the form
\be
f(t) =
e^{-(3 \gamma - \sigma^2)\, t^2/4} \prod_{n \geq 1} 
\Big(1 - \frac{t^2}{z_n^2}\Big)\,e^{-\frac{t^2}{2z_n^2}},
\quad t \in \R.
\en
The terms in this product represent characteristic functions of
random variables $\frac{1}{z_n} X_n$ such that all $X_n$ have
the density $p(x) = x^2 \varphi(x)$. Hence, if 
$\gamma \geq \sum_{n \geq 1} \frac{1}{z_n^2}$, or equivalently
$\gamma \geq \frac{1}{3} \sigma^2$, the function $f(t)$ 
in (23.5) represents the characteristic function of the random variable
$$
X = cZ + \sum_{n \geq 1} \frac{1}{z_n} X_n,
$$
assuming that $X_n$ are independent, and $Z$ is a standard normal
random variable independent of all $X_n$. Necessarily,
$c^2 = \frac{3}{2} \gamma - \frac{1}{2}\sigma^2$.

The condition of Theorem 23.1 can easily be verified for many interesting 
classes including, for example, arbitrary Bernoulli sums and 
(finite or infinite) convolutions of uniform distributions on bounded 
symmetric intervals. It is however not necessary, as illustrated 
by the next generalization of Theorem 23.1.

\vskip5mm
{\bf Theorem 23.2.} {\sl Let $X$ be a subgaussian random variable 
with a symmetric distribution. If all zeros of $f(z)$ with
${\rm Re}(z) \geq 0$ lie in the
cone centered on the real axis defined by
\be
|{\rm Arg}(z)| \leq \frac{\pi}{8},
\en 
then $X$ is strictly subgaussian. Moreover, if $X$ is not normal,
the refining property $(22.3)$ holds true.
}

\vskip5mm
The proof is based upon Hadamard's factorization theorem, cf.
\cite{B-C-G8}.

On the other hand, (23.6) turns out to be a necessary condition for 
the strict subgaussianity for the following subclass of probability 
distributions.

\vskip5mm
{\bf Theorem 23.3} (\cite{B-C-G8}). {\sl Let $X$ be a random variable with 
a symmetric subgaussian distribution. Suppose that $f$ has 
exactly one zero $z = x+iy$ in the positive quadrant
$x,y \geq 0$. Then $X$ is strictly subgaussian, 
if and only if $(23.6)$ holds true.
}

\vskip2mm
As a consequence,
one can partially address the following question from the theory of entire 
characteristic functions (which is one of the central problems in this area): 
What can one say about the possible location of zeros of such functions?

\vskip3mm
{\bf Theorem 23.4} (\cite{B-C-G8}). {\sl Let $(z_n)$ be a finite or 
infinite sequence of non-zero complex numbers in the angle
$|{\rm Arg}(z_n)| \leq \frac{\pi}{8}$ such that
$$
\sum_n \frac{1}{|z_n|^2} < \infty.
$$
Then there exists a symmetric strictly subgaussian distribution
whose characteristic function has zeros exactly at the points
$\pm z_n$, $\pm \bar z_n$. 
}

\vskip5mm
One can show that a random variable $X$ with such distribution 
may be constructed as the sum $X = \sum_n X_n$ of independent 
strictly subgaussian random variables $X_n$ whose characteristic 
functions have zeros at the points $\pm z_n$, $\pm \bar z_n$ 
for every $n$ 
(and only at these points). Moreover, one may require that
$$
\Var(X) = \Lambda \sum_n \frac{1}{|z_n|^2}
$$
with any prescribed value $\Lambda \geq \Lambda_0$ where
$\Lambda_0$ is a universal constant ($\Lambda_0 \sim 5.83$).

\vskip7mm
\section{{\bf Examples of Strictly Subgaussian Distributions}}
\setcounter{equation}{0}

\noindent
An application of Corollary 22.4 allows to construct a rather rich 
family of strictly subgaussian probability distributions like the ones 
in the next 7 examples from Newman's class $\mathfrak L$.

\vskip2mm
{\bf Examples}

{\bf 24.1.} First of all, if a random variable $X \sim N(0,\sigma^2)$ 
has a normal distribution with mean zero and variance $\sigma^2$, 
then it is strictly subgaussian. In this case,
$$
\E\,e^{tX} = e^{\sigma^2 t^2/2}, \quad t \in \R,
$$
so that the inequality in (22.1) becomes an equality.

{\bf 24.2.} If $X$ has a symmetric Bernoulli distribution, supported 
on two points $a$ and $-a$, then it is strictly subgaussian. 
If, for definiteness, $a = 1$, then $\Var(X) = 1$, and 
the Laplace transform of the distribution of $X$ is given by 
$$
\E\,e^{tX} = \cosh(t) = \frac{e^t + e^{-t}}{2}, \quad t \in \R.
$$

{\bf 24.3.}  If $X$ is an infinite Bernoulli sum, that is, 
$$
X = \sum_{n=1}^\infty a_n X_n, \quad 
\P\{X_n = \pm 1\} = \frac{1}{2}, \quad
\sum_{n=1}^\infty a_n^2 < \infty,
$$
with $X_n$ independent symmetric Bernoulli random 
variables, then it is strictly subgaussian with variance 
$\sigma^2 = \Var(X) = \sum_{n=1}^\infty a_n^2$.
The corresponding Laplace transform and characteristic 
function $f$ of $X$ are  given by 
$$
\E\,e^{tX} = \prod_{n=1}^\infty \cosh(a_n t), \quad 
f(t) = \prod_{n=1}^\infty \cos(a_n t).
$$

{\bf 24.4.}  If the random variable $X$ is uniformly distributed 
on a finite interval $[-a,a]$, $a>0$, then it is strictly subgaussian. 
In this case it may be represented as the sum
$$
X = \sum_{n=1}^\infty \frac{a}{2^n}\, X_n, \quad 
\P\{X_n = \pm 1\} = \frac{1}{2},
$$
with $X_n$ independent symmetric Bernoulli random variables. 
Hence, this case is covered by the previous example. 
The corresponding Laplace transform is given by 
$$
\E\,e^{tX} = \frac{\sinh(at)}{at}.
$$
Recall that the strict subgaussian property in this case was already
mentioned in Section 20, cf. (20.3).

{\bf 24.5.}  If the random variables $X_n$ are independent and 
uniformly distributed on the interval $[-1,1]$, then the infinite sum
$$
X = \sum_{n=1}^\infty a_n X_n \quad {\rm with} \ \ 
\sum_{n=1}^\infty a_n^2 < \infty
$$
represents a strictly subgaussian random variable. 
The corresponding Laplace transform is given by 
$$
\E\,e^{tX} = \prod_{n=1}^\infty \frac{\sinh(a_n t)}{a_n t}.
$$

{\bf 24.6.}  Suppose that $X$ has density $p(x) = x^2 \varphi(x)$.
Then $\E X = 0$, $\sigma^2 = \E X^2 = 3$, and the Laplace 
transform satisfies
$$
\E\,e^{tX} = (1 + t^2)\,e^{t^2/2} \leq e^{3t^2/2}.
$$
Hence, $X$ is strictly subgaussian.

{\bf 24.7.}  More generally, if $X$ has a  density of the form
$$
p(x) = \frac{1}{(2d-1)!!}\,x^{2d} \varphi(x), \quad 
x \in \R, \ d = 1,2,\dots,
$$ 
then $\E X = 0$, $\sigma^2 = \E X^2 = 2d+1$, and 
the Laplace transform satisfies
$$
\E\,e^{tX} = \frac{1}{(2d-1)!!}\,H_{2d}(it)\,e^{t^2/2} \leq 
e^{(2d+1)\,t^2/2}, \quad t \in \R.
$$
Hence, $X$ is strictly subgaussian. The last inequality follows from
Theorem 23.1, since the Chebyshev-Hermite polynomials have real 
zeros, only. Note that the characteristic function of $X$ is given by
$$
\E\,e^{itX} = \frac{1}{(2d-1)!!}\,H_{2d}(t)\,e^{-t^2/2}.
$$

\vskip2mm
{\bf 24.8.}
In connection with the problem of location of zeros,
one may examine probability distributions with characteristic
functions of the form
\be
f(t) = e^{-t^2/2}\,(1 - \alpha t^2 + \beta t^4),
\en
where $\alpha,\beta \in \R$ are parameters. When $\beta = 0$, 
we obtain a characteristic function, if and only if $0 \leq \alpha \leq 1$.  
In the general case,
it is necessary that $\beta \geq 0$ for $f(t)$ to be a characteristic function
(although negative values of $\alpha$ are possible for small $\beta$).
The equality $(24.1)$ defines a characteristic function, if and only if 
the point $(\alpha,\beta)$ belongs to one of the following two regions:
$$
4\beta - 2\sqrt{\beta (1 - 2 \beta)} \leq \alpha \leq 3\beta + 1, \qquad 
0 \leq \beta \leq \frac{1}{3}, 
$$
or
$$
4\beta - 2\sqrt{\beta (1 - 2 \beta)} \leq \alpha \leq 
4\beta + 2\sqrt{\beta (1 - 2 \beta)}, \quad 
\frac{1}{3} \leq \beta \leq \frac{1}{2}.
$$
Moreover, given $\beta \geq 0$, a random variable $X$
with characteristic function of the form $(24.1)$ is strictly subgaussian, 
if and only if $\alpha \geq \sqrt{2\beta}$ (cf. \cite{B-C-G8}).

\vskip2mm
{\bf 24.9.} One may illustrate the previous characterization 
by the following simple example. For $\beta = \frac{1}{3}$, 
admissible values of $\alpha$ cover the interval 
$\sqrt{2/3} \leq \alpha \leq 2$.
Choosing $\alpha = \sqrt{2/3}$, we obtain the characteristic function 
$$
f(t) = e^{-t^2/2}\,
\Big(1 - \sqrt{\frac{2}{3}}\, t^2 + \frac{1}{3} t^4\Big)
$$
of a strictly subgaussian random variable. It has four distinct 
complex zeros $z_k$ defined by $z^2 = r^2\,(1 \pm i)$ with
$r^2 = \frac{1}{3}\sqrt{2/3}$, so 
$$
z_{1,2} = (2r)^{1/4}\,e^{\pm i\pi/8},  \quad
z_{3,4} = (2r)^{1/4}\,e^{\pm 7i\pi/8}.
$$
Note that $|{\rm Arg}(z_{1,2})| = \frac{\pi}{8}$. As stated in 
Theorem 23.3, it is necessary that 
$|{\rm Arg}(z)| \leq \frac{\pi}{8}$ for all zeros with 
${\rm Re}(z) > 0$ in the class of all strictly subgaussian probability 
distributions with characteristic functions of the form (24.1).

\vskip2mm
{\bf 24.10.} 
In order to describe the possible location of zeros, let us refine
the characterization in Example 24.8 in the class of functions
\be
f(t)  = e^{-t^2/2}\,(1 - w t)(1+w t)(1-\bar w t)(t+\bar w t), \quad
t \in \R,
\en
with $w = a+bi$. Thus, in the complex plane $f(z)$ has two or 
four distinct zeros $z = \pm 1/w$, $z = \pm 1/\bar w$ depending 
on whether $b = 0$ or $b \neq 0$. Note that 
$$
|{\rm Arg}(z)| = |{\rm Arg}(w)|
$$ 
when $z$ and $w$ are taken from the half-plane 
${\rm Re}(z) > 0$ and ${\rm Re}(w) > 0$.

Assuming for definiteness that $a > 0$, it was shown in \cite{B-C-G8}
that the function $f(t)$ in $(24.2)$ represents a characteristic function 
of a strictly subgaussian random variable, if and only if 
$$
a \leq 2^{-1/4} \sim 0.8409,
$$ 
while 
$|b|$ is sufficiently small. More precisely, this is the case whenever
$|b| \leq b(a)$ with a certain function $b(a) \geq 0$ such that
$b(2^{-1/4}) = 0$ and $b(a)>0$ for $0 < a < 2^{-1/4}$.
Moreover, there exists a universal 
constant $0 < a_0 < 2^{-1/4}$, $a_0 \sim 0.7391$, such that 
for $0 \leq a \leq a_0$ and only for these $a$-values, the property 
$|b| \leq b(a)$ is equivalent to the angle requirement
${\rm Arg}(w) \leq \frac{\pi}{8}$. As for the values
$a_0 < a \leq 2^{-1/4}$, this angle must be smaller.

\vskip7mm
\section{{\bf Laplace Transforms with Periodic Components}}
\setcounter{equation}{0}

\noindent
Following the previous examples, one may naturally expect 
that in the non-normal case the strict subgaussianity (22.1) can be
strengthened to the strict inequality
\be
\E\,e^{tX} < e^{\sigma^2 t^2/2}, \quad t \neq 0,
\en
with $\sigma^2 = \Var(X)$. However, this turns out to be false, and
moreover, the equality in (25.1) may be attained for an infinite sequence
of points $t_n \rightarrow \infty$. Correspondingly, the angle property
$|{\rm Arg}(z)| \leq \frac{\pi}{8}$ as in Proposition 23.2 
for the location of zeros of the characteristic function $f(z)$ of $X$ is no
longer true in general. It may actually happen that this function has infinitely
many zeros $z_n$ such that ${\rm Arg}(z_n) \rightarrow \frac{\pi}{2}$
as $n \rightarrow \infty$. That is, $z_n$ may be getting close to the imaginary
axis, in contrast to the property that on this axis $f(z)$ becomes
the Laplace transform $f(-it) = L(t) = \E\,e^{tX}$
(which is real, positive, and is greater than 1).

To better realize such a surprising phenomenon, we now turn to another 
interesting class of Laplace transforms that contain periodic components.

\vskip5mm
{\bf Definition 25.1.} We say that the distribution $\mu$ of a random 
variable $X$ is periodic with respect to the standard normal law, 
with period $h>0$, if it has a density $p(x)$ such that the function
$$
q(x) = \frac{p(x)}{\varphi(x)} = \frac{d\mu(x)}{d\gamma(x)},
\quad x \in \R,
$$
is periodic with period $h$, that is, $q(x+h) = q(x)$ for all $x \in \R$.

\vskip4mm
Here, $q$ represents the density of $\mu$ with respect to the standard 
Gaussian measure $\gamma$. We denote the class of all such distributions 
by $\mathfrak F_h$, and say that $X$ belongs to $\mathfrak F_h$.
Let us briefly recall several observations from 
\cite{B-C-G8} about this interesting class of probability distributions. 

\vskip4mm
{\bf Proposition 25.2.} {\sl If a random variable $X$ belongs to 
$\mathfrak F_h$, then it is subgaussian, and the function 
$$
\psi(t) = \E\,e^{tX}\,e^{-t^2/2}, \quad t \in \R,
$$
is $h$-periodic. It may be extended to the complex plane as an entire 
function of order at most 2.
Conversely, if $\psi(t)$ for a subgaussian random variable
$X$ is $h$-periodic, then $X$ belongs to $\mathfrak F_h$, as long as 
the characteristic function $f(t)$ of $X$ is integrable on the real line. 
}

\vskip3mm
If $X$ belongs to the class $\mathfrak F_h$, then for all integers $m$,
$$
\E\,e^{mh X} = e^{(mh)^2/2},
$$
implying that the random variable $X$ is subgaussian. 

Since
$$
f(t) = L(it) = \psi(it)\,e^{-t^2/2},
$$
the integrability assumption in the reverse statement is fulfilled,
as long as $\psi(z)$ has order smaller than 2, that is, when
$|\psi(z)| = O(\exp\{|z|^\rho\})$ as $|z| \rightarrow \infty$ 
for some $\rho<2$.

The periodicity property is stable under convolutions: The normalized 
sum $Z_n$ of $n$ independent copies of $X$ belongs to 
$\mathfrak F_{h\sqrt{n}}$, if $X$ belongs to $\mathfrak F_h$.

The class $\mathfrak F_h$ with $h = 2\pi$ contains probability
distributions whose Laplace transform has the form 
$L(t) = \psi(t)\, e^{t^2/2}$, where $\psi$ is a trigonometric polynomial. 
More precisely, consider the functions of the form
$$
\psi(t) = 1 - c P(t), \quad 
P(t) = a_0 + \sum_{k=1}^N (a_k \cos(kt) + b_k \sin(kt)), 
$$
where $a_k,b_k$ are given real coefficients, and $c \in \R$
is a non-zero parameter. 

\vskip4mm
{\bf Proposition 25.3.} {\sl If $P(0) = 0$ and $|c|$ is small enough, 
then $L(t) = \psi(t)\, e^{t^2/2}$ represents the Laplace transform of 
a subgaussian random variable $X$ with density $p(x) = q(x) \varphi(x)$, 
where $q(x)$ is a non-negative trigonometric polynomial of degree 
at most $N$.
}

\vskip4mm
Using the Fourier inversion formula, the polynomial $q$ can be 
explicitly written as
$$
q(x) = 1 - cQ(x), \quad Q(x) =
a_0 + \sum_{k=1}^N e^{k^2/2}\,(a_k \cos(kt) + b_k \sin(kt)).
$$
Hence, if $|c|$ is small enough, $q(x)$ is bounded away from zero,
so that $p(x)$ is non-negative. Moreover,
the requirement $P(0) = a_0  + \dots + a_N = 0$
guarantees that $\int_{-\infty}^\infty p(x)\,dx = 1$. 

Since $q$ is bounded, we also have $T_\infty(p||\varphi) < \infty$.

For further applications to the CLT, there are two more constraints
coming from the assumption that $\E X = 0$ and $\E X^2 = 1$.

\vskip4mm
{\bf Corollary 25.4.} {\sl Suppose that the polynomial $P(t)$ satisfies

\vskip2mm
$1)$ $P(0) = P'(0) = P''(0) = 0$;

$2)$ $P(t) \geq 0$ for $0<t<h$, where $h$ is the smallest period of $P$.

\vskip2mm
\noindent
If $c>0$ is small enough, then $L(t)$ represents the Laplace transform 
of a strictly subgaussian random variable $X$.
}

\vskip4mm
In terms of the coefficients of the polynomial, the moment assumptions
$P'(0) = P''(0) = 0$ are equivalent to
$$
\sum_{k=1}^N k b_k = \sum_{k=1}^N k^2 a_k = 0.
$$
The assumption 2) implies that $0 < \psi(t) \leq 1$, and if $P(t) > 0$ for 
$0<t<h$, then the equation $\psi(t)=1$ has no solution in this interval.

\vskip5mm
{\bf Example 25.5.} Consider the transforms of the form
\be
L(t) = (1 - c \sin^m(t))\, e^{t^2/2}
\en
with an arbitrary integer $m \geq 3$, where $|c|$ is small enough.
Then $\E X = 0$, $\E X^2 = 1$, and the cumulants of $X$ satisfy
$\gamma_k(X) = 0$ for all $3 \leq k \leq m-1$.

Moreover, if $m \geq 4$ is even and $c>0$, the random variable 
$X$ with the Laplace transform (25.2) is strictly subgaussian.
In the case $m=4$, (25.2) corresponds to
$$
P(t) = \sin^4 t = \frac{1}{8}\,(3 - 4 \cos(2t) + \cos(4t)).
$$

The examples based on the trigonometric polynomials may be generalized 
to the setting of $2\pi$-periodic functions represented by Fourier series
$$
P(t) = a_0 + \sum_{k=1}^\infty (a_k \cos(kt) + b_k \sin(kt))
$$
with coefficients satisfying
$\sum_{k=1}^\infty e^{k^2/2} (|a_k| + |b_k|) < \infty$.

\vskip5mm
{\bf Remark 25.6.} Suppose that a non-normal random variable $X$ 
belongs to $\mathfrak F_h$. By analyticity and $h$-periodicity of 
$\psi(t)$ on the real line, we have
\be
\psi(z+h) = \psi(z) \quad {\rm for \ all} \ z \in \C.\label{10.6}
\en
The characteristic function
$$
f(z) = L(iz) = \psi(iz)\,e^{-z^2/2}, \quad z \in \C,
$$
must have at least one zero in the complex plane, say $z_0$. But then, according
to (25.3), all numbers $z_n = z_0 + i h n $, $n \in \Z$, will be zeros
as well. Moreover, for this sequence
$|{\rm Arg}(z_n)| \rightarrow \frac{\pi}{2}$ as
$|n| \rightarrow \infty$.

\vskip7mm
\section{{\bf Richter's Theorem and its Refinement}}
\setcounter{equation}{0}

\noindent
We can now return to the field of local limit theorems for
strong distances and focus on the R\'enyi and Tsallis distances
of infinite order. Recall that
\be
T_\infty(p_n||\varphi) = 
{\rm ess\,sup}_x\, \frac{p_n(x) - \varphi(x)}{\varphi(x)},
\en
where $p_n$ denote the densities of the normalized
sums 
$$
Z_n = \frac{X_1 + \dots + X_n}{\sqrt{n}}
$$
of $n$ independent copies of a random variables $X$ with mean 
zero and variance one (assuming that these densities exist
and bounded for large $n$).

The closeness of $p_n(x)$ to $\varphi(x)$ on growing intervals 
$|x| \leq T_n$ is governed by several limit theorems. For example, 
it follows from Theorem 16.1 that, under the moment assumption 
$\E\,|X|^k < \infty$ with some integer $k \geq 3$, we have
$$
\sup_{|x| \leq T_n} \frac{|p_n(x) - \varphi(x)|}{\varphi(x)}
\rightarrow 0 \quad {\rm as} \ n \rightarrow \infty,
$$
where $T_n = \sqrt{(k-2) \log n}$. An asymptotic behavior of 
$p_n(x)$ in the larger region $|x| = o(\sqrt{n})$ is 
governed under a stronger moment-type assumption
by a theorem due to Richter \cite{Ri}.

\vskip5mm
{\bf Theorem 26.1.} {\sl Let $\E\,e^{c|X|} < \infty$ for some 
$c>0$, and let $Z_n$ have a bounded density for some 
$n = n_0$. Then $Z_n$ with large $n$ have bounded continuous
densities $p_n$ satisfying
\be
\frac{p_n(x)}{\varphi(x)} = \exp\Big\{\frac{x^3}{\sqrt{n}}\,
\lambda\Big(\frac{x}{\sqrt{n}}\Big)\Big\}\,
\Big(1 + O\Big(\frac{1+|x|}{\sqrt{n}}\Big)\Big)
\en
uniformly for $|x| = o(\sqrt{n})$. Here the function
$\lambda(z)$ is represented as an infinite power series in $z$ 
which is absolutely convergent in a neighborhood of the point $z=0$.
}

\vskip2mm
The proof of this theorem may also be found in the book by
Ibragimov and Linnik \cite{I-L}, Theorem 7.1.1, where it 
was additionally assumed that $X$ has a continuous bounded density.
The representation (26.2) was further investigated
there for zones of normal attraction of the form
$|x| = o(n^\alpha)$, $0 < \alpha < \frac{1}{2}$.

One immediate consequence of (26.2) is that
\be
\frac{p_n(x)}{\varphi(x)} \rightarrow 1 \quad 
{\rm as} \ n \rightarrow \infty
\en
uniformly in the region $|x| = o(n^{1/6})$. However,
in general this is no longer true outside this region.
To better understand the possible behavior of the densities,
one needs to involve the information about the coefficients 
in the power series representation
$$
\lambda(z) = \sum_{k=0}^\infty \lambda_k z^k,
$$
which is called Cramer's series. As was mentioned in \cite{I-L},
$$
\lambda_0 = \frac{1}{6}\,\gamma_3, \quad
\lambda_1 = \frac{1}{24}\,(\gamma_4 - 3\gamma_3^2).
$$

However, in order to judge the behavior $\lambda(z)$ for small
$z$, one should describe the leading term in this series.
The analysis of the saddle point associated to the
log-Laplace transform of the distribution of $X$ shows that
\be
\lambda(z) = 
\frac{\gamma_m}{m!}\,z^{m-3} + O(|z|^{m-2}) \quad
{\rm as} \ z \rightarrow 0,
\en
where $\gamma_m$ denotes the first non-zero cumulant
of $X$ (when $X$ is not normal). Equivalently, $m$ is 
the smallest integer such that $m \geq 3$ and
$\E X^m \neq \E Z^m$, where $Z$ is a standard normal 
random variable. In this case
$$
\gamma_m = \E X^m - \E Z^m.
$$

Using (26.4) in (26.2), we obtain a more informative representation
\be
\frac{p_n(x)}{\varphi(x)} = \exp\bigg\{\frac{\gamma_m}{m!}\, 
\frac{x^m}{n^{\frac{m}{2} - 1}} + 
O\Big(\frac{x^{m+1}}{n^{\frac{m}{2}}}\Big)\bigg\}
\Big(1 + O\Big(\frac{1+|x|}{\sqrt{n}}\Big)\Big),
\en
which holds uniformly for $|x| = o(\sqrt{n})$. With this 
refinement, we see that the convergence in (26.3) 
holds true uniformly over all $x$ in the potentially larger region
$$
|x| \leq \ep_n\,n^{\frac{1}{2} - \frac{1}{m}} \quad
(\ep_n \rightarrow 0).
$$ 
For example, if the distribution of $X$ is symmetric about the origin,
then $\gamma_3 = 0$, so that necessarily $m \geq 4$.

For an application to the $D_\infty$-distance, it is desirable
to get some information for larger intervals such as 
$|x| \leq \tau_0\sqrt{n}$ and in particular to replace the term 
$O(\frac{|x|}{\sqrt{n}})$ in (26.2) with an explicit $n$-dependent quantity. 
For this aim, the following refinement of Theorem 26.1 was 
recently proposed in \cite{B-C-G7}.

\vskip5mm
{\bf Theorem 26.2.} {\sl Let the conditions of Theorem $26.1$
be fulfilled. There is a constant $\tau_0 > 0$ with the following property.
Putting $\tau = x/\sqrt{n}$, we have for $|\tau| \leq \tau_0$
\be
\frac{p_n(x)}{\varphi(x)} = e^{n\tau^3 \lambda(\tau) - \mu(\tau)}\,
\big(1 + O(n^{-1} (\log n)^3)\big),
\en
where $\mu(\tau)$ is an analytic function in $|\tau| \leq \tau_0$
such that $\mu(0)=0$.
}

\vskip5mm
Here, similarly to (26.4),
$$
\mu(\tau) = 
\frac{1}{2(m-2)!}\,\gamma_m \tau^{m-2} + O(|\tau|^{m-1}).
$$

As a consequence of (26.6), which cannot be obtained on the basis
of (26.2) or (26.5), we have the following assertion which was also 
obtained in \cite{B-C-G7}.

\vskip5mm
{\bf Corollary 26.3.} {\sl Under the same conditions, suppose that
the first non-zero cumulant $\gamma_m$ of $X$ is negative with
$m \geq 4$ being even. There exist constants $\tau_0 > 0$ and $c>0$ with the 
following property. If $|\tau| \leq \tau_0$, $\tau = x/\sqrt{n}$, then, for all $n$
large enough,
\be
\frac{p_n(x) - \varphi(x)}{\varphi(x)} \leq \frac{c(\log n)^3}{n}.
\en
}

\vskip2mm
{\bf Remark 26.4.} The hypothesis on cumulants in Corollary 26.3 is
always fulfilled for strictly subgaussian distributions. Indeed, since necessarily 
$\E X^3 = 0$, the log-Laplace transform admits 
a power series representation
$$
\log \E\,e^{tX} = \frac{1}{2}\,t^2 + \sum_{k=4}^\infty 
\frac{\gamma_k}{k!}\,t^k,
$$
which is absolutely convergent in some interval
$|t| \leq t_0$ ($t_0 > 0$). Hence, as $t \rightarrow 0$,
$$
\log \E\,e^{tX} = \frac{1}{2}\,t^2 +  
\frac{\gamma_m}{m!}\,t^m + O(t^{m+1}).
$$
The strict subgaussianity $\log \E\,e^{tX} \leq \frac{1}{2}\,t^2$,
$t \in \R$, implies that $m$ must be even, and $\gamma_m < 0$.

\vskip7mm
\section{{\bf CLT in $D_\infty$ with Rate of Convergence}}
\setcounter{equation}{0}

\noindent
As before, let $(X_n)_{n \geq 1}$ be independent copies of a random variable
$X$ with mean zero and variance one, and let $p_n$ denote the densities of 
the normalized sums
$$
Z_n = \frac{X_1 + \dots + X_n}{\sqrt{n}},
$$
assuming that such densities exist for large $n$. Recall that,
for the convergence of  the Tsallis distances
\be
T_\infty(p_n||\varphi) = 
{\rm ess\,sup}_x\, \frac{p_n(x) - \varphi(x)}{\varphi(x)}
\en
to zero, it is necessary that $X$ be strictly subgaussian.
We will now require that a slightly stronger property holds true,
\be
\sup_{|t| \geq t_0} \big[e^{-t^2/2}\, \E\,e^{tX} \big] < 1 \quad
{\rm for \ all} \ t_0>0
\en
(note that it is weaker in comparison with the
properties (22.2)-(22.3) from Proposition 22.6).
In that case, the inequality (26.7) can be extended to the whole
real line, and we arrive at the following statement proved in \cite{B-G2}.

\vskip4mm
{\bf Theorem 27.1.} {\sl Let $X$ be a non-normal random variable
satisfying the condition $(27.2)$. If $T_\infty(p_n||\varphi) < \infty$ 
for some $n$, then
\be
T_\infty(p_n||\varphi) = O\Big(\frac{1}{n}\,(\log n)^3\Big)
\quad {\sl as} \ n \rightarrow \infty.
\en
}

\vskip2mm
Let us recall that the condition (27.2) is fulfilled, if the characteristic 
function $f(z)$ of $X$ has only real zeros in the complex plane 
(the Newman class). Moreover, according to Theorem 23.2, it is 
fulfilled under a weaker assumption that $X$ has a symmetric
distribution, and that all zeros with ${\rm Re}(z) \geq 0$ lie in the
cone $|{\rm Arg}(z)| \leq \frac{\pi}{8}$. Hence, Theorem 27.1
is applicable to all previous examples except for those which we 
discussed in Section 25 about the
Laplace transforms with periodic components.

In the proof of (27.3), the supremum in (27.1) should be first
restricted to the interval $|x| \leq \tau_0 \sqrt{n}$ with a constant
$\tau_0$ taken from Corollary 26.3. It may be applied as
explained in Remark 26.4, thus leading to the desired upper bound (26.7).

The extension of (27.1) to the regions of the form $|x| \geq \tau \sqrt{n}$
is based on the following assertion of
independent interest, which we state in terms of the function
$$
A(t) = \frac{1}{2} t^2 - K(t), \quad K(t) = \log \E\,e^{tX}, \quad
t \in \R.
$$
Note that the strict subgaussianity means that $A(t) \geq 0$
for all $t \in \R$ (when $\Var(X)=1$ which is however not assumed
below).

\vskip4mm
{\bf Proposition 27.2} (\cite{B-G2}). {\sl Let $p_n$ denote the density 
of $Z_n$ constructed for a subgaussian random variable
$X$ whose density $p$ has finite R\'enyi distance of infinite order to 
the standard normal law. Then, for almost all $x \in \R$,
\be
\frac{p_n(x\sqrt{n})}{\varphi(x\sqrt{n})} \leq 
c\sqrt{2}\,e^{- (n-1) A(x)},
\en
where $c = 1 + T_\infty(p||\varphi)$.
}

\vskip4mm
{\bf Corollary 27.3.} {\sl If $\E X = 0$, $\Var(X) = 1$, and 
$X$ is strictly subgaussian, then
$$
T_\infty(p_n||\varphi) \leq 
\sqrt{2}\,\big(1 + T_\infty(p||\varphi)\big) - 1.
$$
}

\vskip4mm
Thus, the finiteness of the Tsallis distance $T_\infty(p||\varphi)$ for 
a strictly subgaussian random variable $X$ with density $p$ ensures 
the boundedness of $T_\infty(p_n||\varphi)$ for all normalized sums $Z_n$.

If $A(x)$ is bounded away from zero, the inequality (27.4) shows that
the ratio on the left-hand side is exponentially small for growing $n$.
In particular, this holds for any non-normal random variable $X$ satisfying 
the separation property (27.2). Then we immediately obtain:

\vskip4mm
{\bf Corollary 27.4.} {\sl Suppose that $X$ has a density $p$ with finite 
$T_\infty(p||\varphi)$. Under the condition $(27.2)$, for any $\tau_0 > 0$, 
there exist $A > 0$ and $\delta \in (0,1)$ such that the densities $p_n$ 
of $Z_n$ satisfy
\be
p_n(x) \leq A\delta^n \varphi(x), \quad
|x| \geq \tau_0\sqrt{n}.
\en
}

As a by-product, this assertion implies that
$$
\liminf_{n \rightarrow \infty} \, \sup_{x \in \R} \,
\frac{|p_n(x) - \varphi(x)|}{\varphi(x)} \geq 1.
$$
Therefore, one cannot hope to strengthen the Tsallis distance by
introducing a modulus sign in the definition (27.1).

Thus, combining (27.6) with (26.7), we arrive at the desired rate
in (27.3).

In particular, if the random variable $X$ is 
bounded and has a bounded density $p$, and if it satisfies $(27.2)$, 
the conditions of Theorem 27.1 are fulfilled.

The next corollary from \cite{B-G2} describes more examples.

\vskip5mm
{\bf Corollary 27.5.} {\sl Assume that $X$
satisfies $(27.2)$ and is represented as
$$
X = c_0 \eta_0 + c_1 \eta_1 + c_2 \eta_2, \quad
c_0^2 + c_1^2 + c_2^2 = 1, \ \ c_1,c_2 > 0, 
$$
where the independent random variables $\eta_k$
are strictly subgaussian with variance one and satisfy
$T_\infty(\eta_k||\varphi) < \infty$ for $k = 1,2$.
Then we have the CLT with rate $(27.3)$.
}

\vskip2mm
As an interesting subclass, one may consider infinite
weighted convolutions, that is, random variables of the form
$$
X = \sum_{k=1}^\infty a_k \xi_k, \quad 
\sum_{k=1}^\infty a_k^2 = 1.
$$

{\bf Corollary 27.6.} {\sl Assume that the i.i.d. random variables 
$\xi_k$ are strictly subgaussian and have a bounded, compactly 
supported density with variance $\Var(\xi_1) = 1$. If $\xi_1$ 
satisfies $(27.2)$, then the CLT $(27.3)$ holds true.
}

\vskip4mm
This statement includes, for example, infinite weighted 
convolutions of the uniform distribution on a bounded symmetric interval.

\vskip7mm
\section{{\bf Action of Esscher Transform on Convolutions}}
\setcounter{equation}{0}

\noindent
While the strengthened variant of the strict subgaussianity via 
the separation property (27.2) guarantees a good rate of normal 
approximation in $T_\infty$, it is also natural to ask about necessary 
and sufficient conditions for the validity of the CLT with respect to 
the R\'enyi distance $D_\infty$ in full generality without specification
of the rate of convergence. An approach to this rather sophisticated 
question has been recently proposed in \cite{B-G2}. It is based on 
the careful analysis of the 
Esscher transform, which generates the semigroup of probability densities
$$
Q_h p(x) = \frac{1}{L(h)}\,e^{hx} p(x), \quad x \in \R,
$$
with parameter $h \in \R$. Here $p$ is a density of a subgaussian
random variable $X$, and
$$
(L p)(t) = L(t) = \E\,e^{tX} = \int_{-\infty}^\infty e^{tx} p(x)\,dx
$$
is the Laplace transform associated to $p$. We call the distribution 
with density $Q_h p$ the shifted distribution of $X$ at step $h$, 
to emphasize the identity $Q_h \varphi(x) = \varphi(x+h)$
for the standard normal density.

The early history of this transform goes back to the works by
Esscher \cite{E} in actuarial science, by Khinchin \cite{Kh} in statistical 
mechanics, and by Daniels \cite{D} in statistics. It has a number of
remarkable properties. In addition to the semi-group property
$$
Q_{h_1}(Q_{h_2} p) = Q_{h_1 + h_2} p, \quad h_1,h_2 \in \R,
$$
one should emphasize its multiplicativity with respect 
to convolutions, i.e.
\be
Q_h p = Q_h q_1 * \dots * Q_h q_n,
\en
whenever $p = q_1 * \dots * q_n$ (similarly to the Laplace transform
with the difference that the convolution in the conclusion should be 
replaced with the product). As a consequence, the Esscher transform 
appears naturally in the following density representation.

\vskip4mm
{\bf Proposition 28.1.} {\sl Let $p_n$ denote the
density of the normalized sum $Z_n$ of $n$ independent
copies of a subgaussian random variable $X$ with density $p$.
Putting $x_n = x\sqrt{n}$, $h_n = h\sqrt{n}$ $(x,h \in \R)$, we have
\be
\frac{p_n(x\sqrt{n})}{\varphi(x\sqrt{n})} \, = \,
\sqrt{2\pi}\,e^{\frac{n}{2}\, (x - h)^2 - n A(h)}\, Q_{h_n} p_n(x_n).
\en
}

Here let us recall that
$$
A(h) = \frac{1}{2} h^2 - K(h), \quad K(h) = \log \E\,e^{hX}, \quad
h \in \R.
$$

Using the subadditivity of the maximum-of-density functional
$M(X) = {\rm ess\,sup}_x p(x)$ along convolutions, this allows us
to establish Proposition 27.2. Its upper bound (27.4) can be applied
outside the set of points $x$ where $A(x)$ is bounded away from zero,
more precisely -- outside the critical zone
$$
A_n(a) = \Big\{x \in \R: A(x) \leq \frac{a}{n-1}\Big\}, \quad a>0.
$$
Then (27.4) leads to
\be
\frac{p_n(x\sqrt{n})}{\varphi(x\sqrt{n})} \leq c\sqrt{2}\,e^{-a}, \quad
x \notin A_n(a),
\en
which is effective as long as $c = 1 + T_\infty(p||\varphi)$ is finite.
If $a$ is large, this upper bound may be used in the proof of the CLT with
respect to the distance $T_\infty$ restricted to the complement of the 
critical zone. 

As for the points $x \in A_n(a)$, we need to study the term 
$Q_{h_n} p_n(x_n)$ in (28.2) by different tools, which requires to 
involve a variant of the uniform local limit theorem (2.6) with a quantitative 
error term, as stated below.

\vskip4mm
{\bf Proposition 28.2.} {\sl Let $(X_n)_{n \geq 1}$ be independent copies 
of a random variable $X$ such that $\E X = 0$, $\Var(X) = 1$,
$\beta_3 = \E\,|X|^3 < \infty$. If $X$ has a density bounded $M$, 
the normalized sums $Z_n$
have continuous densities $p_n$ for all $n \geq 2$ satisfying
$$
\sup_x |p_n(x) - \varphi(x)| \leq \frac{c}{\sqrt{n}}\,M^2 \beta_3
$$
with some absolute constant $c>0$.
}

\vskip2mm
After proper centering and normalization, this statement can be
applied to $Q_{h_n} p_n$, using the property that these densities have
a convolution structure according to (28.1). Namely,
for a subgaussian random variable $X$ with density $p$, denote by 
$X(h)$ a random variable with density $Q_h p$ ($h \in \R$). It is
subgaussian, and has mean and variance
$$
m_h = \E X(h) = K'(h), \quad
\sigma_h^2  = \Var(X(h)) = K''(h).
$$
The last equality shows that necessarily $K''(h) > 0$ for all $h \in \R$,
since otherwise the random variable $X(h)$ would be a constant a.s.
Moreover, if $c = 1 + T_\infty(p||\varphi)$ is finite, it was shown in
\cite{B-G2} that, for all $h \in \R$,
\be
\sigma_h^2 \geq \frac{\pi}{6c^2}\, e^{-2A(h)}.
\en
In addition, if $h \in A_n(a)$ and $n \geq 4(a + 1)$,
the normalized random variables 
$\widehat X(h) = \frac{X(h) - m_h}{\sigma_h}$ have a finite
third absolute moment, and more precisely
$$
\E\,|\widehat X(h)|^3 \leq C\sigma_h^{-3}
$$
up to some absolute constant $C$. This allows one to develop 
an application of Proposition 28.2 with $\widehat X(h)$ in place of $X$ 
and with $h=x$, which leads to the more informative representation 
compared to (28.2). Define
$$
v_x = \frac{x - m_x}{\sigma_x} = \frac{A'(x)}{\sigma_x}.
$$

\vskip3mm
{\bf Proposition 28.3.} {\sl Let $X$ be a strictly subgaussian random variable
with variance one having a density $p$ such that $c = 1 + T_\infty(p||\varphi)$
is finite. Then, for all $x \in A_n(a)$, $n \geq 4(a + 1)$,
\be
\frac{p_n(x\sqrt{n})}{\varphi(x\sqrt{n})} = 
\frac{1}{\sigma_x}\,e^{-nA(x)-nv_x^2/2} + \frac{Bc^4}{\sqrt{n}},
\en
where $B = B_n(x)$ is bounded by an absolute constant.
}

\vskip5mm
It is worthwhile noting that $A''(h) = 1 - K''(h) \leq 1$ which readily implies
$A'(h)^2 \leq 2A(h)$. Hence, by (28.4),
\be
v_x^2 \leq \frac{2A(x)}{\sigma_x^2} \leq
\frac{12}{\pi}\,c^2 e^{A(x)} A(x) \leq 12\,c^2 A(x),
\en
assuming that $x \in A_n(a)$ with $a \leq 1$ and $n \geq 2$ 
in the last step.

\vskip7mm
\section{{\bf Necessary and Sufficient Conditions}}
\setcounter{equation}{0}

\noindent
As before,
suppose that $(X_n)_{n \geq 1}$ are independent copies of the random
variable $X$ with $\E X = 0$ and $\Var(X) = 1$. We assume that:

\vskip4mm
1) \ $Z_n$ has density $p_n$ such that $T_\infty(p_n||\varphi) < \infty$ 
for some $n=n_0$;

2) \ $X$ is strictly subgaussian, that is, $A(t) \geq 0$ for all $t \in \R$.

\vskip4mm
\noindent
Let us now describe a main result which was obtained in \cite{B-G2} and later
refined in \cite{B-G4} towards the question about the CLT with respect to $T_\infty$.
Note that the log-Laplace transform $K(t) = \log \E\,e^{tX}$ represents
a $C^\infty$-smooth function on the real line, so is
$$
A(t) = \frac{1}{2} t^2 - K(t).
$$

\vskip2mm
{\bf Theorem 29.1.} {\sl For the convergence 
$T_\infty(p_n||\varphi) \rightarrow 0$, it is necessary and sufficient 
that the following two conditions are fulfilled:

\vskip3mm
$a)$ \ $A''(t) = 0$ for every point $t \in \R$ such that $A(t) = 0$;

$b)$ \ $\lim_{k \rightarrow \infty} A''(t_k)= 0$ for every sequence
$t_k \rightarrow \pm \infty$ such that $A(t_k) \rightarrow 0$ as 
$k \rightarrow \infty$.
}

\vskip2mm
These conditions may be combined in one as
\be
\lim_{A(t) \rightarrow 0} A''(t) = 0,
\en
which is kind of continuity of the second derivative $A''$ with respect to $A$.

Under the separation property (27.2), the condition $b)$ is fulfilled 
automatically, while the equation $A(t) = 0$ has only one solution $t=0$. 
But near zero, due to the strict subgaussianity, $A(t) = O(t^4)$ and 
$A''(t) = O(t^2)$. Hence, the condition $a)$ is fulfilled as well, and 
we obtain the CLT with respect to $T_\infty$. 

In \cite{B-G2}, the condition $b)$ was stated in a slightly weaker form

\vskip2mm
$b')$ \ $\limsup_{k \rightarrow \infty} A''(t_k) \leq 0$ for every sequence
$t_k \rightarrow \pm \infty$ such that $A(t_k) \rightarrow 0$ as 
$k \rightarrow \infty$.

\vskip2mm
\noindent
Correspondingly, (29.1) was replaced with
\be
\lim_{A(t) \rightarrow 0} \max(A''(t),0) = 0.
\en
However, as was noticed in  \cite{B-G4}, in presence of the strict sub-Gaussianity 
condition 2), there is a lower bound
$$
A''(t) \geq -c A(t)^{1/4}, \quad t \in \R,
$$
holding true with some absolute constant $c>0$,
provided that $0 \leq A(t) \leq 1$. Hence, $\liminf_{A(t) \rightarrow 0} A''(t) \geq 0$,
which shows that the assertions (29.1) and (29.2) are equivalent.

Let us now explain the appearance of the condition (29.2),
assuming for simplicity that $n_0 = 1$. For the sufficiency part,
choose $a = \log(1/\ep)$ for a given $\ep \in (0,1)$, so that, by (28.3),
$$
\sup_{x \notin A_n(a)}\, \frac{p_n(x\sqrt{n})}{\varphi(x\sqrt{n})} \leq 
c\sqrt{2}\,\ep.
$$
In the case $x \in A_n(a)$ with $n \geq 4(a + 1)$, the equality (28.5) 
is applicable and implies
$$
\sup_{x \in A_n(a)}\, \frac{p_n(x\sqrt{n})}{\varphi(x\sqrt{n})} \leq 
\sup_{x \in A_n(a)}\, \frac{1}{\sigma_x} + O\Big(\frac{1}{\sqrt{n}}\Big),
$$
where we recall that $\sigma_x^2 = K''(x)$. Hence,
$$
1 + T_\infty(p_n||\varphi) \leq \sup_{x \in A_n(a)}\, \frac{1}{\sigma_x} + 
c\sqrt{2}\,\ep + O\Big(\frac{1}{\sqrt{n}}\Big).
$$
Thus, a sufficient condition for the convergence 
$T_\infty(p_n||\varphi) \rightarrow 0$ as $n \rightarrow \infty$ is that, 
for any $\ep \in (0,1)$,
$$
\limsup_{n \rightarrow \infty}\, \sup_{x \in A_n(\log(1/\ep))}\, 
\sigma_x^{-2} \leq 1.
$$
Equivalently, $\liminf_{n \rightarrow \infty}\, \inf_{x \in A_n(a)}\, K''(x) \geq 1$
for any $a>0$, that is,
$$
\limsup_{n \rightarrow \infty}\, \sup_{x \in A_n(a)}\, A''(x) \leq 0.
$$
Since $A(x) = O(\frac{1}{n})$ on every set $A_n(a)$, this may be written as
(29.2).

To see that the condition (29.2) is also necessary, let us return to the 
representation (28.5). Assuming that 
$T_\infty(p_n||\varphi) \rightarrow 0$, it implies that, for any $a>0$,
\be
\limsup_{n \rightarrow \infty}\, \sup_{x \in A_n(a)} \frac{1}{\sigma_x}
\exp\Big\{- n \Big(A(x) + \frac{1}{2}v_x^2\Big)\Big\} \leq 1.
\en
Recalling (28.6), we have $v_x^2 \leq 12\,c^2 A(x)$ for all
$x \in A_n(a)$ with $a \leq 1$ and $n \geq 2$. Since  $nA(x) \leq 2a$ 
on the set $A_n(a)$ and $c \geq 1$,
it follows that 
$$
A(x) + \frac{1}{2}v_x^2 \leq 7 c^2 A(x) \leq \frac{14 c^2}{n}\,a,
$$
and (29.3) implies
$$
\limsup_{n \rightarrow \infty}\, \sup_{x \in A_n(a)} \frac{1}{\sigma_x}
\leq e^{14 c^2 a}, \quad 0 < a \leq 1.
$$
Therefore, for all $n \geq n(a)$,
$$
\inf_{x \in A_n(a)} K''(x) \geq e^{-30 c^2 a}.
$$
Since $a$ may be as small as we wish, we conclude that, for any $\ep > 0$, 
there is $\delta>0$ such that 
$A(x) \leq \delta \, \Rightarrow \, K''(x) \geq 1-\ep$, or
$$
A(x) \leq \delta \, \Rightarrow \, A''(x) \leq \ep.
$$ 
But this is the same as the condition (29.2).

It is rather surprising that the proof of Theorem 29.1 does not
use tools based on the Fourier transform (except for the local
limit theorem stated in Proposition 28.2).

\vskip7mm
\section{{\bf Characterization in the Periodic Case}}
\setcounter{equation}{0}

\noindent
One can now apply Theorem 29.1 to the Laplace transforms $L(t)$ with 
\be
\psi(t) = L(t)\,e^{-t^2/2} = \E\,e^{tX}\,e^{-t^2/2}, \quad t \in \R,
\en
being periodic with the smallest period $h>0$.
Assume that $\E X = 0$, $\E X^2 = 1$, and

\vskip2mm
1) $Z_n$ has density $p_n$ for some $n=n_0$ such that 
$T_\infty(p_n||\varphi) < \infty$;

2) $X$ is strictly subgaussian, i.e. $L(t) \leq e^{t^2/2}$,
or equivalently $\psi(t) \leq 1$ for all $t \in \R$.

\vskip3mm
{\bf Theorem 30.1} (\cite{B-G2}). {\sl For the convergence
$T_\infty(p_n||\varphi) \rightarrow 0$ as $n \rightarrow \infty$,
it is necessary and sufficient that, for every $0 < t < h$,
\be
\psi(t) = 1 \, \Rightarrow \, \psi''(t) = 0.
\en
Moreover, if the equation $\psi(t) = 1$ has no solution in $0 < t < h$, 
then
\be
T_\infty(p_n||\varphi) = O\Big(\frac{1}{n}\,(\log n)^3\Big)
\quad {\sl as} \ n \rightarrow \infty.
\en
}
\vskip2mm
Indeed, due to $\psi(t)$ being analytic, the equation
$\psi(t) = 1$ has finitely many solutions in the interval $[0,h]$,
including the points $t = 0$ and $t = h$ (by the periodicity).
Hence, the condition $b)$ in Theorem 29.1 may be ignored, and we obtain
that $T_\infty(p_n||\varphi) \rightarrow 0$ as $n \rightarrow \infty$,
if and only if 
\be
A''(t) = 0 \ {\rm for \ every \ point} \ t \in [0,h] \ {\rm such \ that} \ 
A(t) = 0.
\en
Here one may exclude the endpoints, since $A''(0) = A''(h) = 0$, by 
the strict subgaussianity and periodicity. As for the interior points 
$t \in (0,h)$, note that $A(t) = -\log \psi(t)$ has the second derivative
$$
A''(t) = \frac{\psi'(t)^2 - \psi''(t) \psi(t)}{\psi(t)^2} = -\psi''(t)
$$
at every point $t$ such that $\psi(t) = 1$ (in which case $\psi'(t) = 0$
due to the property $\psi \leq 1$). This shows that (30.4) is reduced to the
condition (30.2).

As for the conclusion (30.3) about the rate of convergence, it is a full
analogue of Theorem 27.1, and its proof is based on Corollary 26.3.

For an illustration of Theorem 30.1, let us return to the setting of
Section 25, where we considered the Laplace transforms (30.1) with
$$
\psi(t) = 1 - cP(t),
$$
where $P(t)$ is a trigonometric polynomial satisfying

\vskip2mm
$a)$ $P(0) = P'(0) = P''(0) = 0$;

$b)$ $P(t) \geq 0$ for $0<t<h$, where $h$ is the smallest period of $P$.

\vskip3mm
\noindent
As was emphasized before, if $c>0$ is small enough (which is assumed below), 
then $L(t)$ is the Laplace transform of a strictly subgaussian random variable 
$X$ with variance one and such that $T_\infty(p||\varphi) < \infty$, where
$p$ is a density of $X$. Hence, the conditions 1)-2) are fulfilled with $n_0=1$.
Combining Theorem 30.1 with Corollary 25.4, we obtain:

\vskip3mm
{\bf Corollary 30.2} {\sl Under the above conditions $a)-b)$, 
$T_\infty(p_n||\varphi) \rightarrow 0$ as $n \rightarrow \infty$,
if and only if, for every $0 < t < h$,
\be
P(t) = 0 \, \Rightarrow \, P''(t) = 0.
\en
Moreover, if $P(t) > 0$ for all $0<t<h$, then the convergence
rate $(30.3)$ holds true.
}

\vskip2mm
{\bf Example 30.3.} Returning to Example 25.5, consider the transforms 
of the form
$$
L(t) = (1 - c \sin^m(t))\, e^{t^2/2}
$$
with an arbitrary even integer $m \geq 4$.
In this case, the conditions in Corollary 30.2 are met, and we obtain
the statement about the R\'enyi divergence of infinite order.

\vskip2mm
{\bf Example 30.4.} Put
$$
P(t) = (1 - 4\sin^2 t)^2 \, \sin^4 t.
$$
Then, $P(t) = O(t^4)$, implying that $P(0) = P'(0) = P''(0) = 0$. 
Note that $P(t)$ is $\pi$-periodic, and $h = \pi$ is the smallest period, 
although
$$
P(0) = P(t_0) = P(\pi) = 1, \quad t_0 = \pi/6.
$$

All the assumptions of Corollary 30.3 are fulfilled for sufficiently 
small $c>0$ with $h = \pi$, and we may check the condition (30.5). 
In this case, 
$$
P(t) = Q(t)^2, \quad Q(t) = (1 - 4\sin^2 t) \, \sin^2 t = 
\sin^2 t - 4\sin^4 t,
$$
so that $P''(t) = Q'(t)^2$ at the points $t$ such that $Q(t) = 0$, 
that is, for $t = t_0$. Hence $P''(t) = 0 \Leftrightarrow P'(t) = 0$. 
In our case,
$$
Q'(t) = 2 \sin t \cos t - 16 \sin^3 t \cos t = \sin(2t)\,(1 - 8 \sin^2 t),
$$
and $Q'(t_0) = -\frac{1}{2}\sqrt{3} \neq 0$. Hence $P''(t_0) \neq 0$, 
showing that the condition (30.2) is not fulfilled. Thus, the CLT 
with respect to $T_\infty$ does not hold in this example.

\vskip2mm
These two examples show that the continuity condition of $A''$ 
with respect $A$ in Theorem 29.1 may or may not be fulfilled in 
general in the class of strictly subgaussian distributions.
In other words, the convergence in $T_\infty$ is (strictly) stronger
than the convergence in all $T_\alpha$ simultaneously.

\vskip7mm
\section{{\bf The Multidimensional Case}}
\setcounter{equation}{0}

\noindent
Theorem 29.1 has a natural generalization to the multidimensional setting.
Let 
$$
Z_n = \frac{X_1 + \dots + X_n}{\sqrt{n}}
$$
denote the normalized sum of the first $n$ independent copies $X_k$ of 
a (sub-Gaus\-sian) random vector $X$ in $\R^d$ with mean zero and identity 
covariance matrix. As in dimension one, the log-Laplace transform 
$$
K(t) = \log \E\,e^{\left<t,X\right>}, \quad 
A(t) = \frac{1}{2} |t|^2 - K(t) \quad (t \in \R^d)
$$ 
represent $C^\infty$-smooth functions. We denote by $A''(t)$ the Hessian, 
that is, the matrix of second order partial derivatives of $A$ at the point $t$. Thus, 
$$
A''(t) = I_d - K''(t).
$$
Note that the matrix $K''(t)$ is positive definite, so that its determinant
${\rm det} K''(t)$ is positive.

We assume that:

\vskip4mm
1) \ $Z_n$ has density $p_n$ such that $T_\infty(p_n||\varphi) < \infty$ 
for some $n=n_0$;

2) \ $X$ is strictly subgaussian, that is, $A(t) \geq 0$ for all $t \in \R^d$.

\vskip4mm
\noindent
According to Theorem 17.2 and Corollary 21.2 on the necessary and sufficient 
conditions for the convergence in all $D_\alpha$ simultaneously with finite $\alpha$,
the conditions 1)-2) are necessary for the convergence $T_\infty(p_n||\varphi) \rightarrow 0$
or equivalently  $D_\infty(p_n||\varphi) \rightarrow 0$.
The following characterization was recently obtained in \cite{B-G4}.

\vskip5mm
{\bf Theorem 31.1.} {\sl For the convergence 
$T_\infty(p_n||\varphi) \rightarrow 0$, it is necessary and sufficient 
that the following two conditions are fulfilled:

\vskip3mm
$a)$ \ $A''(t) = 0$ for every point $t \in \R^d$ such that $A(t) = 0$;

$b)$ \ $\lim_{k \rightarrow \infty} A''(t_k)= 0$ for every sequence $t_k \in \R^d$
such that $|t_k| \rightarrow \infty$ and $A(t_k) \rightarrow 0$ as $k \rightarrow \infty$.
}

\vskip4mm
Similarly to (29.1), the conditions $a)-b)$ may be combined in the requirement
$$
\lim_{A(t) \rightarrow 0} A''(t) = 0 \ \ {\rm or \ equivalently} \ \ 
\lim_{A(t) \rightarrow 0} K''(t) = I_d.
$$
In \cite{B-G4} it was shown that these conditions may be stated in a formally weaker form

\vskip2mm
$a')$ \ ${\rm det} K''(t) = 1$ for every point $t \in \R^d$ such that $A(t) = 0$;

$b')$ \ $\lim_{k \rightarrow \infty} {\rm det} K''(t_k) = 1$ for every sequence $t_k \in \R^d$
such that $|t_k| \rightarrow \infty$ and $A(t_k) \rightarrow 0$ as $k \rightarrow \infty$.

\vskip3mm
It was already explained in the one dimensional situation that, assuming 
the strict sub-Gaussianity 2), the conditions $a)-b)$ may or may not hold in general. 
This shows that the convergence in $D_\infty$ is stronger than the convergence in
$D_\alpha$ simultaneously for all finite $\alpha$. Nevertheless, for a wide class of 
strictly sub-Gaussian distributions the Laplace transform possesses a separation-type property
\be
\sup_{|t| \geq t_0} \big[e^{-|t|^2/2}\, \E\,e^{\left<t,X\right>} \big] < 1 \quad
{\rm for \ all} \ t_0>0,
\en
which is a multidimensional generalization of the condition (27.2).
This is a strengthened form of condition 2), which entails properties $a)-b)$.

\vskip5mm
{\bf Corollary 31.2} (\cite{B-G4}). {\sl If a random vector X with mean zero and identity covariance
matrix satisfies $(31.1)$, then $T_\infty(p_n||\varphi) \rightarrow 0$ as $n \rightarrow \infty$.
}

\vskip5mm
On the other hand, the case of equality in the sub-Gaussian bound (31.1) is quite
possible, and one can observe new features in the multidimensional case. While in
dimension one, an equality 
$$
L(t) = \E\,e^{\left<t,X\right>} = e^{|t|^2/2}
$$
is only possible for a discrete set of points $t$, in higher dimensions the set of points where 
this equality holds may have dimension $d-1$. In order to clarify this behavior, 
one may consider the class of Laplace transforms which contain periodic components. 
Specializing Theorem 31.1 to this class, the general characterization may be simplified
in full analogue with Theorem 30.1.

\vskip5mm
{\bf Corollary 31.3} (\cite{B-G4}). {\sl Suppose that the function $\psi(t) = L(t) e^{-|t|^2/2}$ is
$h$-periodic for some vector $h \in \R^d_+$ $(h \neq 0)$. For the convergence
$T_\infty(p_n||\varphi) \rightarrow 0$ as $n \rightarrow \infty$, it is necessary
and sufficient that, for every $t \in [0,h]$,
$$
\psi(t) = 1 \, \Rightarrow \, \psi''(t) = 0.
$$
}

\vskip2mm
As for the proof of Theorem 31.1, let us only mention that the multidimensional 
situation turns out to be more complicated, since it requires a careful treatment of eigenvalues 
of the matrix $K''(t)$, when $A(t)$ approaches zero. Another ingredient in the proof 
is a quantitative version of the uniform local limit theorem -- the multidimensional extension 
of Proposition 28.2, which was recently developed in \cite{B-G3}.

\vskip5mm
{\bf Acknowledgement.}
The research has been supported by the NSF grant DMS-2154001 and 
the GRF -- SFB 1283/2 2021 -- 317210226.

\end{document}